\documentclass[fleqn,usenatbib]{mnras}

\usepackage{newtxtext,newtxmath}
\usepackage{booktabs}
\usepackage{wrapfig}
\usepackage[table,xcdraw]{xcolor}
\usepackage{multirow}
\usepackage{rotating,tabularx}

\usepackage[T1]{fontenc}
\DeclareRobustCommand{\VAN}[3]{#2}
\let\VANthebibliography\thebibliography
\def\thebibliography{\DeclareRobustCommand{\VAN}[3]{##3}\VANthebibliography}

\usepackage{graphicx}	
\usepackage{amsmath}	
\usepackage{tablefootnote}

\title[CoRoT-7 Planet system]{The impact of two non-transiting planets and stellar activity on mass determinations for the super-Earth CoRoT-7b}

\author[0000-0002-1715-6939]{
Ancy Anna John,$^{1,2}$\thanks{E-mail: aaj1@st-andrews.ac.uk }
Andrew Collier Cameron, $^{1,2}$
Thomas G. Wilson $^{1,2}$
\\
\\
$^{1}$SUPA, School of Physics \& Astronomy, University of St Andrews, North Haugh, St Andrews, KY169SS, UK\\
$^{2}$Centre for Exoplanet Science, University of St Andrews, North Haugh, St Andrews, KY169SS, UK\\
}

\pubyear{2021}

\begin{document}
\label{firstpage}
\pagerange{\pageref{firstpage}--\pageref{lastpage}}
\maketitle

\begin{abstract}
CoRoT-7 is an active star, whose orbiting planets and their masses have been under debate since their initial detection. In the previous studies, CoRoT-7 was found to have two planets, CoRoT-7b and CoRoT-7c with orbital periods 0.85 and 3.69 days, and a potential third planet with a period$\sim$9 days. The existence of the third planet has been questioned as potentially being an activity-induced artefact. Mass of the transiting planet CoRoT-7b has been estimated to have widely different values owing to the activity level of the parent star, the consequent RV “jitter”, and the methods used to rectify this ambiguity. Here we present an analysis of the HARPS archival RV (RV) data of CoRoT-7 using a new wavelength-domain technique, {\sc scalpels}, to correct for the stellar activity-induced spectral line-shape changes. Simultaneous modelling of stellar activity and orbital motions, identified using the ${\ell_1}$- periodogram, shows that {\sc scalpels} effectively reduce the contribution of stellar variability to the RV signal and enhance the detectability of exoplanets around active stars. Using {\sc kima} nested-sampling package, we modelled the system incorporating a Gaussian Process together with {\sc scalpels}. The resultant posterior distributions favoured a three-planet system comprising two non-transiting planets, CoRoT-7c and CoRoT-7d with orbital periods 3.697±0.005 and 8.966±1.546 days, in addition to the known transiting planet. The transiting planet CoRoT-7b is found to be a rocky super-Earth with a mass of M$_{b}$=6.06±0.65 $M_{\earth}$. The determined masses of M$ _{c}$=13.29±0.69 $M_{\earth}$ and M$_{d}$=17.14±2.55 $M_{\earth}$ suggest the non-transiting planets CoRoT-7c and CoRoT-7d to be structurally similar to Uranus and Neptune.

\end{abstract}




\section{Introduction}

Orbiting planets exert a gravitational pull on their host star, which results in a periodic variation of the star's velocity along the line of sight of observation. This wobbling effect can be inferred as radial velocity (RV) information, which is a measure of the associated wavelength shifts. Shortly after the detection of first exoplanet 51 Pegasi-b, \citet{1997ApJ...485..319S} recognized that stellar activity severely impacts the measurement of RVs. 
If the variable stellar surface phenomena influence the RV measurement (by modifying the line shape or its position), then exoplanet detectability and characterization are also affected.
These include stellar oscillations, granulation, spots, and faculae/plages, and long-term magnetic activity cycles\citep{1997ApJ...485..319S, 2011IAUS..276..527D}. The most confusing signals are those caused by the presence of active regions on the stellar
surface, which can show periodicities and amplitudes similar to the
ones generated by real planetary signals. Studies by \citet{2010A&A...513L...8F} and  \citet{2015ApJ...805L..22R} show that these signals may be harder to disentangle.

Although the stellar signal significantly impacts the detection and characterization, RVs have been widely used to detect exoplanets, especially for low mass planets \citep{2021arXiv210406072M}. 
  These low mass and/or long period planets induce RV semi-amplitudes similar to or smaller than those produced by stellar activity  \citep{2016AJ....152..204L,2017AJ....154..226D,2018AJ....155..203H}. Therefore, it is highly important to put effort in understanding and correcting for stellar activity, as it is essential for accurate RV detection and hence for characterizing interesting new exoplanets.
Instrument stability  is no longer the dominant obstacle for the precise and accurate mass determination of low-mass planets orbiting bright stars, considering the advanced latest generation of instruments. However, it is often limited by the intrinsic stellar variability \citep{1997ApJ...485..319S, 2001A&A...379..279Q,2014MNRAS.443.2517H}. So, caution is needed while characterizing planets discovered with the RV method around stars that are active/potentially active, as this method is highly prone to the uncertainty arising from the host star's activity.

Various analysis techniques have been developed to understand and mitigate the impact of stellar activity effects. The early efforts include decorrelation against proxy indicators of activity, or simply rejecting a planet candidate if the same periodicity was found in the FWHM (Full Width Half Maximum), BIS (Bisector Inverse Slope)
or in the chromospheric Ca II H\&K emission flux, as in the RV \citep[e.g.,][]{2010A&A...520A..93H}. \citet{2012MNRAS.419.3147A} used well-sampled light curves to predict the stellar variability expected in RV measurements. Known as the FF$^\prime$ method,  this approach uses the light curve and its first derivative to approximate the effects of spots and faculae, and requires no information about the rotation period or spot modelling. It does, however, require simultaneous high-precision photometry.
 Recently, Gaussian Process regression has been used to model the correlated noise induced by stellar activity in the RV signal, by fitting the orbital signal simultaneously \citep{2014MNRAS.443.2517H}. Although GP regression with a quasi-periodic kernel is successful at fitting rotationally modulated activity, other signals are also present.

 As noted by \citet{ 2015ApJ...814L..21D} and  \citet{2021A&A...653A..43C}, different systematics on the spectra such as micro-tellurics, tellurics, colour variation due to airmass and detector stitching can also affect the RV, on timescales very different from those modelled with a single GP kernel. YARARA \citep{2021A&A...653A..43C} and WOBBLE \citep{2019AJ....158..164B} were developed on this ground as post-processing pipelines to cope with these features in the spectra. YARARA uses Principal Component Analysis on HARPS spectral time series to correct for the known non-planetary systematics without using models for the different effects. WOBBLE is a data-driven method that can be used to infer the stellar spectra, telluric spectra and RVs simultaneously from the data \citep{2019AJ....158..164B}.


A complementary approach is to model the modes of variation present in the spectral line profile independently of their temporal behaviour, as explored by YARARA. While a planet produces a shift of all spectral lines, the stellar activity affects the shape of the spectral lines. The FourIEr phase SpecTrum Analysis (FIESTA, a.k.a. $\Phi$ESTA) \citep[e.g.,][]{2020MNRAS.491.4131Z, 2022arXiv220103780Z} uses the real and imaginary parts of the Fourier transform of the line profile to disentangle apparent RV shift due to spectral line-profile variability from  true Doppler shift.  Concurrently, \citet{2021AAS...23733204D} also developed machine learning techniques such as linear regression and neural networks to separate activity signals from true center-of-mass RV shifts using the average spectral line-shape variations, without any time-domain information. Another recent investigation \citep{2022MNRAS.512.5067K} employs Doppler imaging to simultaneously model the activity-induced distortions and the planet-induced shifts in the line profiles. It is challenging to differentiate between the effects induced  by a planet and stellar activity, as the latter has varied effects on individual spectral lines \citep{2015ApJ...814L..21D}. 

Keeping this in mind, a new algorithm ({\sc scalpels}- {\em Self-Correlation Analysis of Line Profiles for Extracting Low-amplitude Shifts }) has been developed by \citet{2021MNRAS.505.1699C} to decouple the RV information from the time domain aiming to separate the Doppler shifts caused by orbital motion from the apparent shifts caused by spectral line-shape variability triggered by stellar activity. By analysing orthogonal modes of variation in the translationally-invariant autocorrelation function of the line profile, this method of shape-shift separation offers precise Doppler detection and characterization of exoplanets around well-observed, bright main-sequence stars across a wide range of orbital periods, especially for lower-mass planets. 

\citet{2021MNRAS.505.1699C} demonstrated the algorithm using observations of the solar spectrum from HARPS-N. They successfully verified and validated that the algorithm can accurately detect multiple simulated planets injected into heliocentric solar observations, spanning a wide range of orbital periods. \citet{2022MNRAS.511.1043W} also used the {\sc scalpels} to clean up HARPS RVs for the stronger detection of two sub-Neptunes in the TOI-1064 system.

Here we present the first in-depth investigative analysis to study the potential of this approach to “clean-up” the RV of a stellar target  other than the Sun and thus to improve the fidelity of RV measurements. We apply {\sc scalpels} to the archival HARPS data on the CoRoT-7 system, which has been studied extensively with time-domain methods, but for which the precise system architecture remains a subject of debate. In Section \ref{Obs} we summarize the history of investigations of this system. In Section \ref{sec:scalpels} we describe the version of {\sc scalpels} employed here and its joint use with the ${\ell_1}$ periodogram method and {\sc kima} in the following subsections. In subsequent subsections of Section \ref{sec:analysis} we examine the evidence for a third planet in the system and the impact of the number of non-transiting planets on the planetary mass of transiting planet CoRoT-7b extensively, followed by a discussion and conclusion in Section \ref{sec:discussion} and \ref{sec:conclusion}.


\section{Target and Observations}
\label{Obs}

  
CoRoT-7 is a G9V type main-sequence star of V magnitude = 11.7, slightly cooler ($T_{\rm eff}$ = 5250 ± 60 K) and younger (1.2–2.3 Gyr) than our Sun \citep{2010A&A...520A..93H}. Here, we revisit the published HARPS RV measurements of CoRoT-7. \citet{2009A&A...506..287L} reported the detection of the first ever known transiting super-Earth, CoRoT-7b with an orbital period of 0.85 days and a measured radius of 1.68 ± 0.09$R_{\earth}$, which had the smallest exoplanetary radius at that time. Following this discovery, a 4-month intensive HARPS RV follow-up campaign was launched in order to measure the mass of CoRoT-7b. These observations were carried out under ESO Program IDs 082.C-0120, 082.C-0308(A) and 282.C-5036(A) using the HARPS instrument on ESO 3.6 m telescope at La-Silla. CoRoT-7 was observed for 3 nights in 2008 March and then observed continuously from 2008 November to 2009 February, spanning over 4 months.  In immediate course, \citet{2009A&A...506..303Q} revealed CoRoT-7c, the second planet with an orbital period of 3.69 d.

From the large amplitude of photometric modulation, the authors anticipated the variations in RV to be highly perturbed by the activity of the host star. Aiming to remove the RV variations caused by stellar activity, \citet{2009A&A...506..303Q} applied a harmonic decomposition to the data, preceded by a pre-whitening procedure. In this 'cleaning' procedure, the period of the stellar rotation signal is identified by means of a Fourier analysis, and a sine wave fitted with this period is subtracted from the data. The process of the removal of the next strongest signals (which are the sinusoids at integer multiples of the stellar rotation frequency) from the residuals was repeated until the noise limit was achieved.  The 0.85d signal associated with CoRoT-7b  was found to agree with the CoRoT transit ephemeris, and thus confirmed its planetary nature. In this significant analysis, the author claims, apart from the 0.85d and 3.69d signals, all other resultant signals to be linked to the harmonics of stellar rotation period. In the study which revealed the discovery of CoRoT-7b,  \citet{2009A&A...506..287L} performed a harmonic decomposition of the rotational period and up to the first three harmonics to filter out the activity signal from RV variations caused by the orbiting planet.
  
     
     
 \subsection{The second non-transiting planet candidate }
\label{sec:whycorot}

\begin{table*}
  \caption{A timeline of wide range of reported planet masses, especially for CoRoT-7b. It is evident that the mass of CoRoT-7c is nearly consistent. The measured orbital periods for all (proposed and confirmed) planets are also included.}
  \label{tab:parameters}
  \footnotesize
  
  \begin{tabular*}{\textwidth}{|c|c|c|c|c|c|c|}
 \hline
  \multicolumn{1}{|c}{} &  \multicolumn{3}{|c|}{Planetary masses}& \multicolumn{3}{c|}{Orbital periods} \\
   \hline
  Authors & CoRoT-7 b & CoRoT-7 c & CoRoT-7 d  & CoRoT-7 b & CoRoT-7 c & CoRoT-7 d \\
   \hline
  Leger et al.(2009)$^a$ & < 21$M_{\earth}$     &  --  &    --& 0.85 days  &--&--\\
 
 Queloz et al.(2009)$^b$&   4.8 ± 0.8$M_{\earth}$   & 8.4 ± 0.9$M_{\earth}$   & -- & 0.85 days & 3.69 days& --\\
 
 Hatzes et al. (2010)$^b$& 6.9 ± 1.4$M_{\earth}$&12.4 ± 0.42$M_{\earth}$ &16.7 ± 0.42$M_{\earth}$&0.85 days & 3.69 days& 9.02 days\\
 
 Ferraz-Mello et al. (2011)$^b$&8.0 ± 1.2$M_{\earth}$&13.6 ± 1.4$M_{\earth}$&--&0.85 days & 3.69 days&-- \\
 
 Boisse et al. (2011)$^b$ &  5.7 ± 2.5$M_{\earth}$&13.2 ± 4.1$M_{\earth}$ &--&0.85 days & 3.69 days& -- \\
 
Pont, Aigrain, \& Zucker (2011)$^b$ & 2.3 ± 1.8$M_{\earth}$&--&--&0.85 days& -- & --\\
 
Hatzes et al. (2011)$^b$&  7.4 ± 1.2$M_{\earth}$&--&--&0.85 days & 3.69 days& 9.02 days\\

Haywood et al. (2014)$^c$& 4.7 ± 0.9$M_{\earth}$ & 13.5 ± 1.08$M_{\earth}$&--&0.85 days & 3.69 days&8.58 days (deemed as activity)\\
 
Tuomi et al. (2014)$^d$& 4.8 ± 2.3$M_{\earth}$& 11.8 ± 4.1$M_{\earth}$&15.4 ± 6.1 $M_{\earth}$ & 0.85 days & 3.71 days & 8.89 days (unsure about origin)\\
Barros et al.(2016)$^e$& 5.5 ± 0.8$M_{\earth}$&-- & --& 0.85 days & -- & --\\
 
 Faria et al.(2016)$^c$& 5.5 ± 0.8$M_{\earth}$&12.6 ± 0.7$M_{\earth}$&--&0.85 days & 3.69 days&8.58 days (deemed as activity)\\
  \hline
  
  \end{tabular*}
  \footnotesize{Data used : $^a$ CoRoT photometric LRa01\& SOPHIE RV data,    $^b$ HARPS RV 2008-9},     $^c$ HARPS RV 2012,   $^d$ HARPS-TERRA RV 2008-9, $^e$ CoRoT photometric LRa06\\
  \end{table*}
  Subsequently, the presence of a potential third planet, CoRoT-7d, with an orbital period of 9.02d and a mass of 16.7±0.42 $M_{\earth}$ was disclosed by \citet{2010A&A...520A..93H}, making CoRoT-7 a compact system. \citet{2010A&A...520A..93H} applied the pre-whitening procedure to the BIS, FWHM and CaII H \& K emission lines (spectral quantities used as measures of the host star's activity) owing to their in-dependency to the planetary dynamics.
  As no significant signal came up at 0.85 and 3.69 days in any of these indicators, the planetary candidacies of CoRoT-7b \& CoRoT-7c were strengthened. 
  
  However, the calculated masses were not in good agreement with the previous studies. This discrepancy in the mass calculation could possibly be rooted in the difference in the methods adopted to cancel out the effect of stellar activity signals from the RV data.  Furthermore, no signal was found in the periodogram of activity indicators at 9.02 days, and thus this RV signal was attributed to a third-planetary companion in the system, CoRoT-7d. This signal was previously detected by \citep{2009A&A...506..303Q} as well, but had been attributed to a ‘two frequency beating mode’ arising from an amplitude modulation of a 61 days signal. It was therefore deemed to be associated with stellar activity, as this 61 days signal is quite close to twice the stellar rotation period. 
  Subsequently, \citet{2011MNRAS.411.1953P}  argued that the data cannot be used to search for additional (non-transiting) planets in the 3–10 d period range and that claims of the detection of such planets (‘CoRoT-7c’ and ‘CoRoT-7d’) do not withstand scrutiny. 
  
 Aiming to settle these arguments, simultaneous photometric and spectroscopic observations were obtained in 2012 from CoRoT and HARPS (ESO Program ID 088.C-0323) for 26 consecutive clear nights in a row from 2012 January 12 to February 6, with multiple well-separated measurements on each night adding up to 77 observations in total \citep{2014MNRAS.443.2517H}. These RV data were reprocessed in the way as the 2008–2009 data \citep{2009A&A...506..303Q} using the HARPS data analysis pipeline. Altogether, the system was observed for RV studies using HARPS in 2008, 2009 and 2012 with a total of 177 measurements. 

The goal of the 2012 observations was to observe the system with CoRoT and HARPS simultaneously, and to use photometry as a proxy for activity-driven RV variations. Along with updating the planetary parameters for CoRoT-7b, \citet{2014MNRAS.443.2517H} analysed this system further for the evidence of the additional planetary companion reported by \citep{2010A&A...520A..93H} at the period of 9.02 days.  Although, they found a strong peak in the 6-10 days range in the Lomb-Scargle periodogram, the marginal likelihoods calculated from their GP regression analysis favoured the two-planet solution over a three-planet one including the 9-day signal.  They concluded that this signal was more likely associated with the second harmonic of the stellar rotation at $\sim $7.9 d. However, they successfully confirmed CoRoT-7c along with CoRoT-7b and  improved their planetary parameters. \citet{2016A&A...588A..31F} carried out a model-comparison study employing GP regression to model the activity alongside simultaneous orbit fitting in a nested-sampling scheme using the full CoRoT-7 RV data set. While they found evidence for a weak signal at 9 days, the odds' ratio again favoured the 2-planet solution. \citet{2014arXiv1405.2016T} presented an analysis of the HARPS- {\sc TERRA} (Template-Enhanced RV Re-analysis Application) velocities of CoRoT-7 that suggested a system of two planetary companions (CoRoT-7b \& CoRoT-7c), possibly three (CoRoT-7d).

\subsection{The mass of transiting planet }
 
Table \ref{tab:parameters} lists the reported planet masses for CoRoT-7b spanning a wide range from 2.3$M_{\earth}$ to 8$M_{\earth}$, which is not ideal. This is plausibly due to the activity level of the star that contributes a significant amount of RV “jitter” and how the various methods correct for this ambiguity. It also depends on how many non-transiting Keplerian signals are included in the model. The method of removal of activity signal will also affect the RV amplitude, which in turn can bias the mass determination of companions. All the aforementioned mass determinations used the same HARPS RV data set, which clearly highlights the trouble in planets' mass determination around an active star like CoRoT-7 \citep{2011ApJ...743...75H}.

 Collectively, these studies outline the critical role of stellar activity and non-transiting signals in contaminating the RV observations not only in this particular system but also in general. Therefore, it is crucial to understand this role to enhance our ability to detect low-mass planets and thereby measure their masses precisely. The present study aims to redetermine the planet masses using the entire set of HARPS RV measurements of CoRoT-7 obtained so far. We explore techniques for determining the masses and orbital elements of planets discovered around active stars.
 We expect this study to offer precise RV measurements corrected for stellar activity and hence to resolve the existing debates on the number of companions in CoRoT-7 planetary system and to update their masses with improved precisions. 

\section{Methods}
\label{sec:methods}
\subsection{SCALPELS}
\label{sec:scalpels}

The analyses of the CoRoT-7 system architecture described above employed time-domain methods such as pre-whitening and GP regression to model the contribution of stellar activity to the RV signal. \citet{2021MNRAS.505.1699C} developed the {\sc scalpels} algorithm as an alternative, wavelength-domain method for separating Doppler shifts of dynamical origin from apparent velocity variations arising from variability-induced changes in the stellar spectrum. 


{\sc scalpels} seeks to decouple the effects of genuine dynamical Doppler shifts from spurious variations caused by line shape changes arising from stellar activity and instrumental systematics. 
{\sc scalpels} uses the translational invariance property of the autocorrelation function (ACF; Adler\&Konheim 1962) of the cross-correlation function (CCF) to isolate the effects of shape changes in the CCF from shifts.
In their analysis, the ACF, $A(\delta v)$ has been described as the expectation value of the vector cross-product of the CCF with itself at a sequence of lags $\delta v$:
\begin{equation}
A(\delta v) = {\rm E}(\mathbf{\mathrm{CCF}}(v) \cdot  \mathbf{\mathrm{CCF}}(v+\delta v))
\label{eq:acf}
\end{equation}

\citet{2021MNRAS.505.1699C} carried out a Principal-Component Analysis of the CCF and its ACF, using data from the HARPS-N solar telescope feed. The temporal variability of the CCF contains both line shape changes and Doppler shifts. They found that the majority of time variations of the CCF in general appear similar to those of the shift-invariant profile shape changes probed by the ACF.
The principal modes of variability in the CCF or ACF can be isolated by calculating the Singular-Value Decomposition (SVD) of the ensemble of CCFs or ACFs respectively, and can be expressed as: 
\begin{equation}
\mathbf{C}(v_i,t_j) = \left<\mathbf{C}(v_i)\right> + \mathbf{U}_C(t_j) \cdot {\rm diag}(\mathbf{S}_C)\cdot \mathbf{P}_C(v_i).
\label{eq:ccfSVD}
\end{equation}
\begin{equation}
\mathbf{A}(\delta v_i,t_j) = \left<\mathbf{A}(\delta v_i)\right> + \mathbf{U}_A(t_j) \cdot \rm{diag}(\mathbf{S}_A)\cdot \mathbf{P}_A (\delta \mathit{v_i}).
\label{eq:acfSVD}
\end{equation}
The perturbations resulting from solar activity were isolated by  projecting the RVs on to the time-domain subspace spanned by the amplitude coefficients $\mathbf{U}_A$  of the ACF’s principal components \citep{2021MNRAS.505.1699C}. On the contrary, dynamical shifts due to the planets are preserved when projected onto the orthogonal complement of time-domain subspace.

The variation of the shape-driven component of the RV with time can be obtained from the sum of scaled velocity contributions from all principal components of ACF, $\mathbf{v}_\parallel=\mathbf{U}_A\cdot\hat{\mathbf{\alpha}}$, where $\hat{\mathbf{\alpha}}$ is the vector of response factors obtained by taking the inner product $\mathbf{U}_A^T$ with the mean-subtracted time series of observed velocities. 
The transformation from observed to shape-driven velocities can be considered as a linear projection of the observed velocities into a subspace spanned by the columns of $\mathbf{U}$:
\begin{equation}
    \mathbf{v}_\parallel=\mathbf{U}_A\cdot
    \mathbf{U}_A^T\cdot(\mathbf{v}_{\rm obs}-\left<v\right>_{\rm obs}).
    \label{eq:acfRVproj}
\end{equation}
These shape-driven perturbations are strongly correlated with the observed RVs, as seen from the first panel of Figure \ref{fig:appcorrelation}, which could reliably reproduce the stellar activity dominated long-term and short-term fluctuations \citep{2021MNRAS.505.1699C}. Even though this offers a linear decorrelation, the time-coefficients of the scalpels basis functions exhibit time lags relative to the observed RV signals, which needs to be addressed.

$\mathbf{v}_\perp $ is defined as the residual velocity which lies outside the $\mathbf{U}_A$ subspace, and can also be considered as a projection into the orthogonal complement of the ACF subspace, where the planet signals are sought: 
 \begin{equation}
    \mathbf{v}_\perp =(\mathbf{v}_{\rm obs}-\left<v\right>_{\rm obs}) - \mathbf{v}_\parallel 
    =(\mathbf{I}-\mathbf{U}_A\cdot\mathbf{U}_A^T)\cdot(\mathbf{v}_{\rm obs}-\left<v\right>_{\rm obs})
    \label{eq:perpproj}
\end{equation}

To avoid overfitting, a reduced-rank version of $\mathbf{U}$ is employed. \citet{2021MNRAS.505.1699C} describe  the determination of optimal rank using a leave-one-out cross-validation (LOOCV). This method is found to be efficient in identifying the number of leading columns of $\mathbf{U}$ that contains significant profile information.

In this study, we apply this variant of the {\sc scalpels} method to the CoRoT-7 system together with a GP regression. The host star's high level of stellar activity has been a persistent obstacle to previous efforts using other methods to determine the number of non-transiting planets present, and to determine their masses. The system's intrinsic scientific interest complements its suitability as a test target of the {\sc scalpels} method.

\subsection{ \texorpdfstring{$l_{1}$}{l1}-periodogram}
\label{sec:l1_intro}

The ${\ell_1}$- periodogram \citep{2017MNRAS.464.1220H} is primarily designed to identify the sparsest set of orbital signals that fit the observed RVs. This sparsity is achieved by simultaneously fitting Keplerian signals on an optimally sampled grid of frequencies to the data, regularized by minimizing the $\ell_1$ norm of their amplitudes.
\citet{2017MNRAS.464.1220H} present the algorithm as a tool to search for a representation of the input signal as a sum of a few sinusoidal components, that is, a representation which is sparse in the frequency domain. 

As the primary step, the data is normalized to the mean of the RV data to get an estimate of stellar noise. The selection of frequency grids ${\omega}$, covariance matrix $\mathbf{V}$ and the width of averaging interval ${\eta}$ are done in succession.  This algorithm suggests that RV measurements can be well approximated by the linear combination of a few vectors of $e^{-i {\omega}t}$ and $e^{i\omega t}$, which spans the columns of the $\mathbf{A}$ matrix.

The noise is assumed to be drawn from a Gaussian with the covariance matrix $\mathbf{V}$. 
If the stellar rotation information $P_{\rm rot}$ is given, the covariance matrix uses a quasi-periodic kernel. The contribution of activity to the covariance matrix is then modelled as the product of an exponential correlation term and a quasi-periodic term.

\begin{equation}
    {V}^{({\rm act})}_{ij} = \exp \left(-\frac{(t_{i}-t_{j})^2}{2\tau^2}\right) \left(0.5 + {\rm cos} \left(\pi\frac{(t_{i}-t_{j})}{P_{\rm rot}}\right)\right)
\end{equation}

The full covariance matrix including the noise terms can be then expressed as:
\begin{equation}
    \mathbf{V} = \mathbf{\delta}_{\rm k} + {\sigma_R^2} {V}^{({\rm act})}_{ij} + \mathbf{I} {\sigma_W^2} + {\sigma_C^2}
\end{equation}
where $\mathbf{\delta_{k}}$ is a diagonal matrix spanned by the RV uncertainties and $\sigma_{W}$ is the white noise amplitude, which we considered as 0.1 ms$^{-1}$. $\sigma_{R}$ stands for the red noise, which is 0.5 ms$^{-1}$ here multiplied with an identity matrix $\mathbf{I}$. In this simple covariance model, a block-diagonal calibration-noise covariance component ${\sigma_{C}}$ is also included to account for the night-to-night calibration errors, which are known to be present in HARPS data at the $\sim$0.5 ms$^{-1}$ level.


A remarkable advantage of the method is the use of $\ell_1$ norm weighting to find periodicity in unevenly sampled signals. Even without any prior knowledge of the number of planets, the intrinsic signal information can be efficiently decoded. The algorithm assigns an amplitude to every frequency on the grid. It then tensions the goodness of fit to the data (using $\chi^2$, which is the "$\ell_2$ norm") against the $\ell_1$ norm (median absolute deviation) of all the amplitudes on the frequency grid. The goal is to identify the minimal set of sinusoids that give an optimal fit to the data. The optimal solution is then found iteratively. 

Moreover, we also used its capability to fit a set of externally-determined basis functions simultaneously with the orbit fit. If the columns of the reduced-rank matrix $\mathbf{U}_A$ are used for this purpose, this is equivalent to the simultaneous sinusoidal orbit fitting method described in Section \ref{sec:simultaneous_intro}. This inter-operability with the {\sc scalpels} enables us to perform shape-shift signal separation simultaneously with identifying the minimal set of orbit signals present.


\subsection{Simultaneous sinusoidal planet orbit fitting}
\label{sec:simultaneous_intro}

We employed the shape-signal separation simultaneously with fitting the orbits, given prior knowledge of the orbital periods obtained from the $\ell_1$-periodogram analysis. A simultaneous modelling was chosen to be the best way to figure out the orbital solution, as a simple periodogram has the limitation of fitting only a single sinusoid per frequency sample, so that the interaction between various signals may lead to inaccurate amplitude estimation. To determine the impact of the {\sc scalpels} signal separation, \citep{2021MNRAS.505.1699C} used this approach in the solar RV data, with properly injecting a few planet signals that should be orthogonal to all elements of $\mathbf{U}_A$. This prevented the signals from being partially absorbed during the {\sc scalpels} projection. 

Estimation of parameters and signal separation can be achieved in a single linear calculation, once the candidate signal periods have been determined via periodogram search or through prior knowledge of transits. In the case of CoRoT-7, we have an estimate of the periods of the three candidate signals from the blind periodogram search and $\ell_1$-periodogram, while the period and phase of the CoRoT-7b signal were obtained more precisely from transit observations.

As described by \citet{2021MNRAS.505.1699C}, the net orbital velocity vector $\mathbf{v}_{\rm orb}$ for a set of $n$ planet signals can be modelled as the product of a set of coefficient pairs $\mathbf{\theta}_{\rm orb} = \{A_1, B_1, \cdots, A_n, B_n\}$ with an array of time-domain function pairs
$\mathbf{F} = \{\cos\omega_1 t_j, \sin\omega_1 t_j, \cdots, \cos\omega_n t_j, \sin\omega_n t_j\}$, where $\omega_k$ is the orbital frequency of the $k^{th}$ planet:
\begin{equation}\label{orbit}
\mathbf{v}_{\rm orb} = \mathbf{F}\cdot \mathbf{\theta}_{\rm orb}
\end{equation}

If we consider the planet orbits to be circular, $\mathbf{v}_{\rm orb}$ from Equation \ref{orbit} could then provide a complete model of the RV data as the sum
of shift (planet)-driven velocity variations and the shape (stellar activity)-driven velocity variations \citep{2021MNRAS.505.1699C}. Treating this as a least-square problem enabled us to solve for the unknowns, the amplitudes, and phases of $\mathbf{F}$.  The vector  ${\mathbf{\theta}_{\rm orb}}$ can then be evaluated by minimizing the $\chi^2$. Similar Implementation has been employed by \citep{2022MNRAS.511.1043W} as well.  The detailed algorithm is given in Appendix \ref{sec:appsimul}.

\subsection{Trans-dimensional Nested Sampling using \emph{kima}}

We used the {\sc kima} package of \citet{2018JOSS....3..487F} to sample  from the posterior distribution of the orbital model parameters. {\sc kima} employs a Diffusive Nested sampling algorithm of \citet{2010ascl.soft10029B} for calculating the evidence or fully marginalized likelihood not only for a model with a fixed number of Keplerian signals $N_{p}$, but also after marginalizing over $N_{p}$. The joint posterior distribution was calculated using the prior ${p(\Theta \,| \,M)}$ , likelihood ${p(D\,|\,\Theta,M)}$ and evidence information ${p(D\,|\,M)}$ from the Bayes theorem:
 \begin{equation}
    \mathbf{p}(\Theta \,| \,{D,M}) = \frac{\mathbf{p}(\Theta \,| \,M) \,\,\mathbf{p}(D\,|\,\Theta,M)}{\mathbf{p}(D\,|\,M)}
 \end{equation}
 where $M$ is the model with $\Theta$ as the vector of all considered parameters and $D$ is the RV data series.
Undertaking a similar approach to \citet{2014MNRAS.443.2517H} and  \citet{2016A&A...588A..31F}, we also included a quasi-periodic Gaussian process(GP) to account for the correlated noise with the objective of investigating the improvement in the posterior when using the {\sc scalpels} shift RVs instead of the raw RVs uncorrected for the shape variations. 

The kima covariance matrix also includes a white-noise term. Unlike the ${\ell_1}$- periodogram, kima does not yet model calibration noise by imposing correlations between observations made on the same night. Nonetheless, this white-noise contribution serves a similar purpose and enables us to optimise the amplitude of the calibration noise used in the ${\ell_1}$- periodogram. The quasi-periodic component of the GP is defined by the kernel $k(t,t^\prime)$ (a.k.a. covariance function):

\begin{equation}\label{GPkernel}
{k(t,t^\prime)} = {{\eta_{1}}^2} \rm exp \left(-\frac{(t-t^\prime)^2}{2{\eta_{2}}^2} - \frac{2}{\eta_{4}^2} \rm sin^{2} \left(\frac{\pi(t-t^\prime)^2}{{\eta_{3}}}\right) \right)
\end{equation}
The hyperparameters of $k(t, t^\prime)$ are represented by the variables $\eta_{1}$, $\eta_{2}$, $\eta_{3}$ and $\eta_{4}$. $\eta_{1}$ and $\eta_{2}$ define the amplitude of correlation between RVs at different time separations and the evolution timescale of active regions. The correlation timescale $\eta_{3}$ reflects the stellar rotation period, $\eta_{4}$ governs the variation timescale of GP latent functions in relation to $\eta_{3}$. Smaller values of $\eta_{4}$ indicate more short-scale structures within a single stellar rotation period. We used this GP to model the correlated noise occurring on time-scales of the order of stellar rotation period and its harmonics.

We also computed the False Inclusion Probability (FIP) and True Inclusion Probability (TIP) from the joint posterior distribution of orbital parameters and $N_{p}$. \citet{2021arXiv210506995H} introduces the FIP as the probability that there is no planet with period $P$ in a given frequency interval $I$. On the other hand, the TIP is defined as the counterpart of the FIP, that is, the probability that there is a minimum of one planet with $P$ $\epsilon$ $I$. 
The TIP is  calculated by averaging the planet detection over the possible number of planets, $\frac{N_{p}(i)}{\Sigma \, \,N_{p}}$.  \citet{2021arXiv210506995H} presents this detection criterion as an efficient way to evaluate the reliability of significance levels, by effectively accounting for aliases and favouring to discard the presence of planets under a certain confidence level.

\label{sec:FIP} 

\begin{figure*}
    \centering
    \includegraphics[width=1\columnwidth]{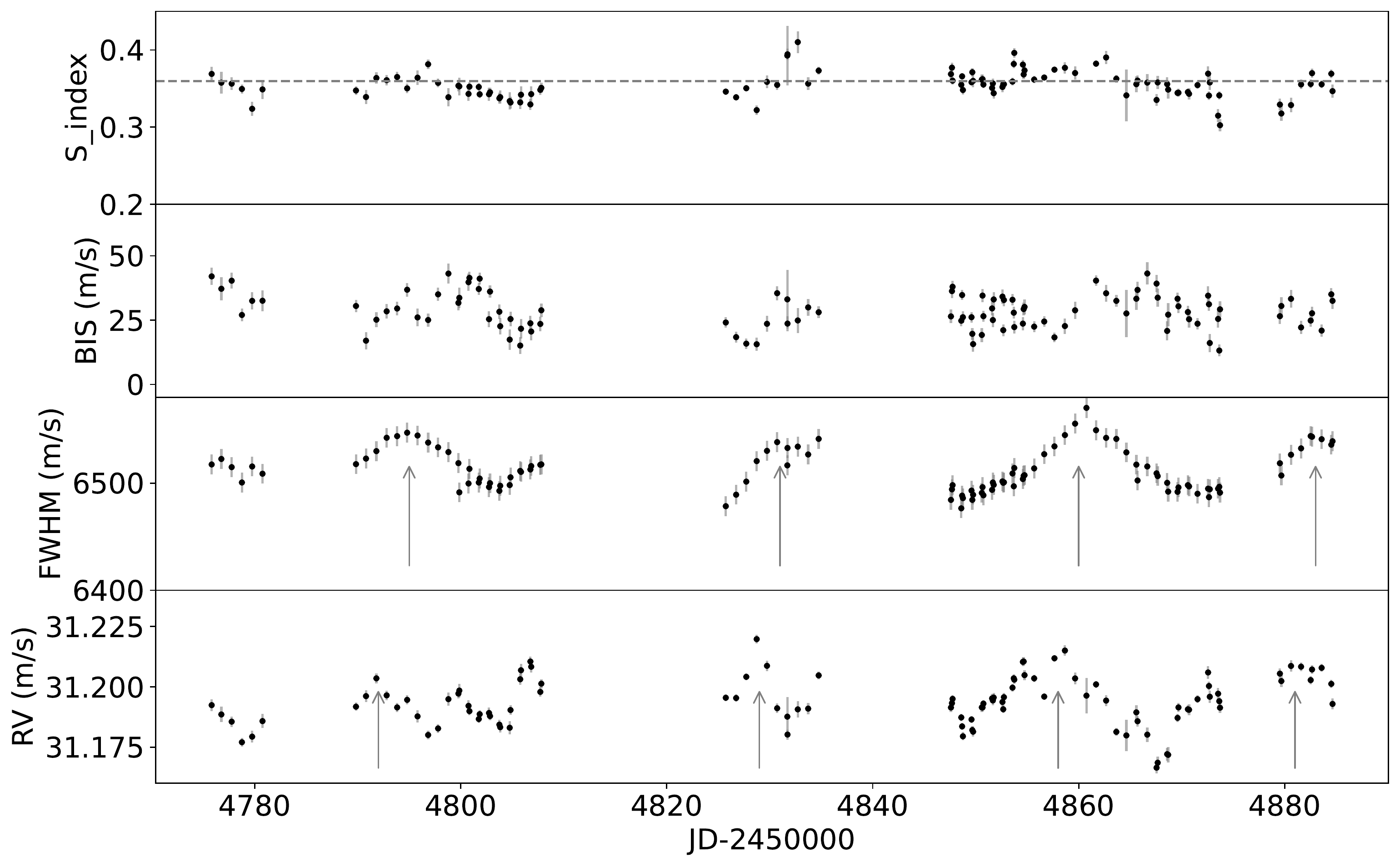}
    \includegraphics[width=1\columnwidth]{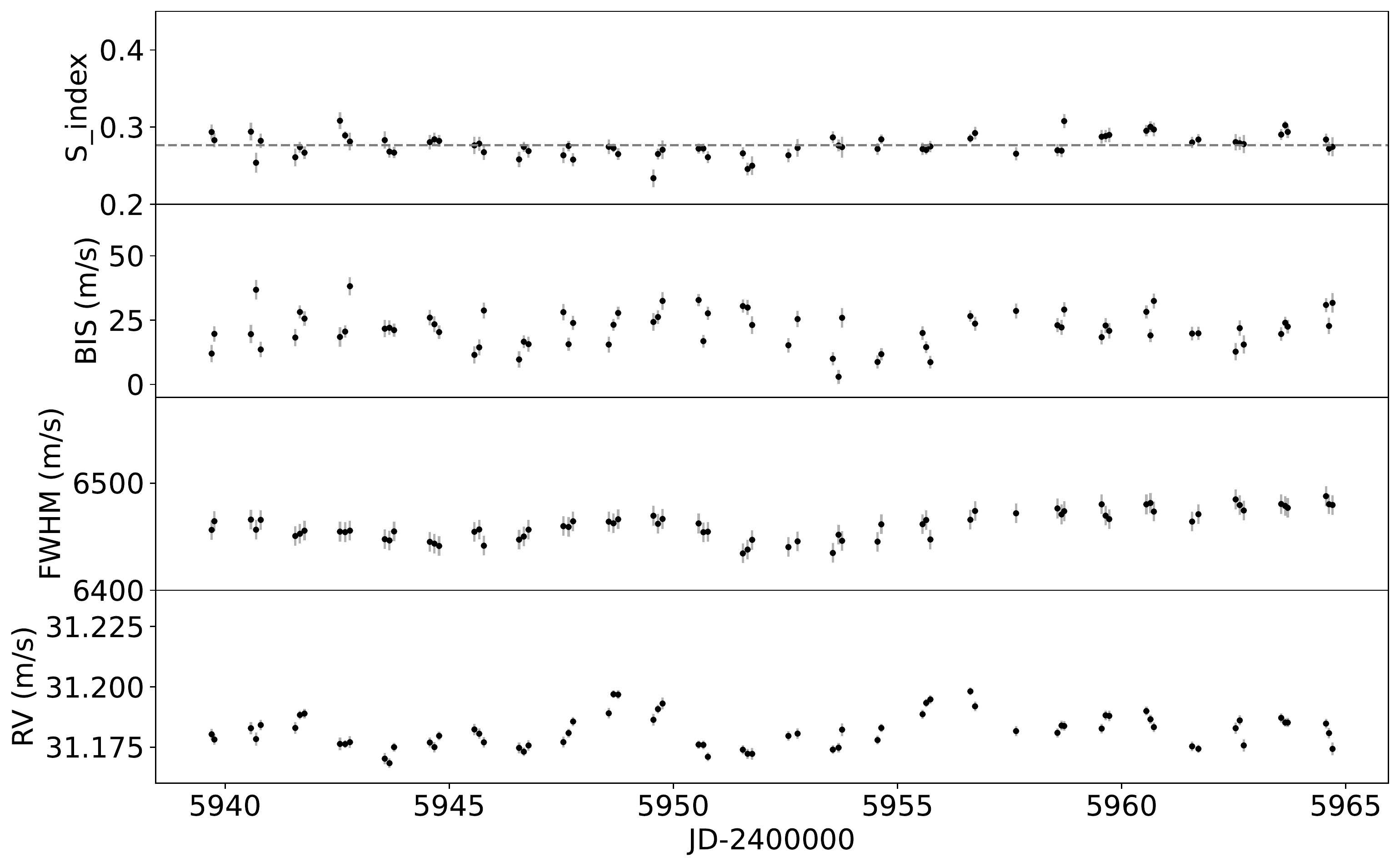}
    \caption{The spectroscopic information measured during the HARPS run spanning over November-2008 and February-2009 is given in the left figure. The RV, FWHM, Area, BIS and S-index data have been plotted against the Barycentric Julian date and shown in each panels from bottom to top. The figure at the right shows the respective representation of spectroscopic data over the span of the 2012 HARPS run. The dashed grey lines running horizontally in the top panels of both figures show the mean values of S-index, indicating the difference in the level of stellar activity during both observing runs. There are arrows given in the RV, FWHM and Area panels of the left figure, pinpointing that the RV peaks occur $\sim$ 2 days before the FWHM and Area peaks. The respective uncertainties are shown using grey error bars.}
    \label{fig:activity}
\end{figure*}

\subsection{Stacked Bayesian General Lomb-Scargle Periodograms}
\label{sec:sbgls_intro} 
 
 To probe the statistical implication of the planet candidate signals, a method of stacking the Bayesian General Lomb-Scargle (BGLS) periodograms has been used. This tool was developed for identifying the periodicities caused by stellar activity, and to show how it can be used to track the signal-to-noise ratio (SNR) of the detection over time \citep{2017A&A...601A.110M}. Adding in more observations, should increase the power or SNR measure of the signal, if the signal is real and coherent. On the contrary, if the signal is incoherent or short-lived, adding more data would cause the SNR or Lomb-Scargle power to decrease \citep[e.g.,][]{2011ApJ...726...73H,2013AN....334..616H,2017A&A...601A.110M}. Hence, signals arising from stellar activity can be identified by monitoring their variable and incoherent nature.   To track the significance of periodic signals, the BGLS periodogram was calculated for a smaller data subset and recalculated repeatedly with subsequent addition of observations. 

\citet{2017A&A...601A.110M} recommend that one should not solely rely on this approach to establish the planetary nature of a signal. This can be used as an additional test to reinforce a finding. We are aiming is to devise a workflow by combining the methods mentioned in Section~\ref{sec:scalpels},  \ref{sec:l1_intro}, \ref{sec:simultaneous_intro}, \ref{sec:FIP} and \ref{sec:sbgls_intro}.

\section{Data analysis and results}
\label{sec:analysis}

\subsection{Time series analysis}
\label{sec:act}

In addition to measuring the RV, the HARPS Data Reduction Software returns several shape diagnostics for the CCF and flux indices for known activity-sensitive spectral lines. The RV data and supplementary information including the S-index, BIS, Area and FWHM from 2008-9 and 2012 campaigns are shown in left and right panels of Figure ~\ref{fig:activity}. The quantity 'Area' is the product of the FWHM of the CCF and its contrast, i.e., the fractional depth of the CCF measured relative to the continuum level. The y-axis values for each parameter in both observing seasons are set to have the same limits for easy comparison. 

The supporting measurements of the 'activity indicators' such as the FWHM and BIS of CCF and the Ca II H and K values from HARPS allow us to examine the intrinsic stellar variability. \citet{2009A&A...506..303Q} suggests that one can check two simple characteristics to look for changes in the shape of CCF: the width parameter FWHM and its BIS, computed following \citet{2001A&A...379..279Q}.  The chromospheric activity is typically quantified by the S-index, as noted by \citet{1978ApJ...226..379W}. The S-index varies linearly with the chromospheric emission flux in the cores of the Ca II H and K lines in the near ultraviolet \citep{2016A&A...596A..31S}.  From both panels of the Figure ~\ref{fig:activity}, the mean value for S-index can be calculated as 0.3569 in the 2008-9 season and  0.2759 in 2012 and is denoted by the grey horizontal dashed line. The lower mean value in 2012 indicates that the star was less active during 2012.

    

Similarly, the RV measurements possessed an RMS scatter of  10 ms$^{-1}$ in the 2008-9 run and a lesser scatter value of about 6.8 ms$^{-1}$ in 2012, showing that the amplitudes of RV variability were lower in 2012 than 2008-9. The mean level of RV amplitudes were comparable (31.193 and 31.182) during the two campaigns. However, both sets exhibit obvious multi-periodic variability structures. 
The FWHM variations exhibit a fairly smooth periodic pattern, with a period of $\sim$23 days in both seasons, as seen in \citet{2009A&A...506..303Q}.

A close inspection of  Figure ~\ref{fig:activity} shows a temporal offset between FWHM, Area and the RV. The RV maxima occur $\sim$2 days prior to those of FWHM and Area. In the Sun, \citet{2021MNRAS.505.1699C} observed similar time-lags between  the RV, FWHM and BIS. Such temporal shifts between RV and several activity proxies were reported in previous studies as well \citep[e.g.,][]{2014A&A...566A..35S,2009A&A...506..303Q}. While  \citet{2009A&A...506..303Q} suggest these time-structured variations between these two parameters to be associated with star-spot related variability in CoRoT-7, \citet{2021MNRAS.505.1699C} note that the same phenomenon is present even when the RV signal
is dominated by facular suppression of the convective blueshift. \citet{2014MNRAS.443.2517H} detail the suppression of convective blueshift by active regions surrounding star-spots to have a much greater impact on RV than the flux blocked by star-spots, particularly in slowly rotating stars like Sun and CoRoT-7.

The BGLS periodograms for the above-mentioned activity indicators over the entire span of observation are given in Figure \ref{fig:bgls_act}.



\subsection{Blind Periodogram search}
\label{sec:blindpgm}

\begin{figure}
    \centering
    \includegraphics[width=1\columnwidth]{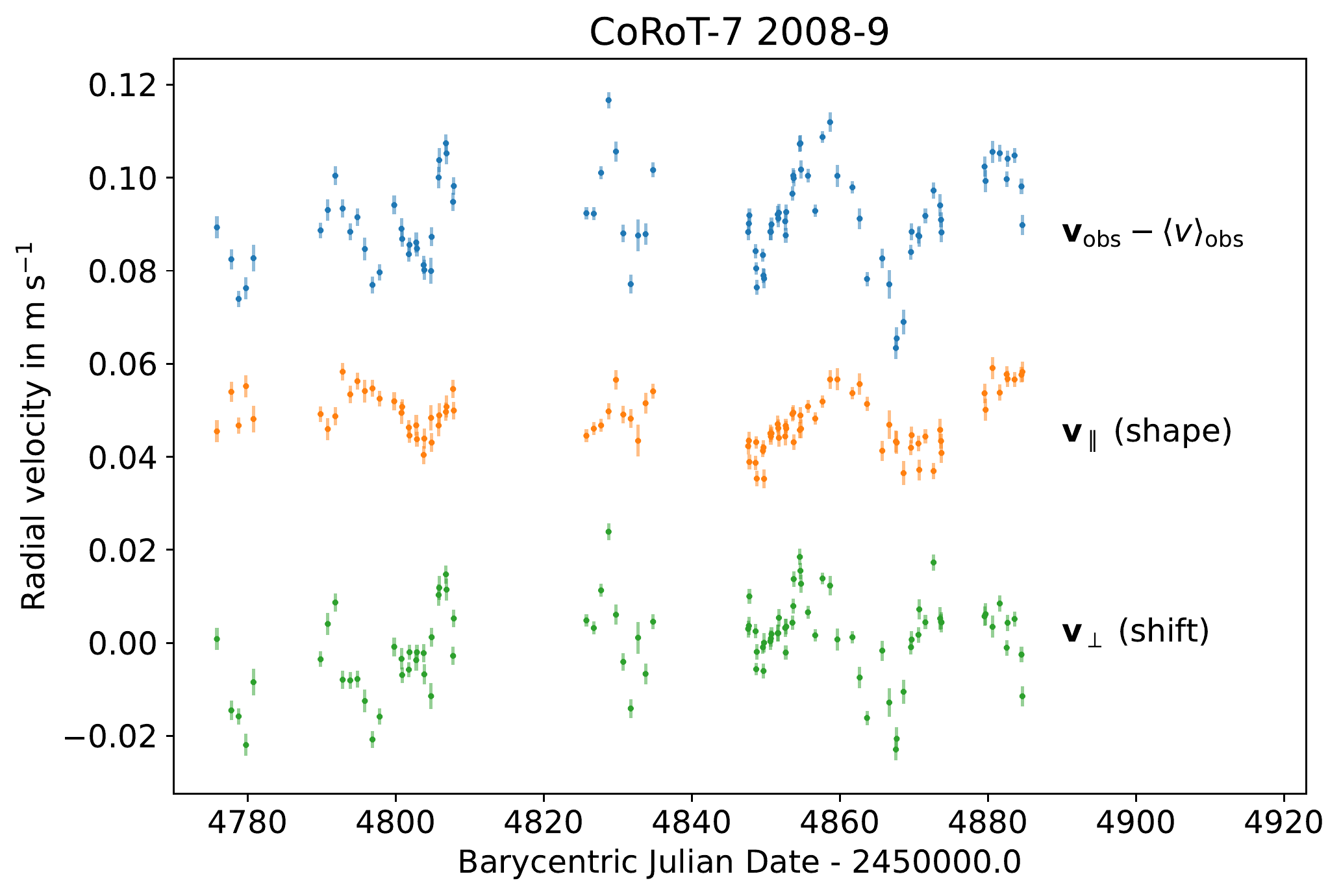}
    \includegraphics[width=0.97\columnwidth]{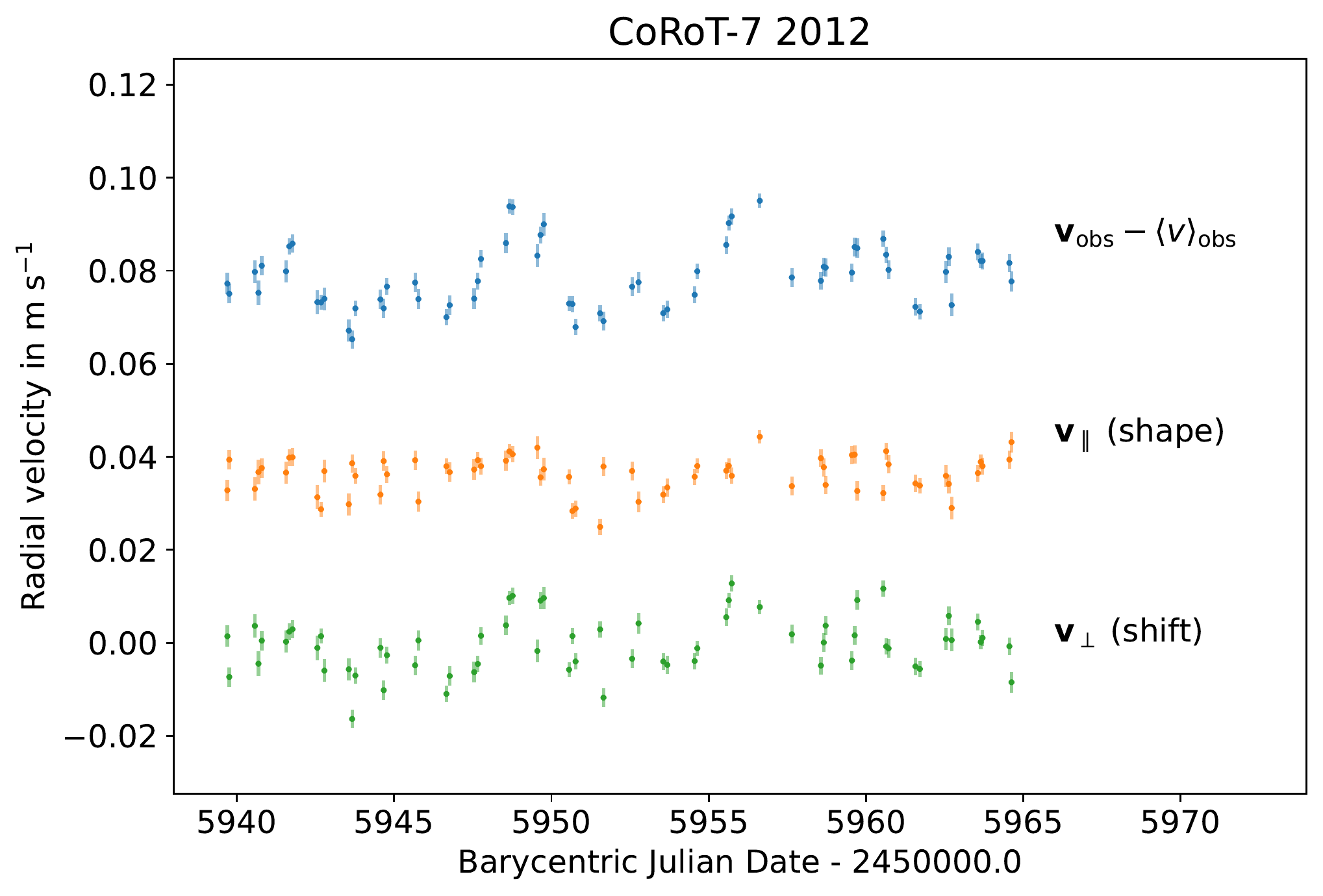}
    \caption{The top panel corresponds to the data from 2008-9 run, while the bottom panel shows 2012 data. In both panels, the blue curve at the top is the barycentric RV subtracted from its own mean. The middle orange curve represents the shape-driven component obtained from the {\sc scalpels} projection, while the green curve manifests the 'cleaned' RV, which is the shift-driven component. Please note that an offset of $\pm 60$~m~s$^{-1}$ is introduced for clarity. The relatively small uncertainties are shown as semi-transparent error bars to avoid overcrowding.} 
    \label{fig:RVjd}
\end{figure}

We applied {\sc scalpels} to the barycentric RVs of CoRoT-7, and examined the results using a Generalized Lomb-Scargle (GLS) periodogram \citep{2009A&A...496..577Z} for different subsets of the data (2008-9, 2012 and combined), given the varied level of activity in both seasons as discussed in Section \ref{sec:act}. The first two curves in Figure~\ref{fig:RVjd}, shown in blue and orange in all panels, show the observed barycentric RVs from which their own mean is subtracted ($\mathbf{v}_{\rm obs}-\left<v\right>_{\rm obs}$), and the shape-driven ($\mathbf{v}_\parallel$ ) RV perturbations derived using the {\sc scalpels} projection respectively. The green curve shows the cleaned shift-driven velocities, obtained when the shape-driven variations are subtracted from the observed RVs. Histograms showing the reduction of RMS scatter after separating the shape-driven signals are presented for all three analyses in Section \ref{sec:appblind}.

\begin{figure}
    \centering
    \includegraphics[width=0.9\columnwidth]{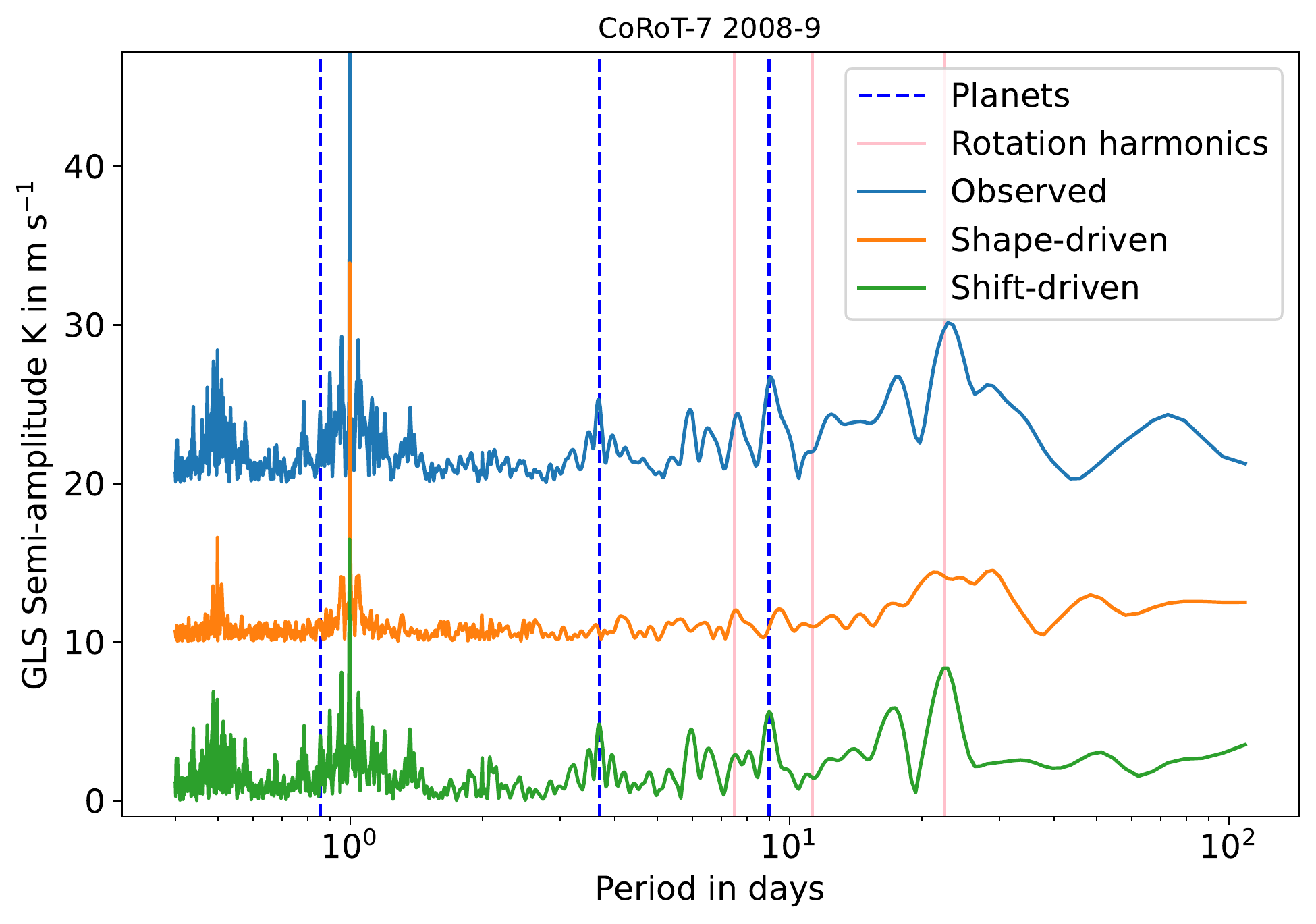}
    \includegraphics[width=0.9\columnwidth]{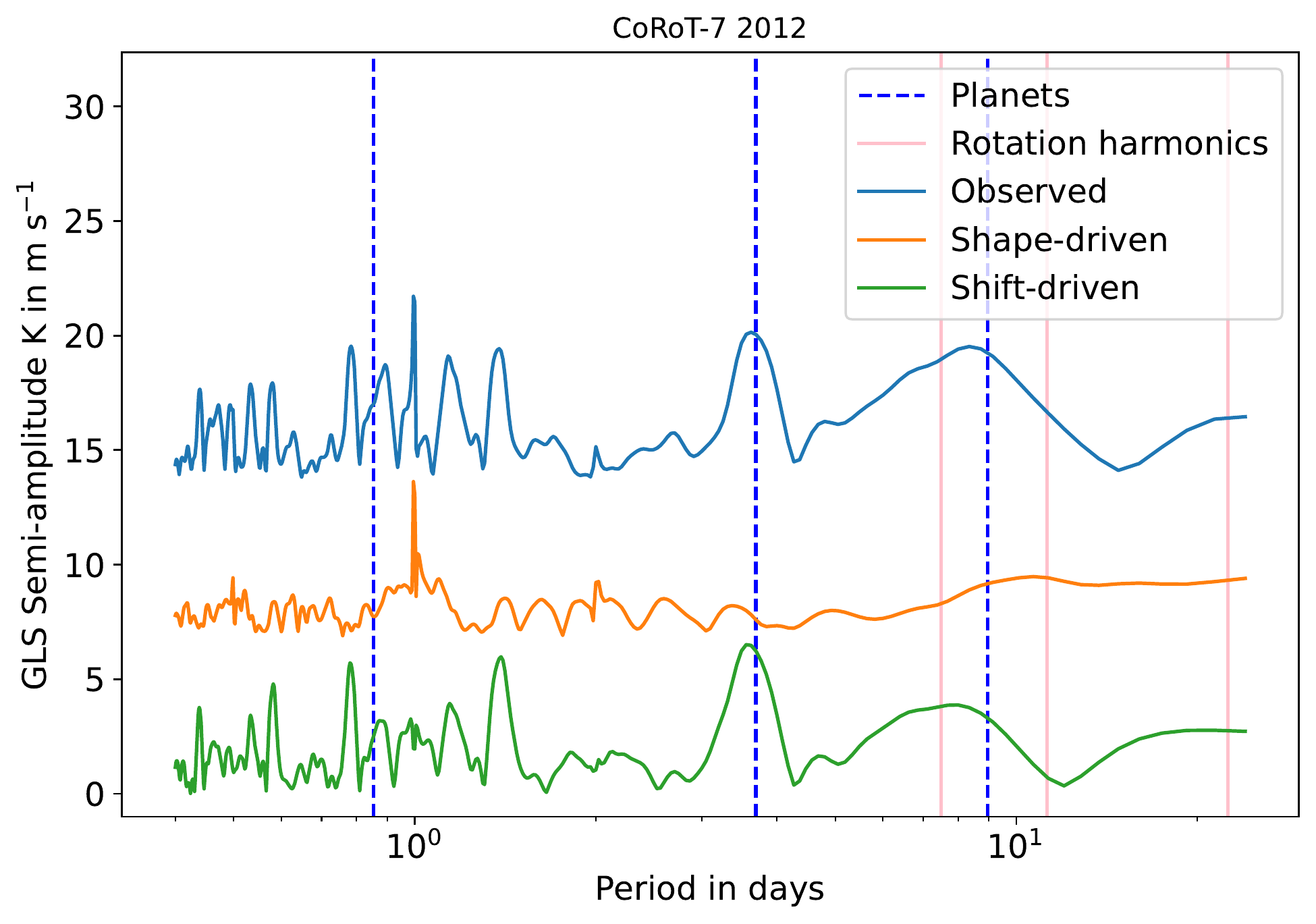}
    \includegraphics[width=0.9\columnwidth]{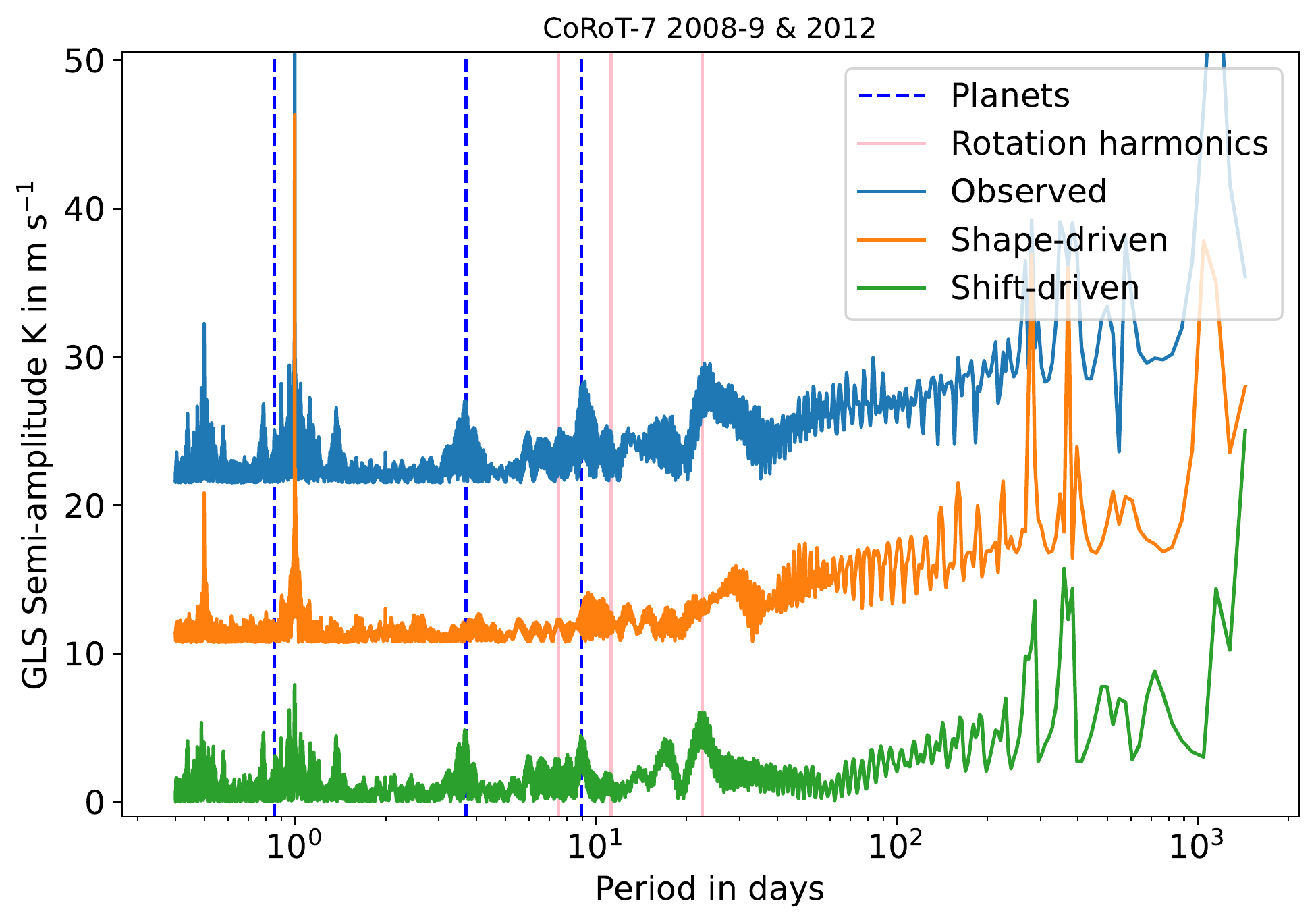}
    \caption{The left, middle and right figures represents 2008-9, 2012 and combined data sets respectively. In each figure, the top periodogram (blue) is for the measured velocities derived from the CCF. The middle trace (orange) is for the shape-driven velocities ($\mathbf{v}_\parallel$ ) obtained from the {\sc scalpels} projection. The third periodogram (green) is the shift-driven velocities ($\mathbf{v}_\perp$ ) remaining after the subtraction of shape-driven velocities from the observations. Light blue dotted vertical lines denotes the orbital periods of all confirmed and potential planetary companions}
    \label{fig:blind}
\end{figure}

Along with the confirmed planet signals with 0.85 day and 3.69 day orbital periods, a third signal with a period of 8.96 days and an amplitude of $\sim$4 ms$^{-1}$ also appears in all shift-driven periodograms, with no significant counter-part in the shape-driven periodogram as seen from Figure \ref{fig:blind}. This signal, noted previously by \citet{2011ApJ...743...75H} as being attributable to a third planet CoRoT-7d, was suppressed in the GP regression analysis of \citet{2014MNRAS.443.2517H}, attributing it to be more likely an activity signal. However, we found no significant signal at 8.96 days period in the activity-induced shape-driven periodogram, despite the dataset chosen. This argues against an activity-related origin. 


A curious feature of these periodograms is that in the 2012 and combined data, the spurious 1-day periodicity seen in in the raw observations is almost entirely accounted for in the shape signal. In the 2012 season, the target was observed 2 or 3 times per night over 26 consecutive clear nights. The target was therefore observed at a range of airmasses an a clear correlation was observed between the airmass and one of the SCALPELS basis vectors ($\mathbf{U}_2$). We consider it possible that chromatic extinction could produce a subtle change in CCF shape with airmass, and hence with local sidereal time, despite the care with which the HARPS DRS fits the blaze function and normalizes the continuum prior to cross-correlation. A detailed investigation is, however, beyond the scope of this study.

A strong peak in the combined observed periodogram shows up at around 23 days with an amplitude of 6.42 ms$^{-1}$, which is the stellar rotation period. Although a shift signal at this period is present, the appearance of a shape-driven signal at the same period casts doubt on a planetary interpretation of this signal. While the presence of a shift signal without a counterpart in the shape signal  points strongly to a planetary origin, we cannot completely ignore the possibility of some form of activity in CoRoT-7 producing a shape change in the profile that closely mimics a shift. Therefore, we have to investigate further the nature of the 23 days signal.

The complex frequency structure of the periodogram for the combined 2008-9 and 2012 data (Figure \ref{fig:blind}) arises from both cycle-count uncertainty and aliasing. In principle, all of this structure should be attributable to a finite number of orbital signals combined with a quasi-periodic activity signal and noise. Given the difficulty in interpreting the system using standard GLS periodograms, specially when the window function is complex and there are multiple, non-harmonic signals present, we recommend performing further analyses to understand similar complex systems to overcome these inherent problems.

\begin{figure}
    \centering
    \includegraphics[width=1\columnwidth]{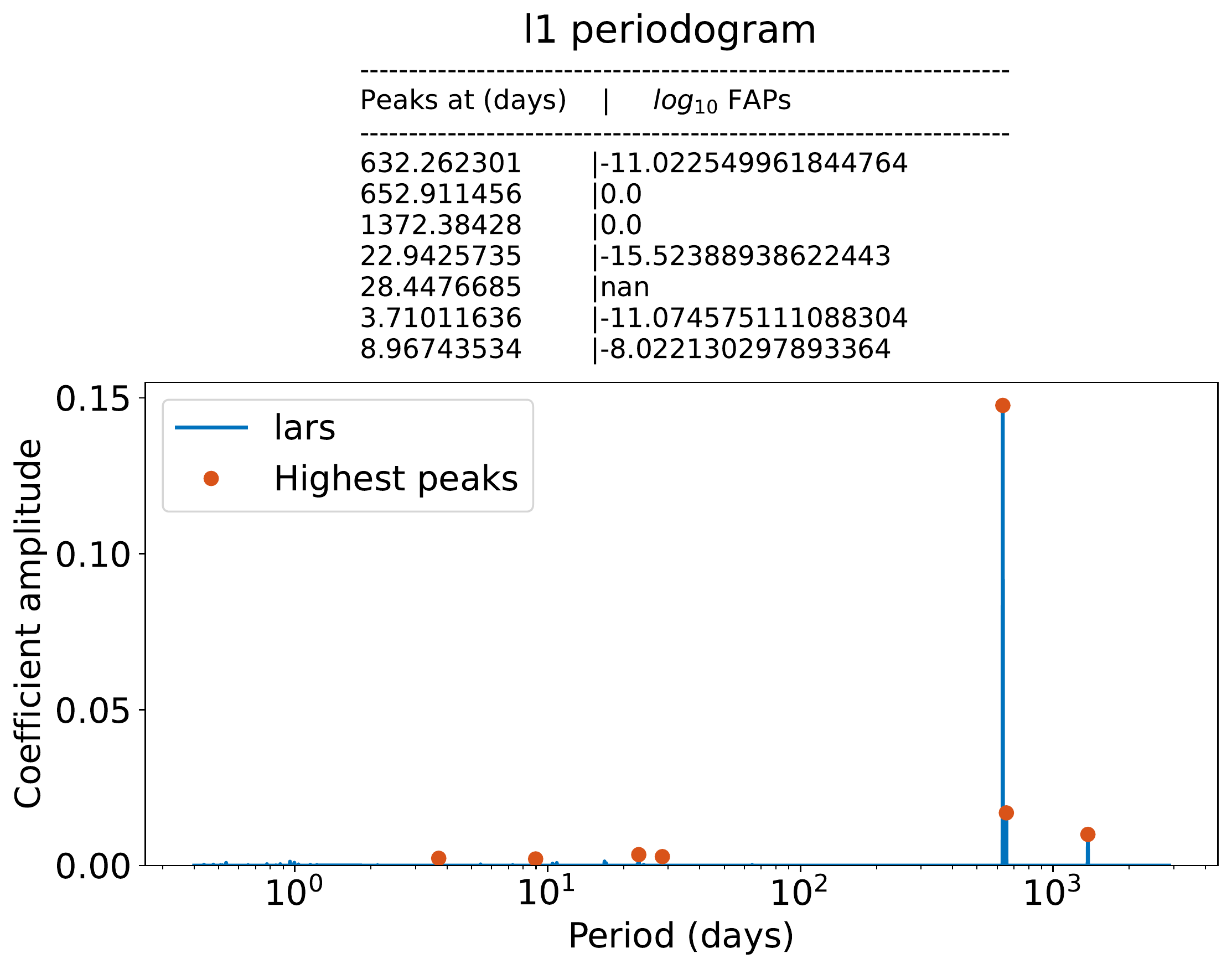}
    \includegraphics[width=1\columnwidth]{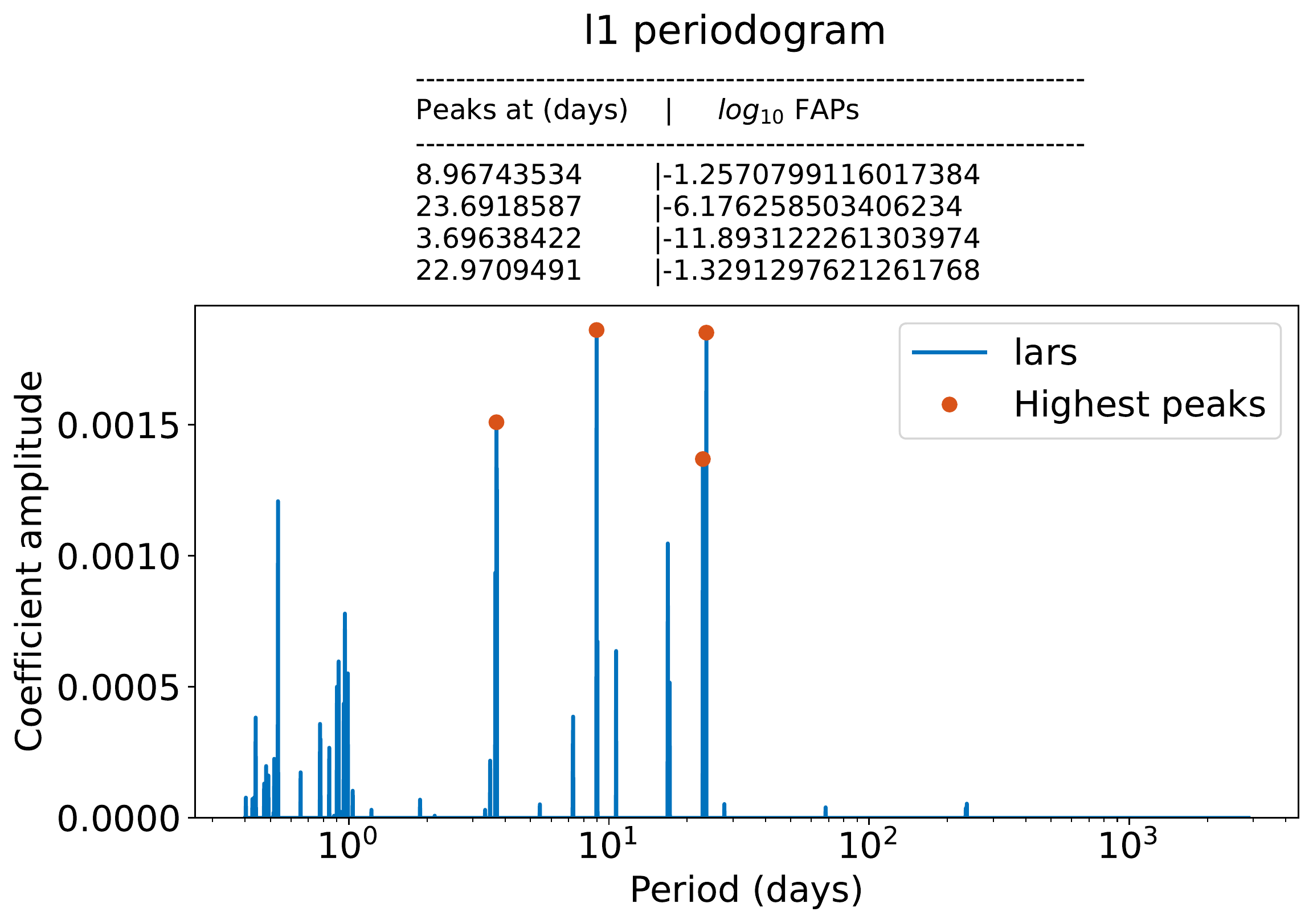}
    \caption{ $\mathbf{Top}$: The ${\ell_1}$- periodogram of the raw RVs of CoRoT-7 after unpenalizing the transiting 0.85 days signal of CoRoT-7b, with the strongest spikes marked with red dots at the top. Traces of peaks for the planetary signals of CoRoT-7c (3.7 days) and  CoRoT-7d (8.9 days) along with the stellar rotation signal ($\sim$23 days) are present after the dominant long-term activity signals. $\mathbf{Bottom}$: The ${\ell_1}$- periodogram for the Shift RVs after correcting for the shape variations, showing the strongest peaks at orbital period of planets (8.96 and 3.71 days) and the stellar rotation period with the activity trends suppressed significantly. The corresponding False Alarm Probabilities (FAP) are also listed above each panel.}
    \label{fig:l1}
    
\end{figure}

\subsection{\texorpdfstring{$l_{1}$}{l1}-periodogram}
\label{sec:l1}


  Figure \ref{fig:l1} shows the ${\ell_1}$- periodograms, after un-penalizing the 0.85-day signal of the transiting planet. In  the ${\ell_1}$- periodogram terms, this means that we are always including the 0.85 days planet in the fit. The top panel represents the ${\ell_1}$- periodogram for the raw RVs, dominated by the long-term activity signals.


An $\ell_1$ periodogram analysis for the Shift RVs is shown in the bottom panel of Figure \ref{fig:l1}. After decorrelating the stellar activity-induced shape variations using the {\sc scalpels} $\mathbf{U}$ vectors,  the two non-transiting planet candidates dominate the periodogram. Now the 4 signals with the greater significance and lowest false alarm probabilities are found at periods 3.70 days, 8.96 days and a closely spaced pair (23.69 and 22.94 days) around the stellar rotation period. The long-term signals are weakened considerably. All other strong peaks that showed up in the periodogram are identifiable as one-day aliases ($|\frac{1}{P} \pm 1|^{-1}$) of these four signals. 

The presence of two closely-spaced signals near 23d in the both RV sets suggest that, in the $\ell_1$ periodogram's sparse representation of the data, the change in the amplitude of this signal between 2008-9 and 2012 is modelled as a 1990 days beat pattern between the two closely-spaced sinusoids.
\begin{figure*}
    \centering
    \includegraphics[width=1.1\columnwidth]{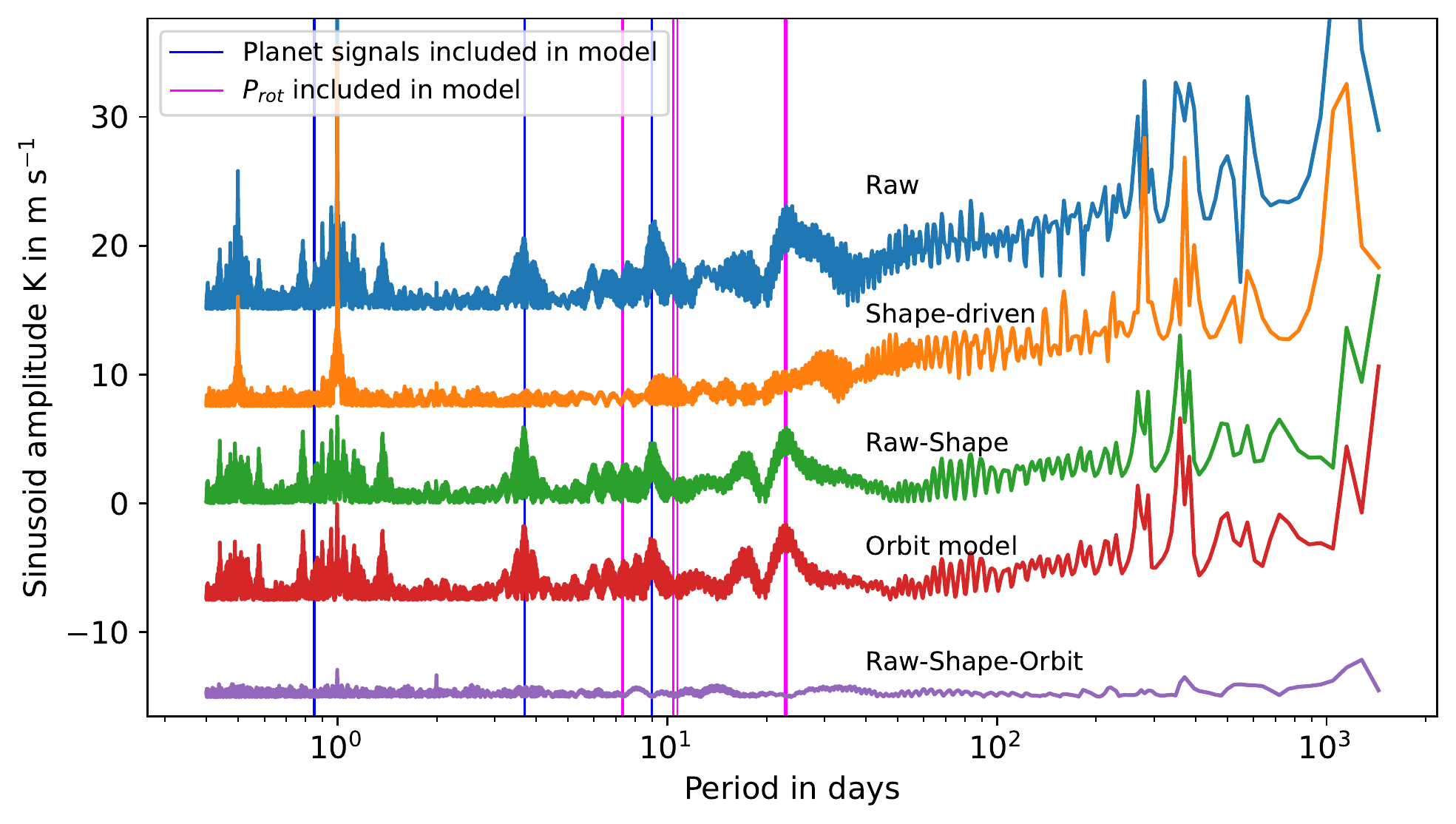}
    \includegraphics[width=0.9\columnwidth]{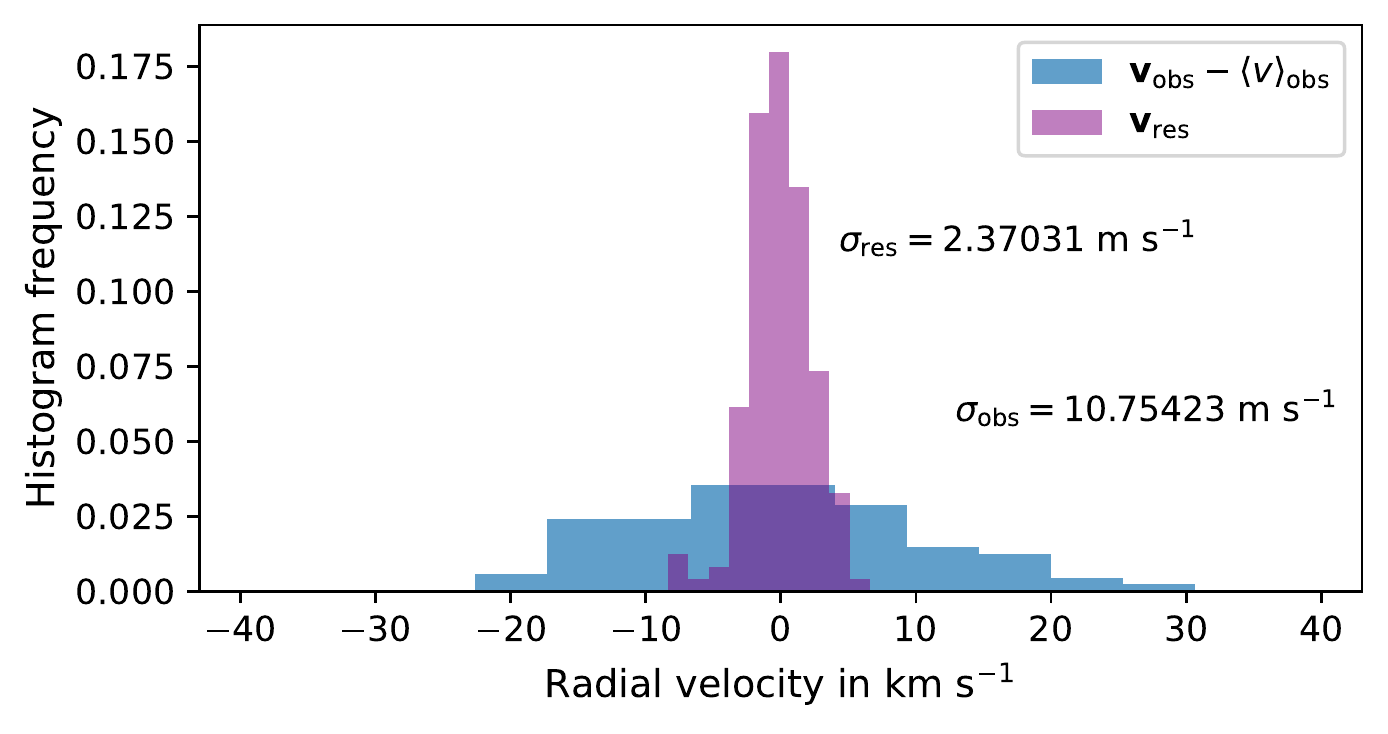}
    
    \caption{ The periodograms of RVs derived from the barycentric CCFs when the signal separation and orbit fitting of the sinusoidal signals of periods derived from the analysis associated with Figure \ref{fig:l1} have been performed simultaneously. 
  $\mathbf{Left}$:  The top three traces are as defined in the caption of Figure \ref{fig:blind}.The fourth trace (red) shows the periodogram of the fitted model of all the signals, including the 3  candidate planet signals (dark blue vertical bars) and 3 pairs of closely spaced beat periods associated with the rotation period and its harmonics (magenta vertical lines) found by the ${\ell_1}$- periodogram. The orbital model signal (red) is now much closer in appearance to the shift signal (green), leaving little residuals in the bottom trace (purple), obtained after subtracting both the shape-driven and orbital RV models from the onserved RVs. Each trace is offset by $\sim 10$ m\,s$^{-1}$ for better illustration. $\mathbf{Right}$:The resultant histogram of residual RVs shows significantly reduced RMS scatter.}
    \label{fig:simultaneous}

\end{figure*}  

\begin{figure*}
    \centering
   \includegraphics[width=1\columnwidth]{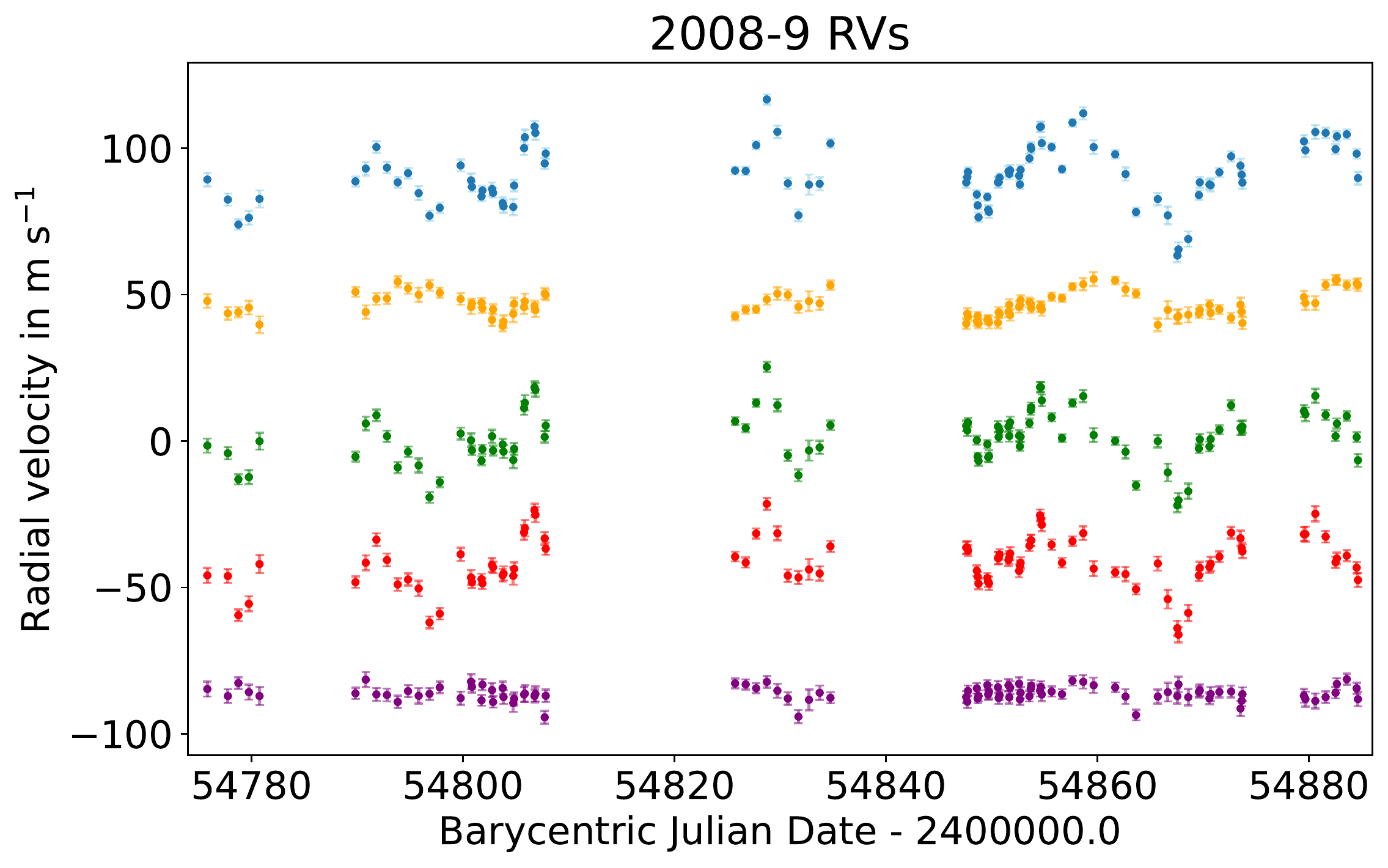}
    \includegraphics[width=1\columnwidth]{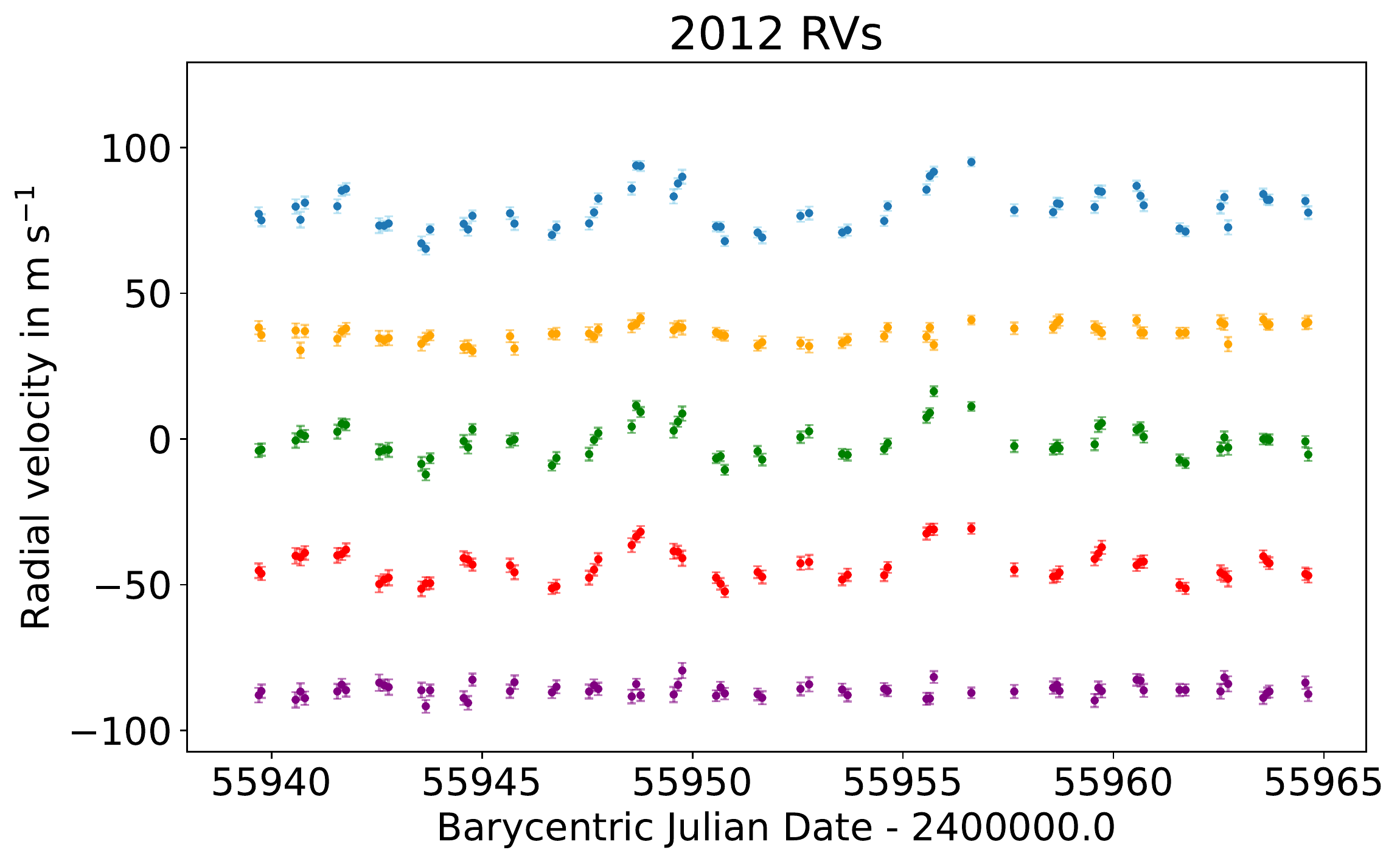}
     \caption{ $\mathbf{Bottom}$: The time-domain RV plots obtained from the simultaneous modelling for 2008-9 and 2012 data series are given in the left and right figures, colour-coded corresponding to each periodograms shown in the previous panels. Semi-transparent error bars are also included in the respective curves.}
\label{fig:simultaneousrvs}

\end{figure*} 
    


The presence of low-amplitude peaks in Figure ~\ref{fig:l1} may also indicate the presence of correlated, non-Gaussian noise, which we expect to be the case here. The ${\ell_1}$- periodogram is clearly successful at shortlisting the most plausible candidate planet signals in our data and suppressing the parasitic signals. To validate the physical origin of these signals,  we computed additional $\ell_1$ periodograms for the shape-driven RVs and the activity-indicators, as shown in Figure \ref{fig:l1_act}. None of these show traces of peaks at or around 0.85, 3.69 and 8.96 days, adding to the case for their planetary origins, especially for the third signal. Having identified the most plausible signals, we now turn to fitting them simultaneously with {\sc scalpels}.

\subsection{Simultaneous sinusoidal planet orbit fitting}
\label{sec:simultaneous}

In Section \ref{sec:blindpgm}, we projected the observed RVs, including planet signals, into the space spanned by the {\sc scalpels} basis vectors. Although, we found that there is hardly any correlation between the shape and the shift RVs (See \ref{fig:appcorrelation}), if any of the planet signals are correlated with any of the basis vectors, they may get partially absorbed in the shape signal. This is problematic, as we cannot guarantee orthogonality between any irregularly sampled orbital superposition with the components of the {\sc scalpels} $\mathbf{U}$ vectors. 
The orbital perturbations of any planet and the scalpels projection process must therefore be modelled self-consistently for the signal separation to recover their semi-amplitudes as reliably as possible. For this reason, \citet{2021MNRAS.505.1699C} solved simultaneously for shape-driven variations and a set of circular Keplerian orbits with known periods. Here we use a similar approach, using the set of orbital frequencies obtained with the ${\ell_1}$- periodogram as inputs. The approach of obtaining a simultaneous solution involved determining the shape-driven variations from the difference between observed RVs and the orbit model velocities \citep{2021MNRAS.505.1699C}. 

The assumption of circular orbits for candidate signals of the three planets can be justified by considering the tidal circularization timescales. According to \citet{2004ApJ...610..464D} we could assume negligible eccentricities for the signals at 0.85 and 3.69 days, having periods less than 6 days and hence accounted for by dissipation of tidal disturbances within their envelopes that are induced by their host stars. For the third signal at a period of 8.96 days, we calculated the circularization timescale as described in \citet{2004ApJ...610..464D}, and found a value of $\sim$610 Gyr for an assumed planetary tidal quality factor $Q_p' = 10^6$. We also relaxed the prior on eccentricity for the analysis in Section \ref{sec:kima} to allow the data to find the feasible value. The obtained value for eccentricity falls around 0.057±0.062 (see Figure \ref{fig:amp_ecc}), again supporting the assumption of circular orbits.

The left panel of Figure \ref{fig:simultaneous} shows the RV periodograms when the signal separation and orbit modelling of all 4 strongest sinusoidal signals having non-zero FAPs from the ${\ell_1}$- periodogram search along with the transiting planet signal and two pairs of periods at $P_{\rm rot}/2$ and $P_{\rm rot}/3$ have been performed simultaneously. The orbital periods of candidate planets are marked by vertical blue lines at 0.85 days, 3.71 days, 8.96 days and beat periods for stellar rotation periods and harmonics by magenta lines at 22.94, 23.75, 10.42, 10.74, 7.30 and 7.35 days. 

The top three traces are as defined in the Section \ref{sec:blindpgm}, with the only difference being the shape-driven periodogram here is obtained by subtracting the orbit model from the raw RVs. The fourth trace (red) shows the periodogram of the fitted orbit-model of the candidate signals. The bottom periodogram in purple shows the residuals after subtraction of both the shape-driven and orbital RV models. It is worth noting the level of matching between both periodograms (green \& red), even with the finest details, in this case.


\begin{table}
\caption{The periods and semi-amplitudes of the strongest signals in the periodograms of raw (${k=0}$) and shape-corrected apparent shift-velocities ($k=k_{\rm max}$) from simultaneous modelling of CCF shape changes and planetary motion, made with prior knowledge of the periods for candidate planets (from the ${\ell_1}$- periodogram) and pairs of beat period at the stellar rotation period and its harmonics are listed below. }


\centering
\begin{tabular}{c c c c c}
\hline
P(days) &K(ms$^{-1}$) &$\sigma_K$(ms$^{-1}$)&K(ms$^{-1}$)& $\sigma_K$(ms$^{-1}$)\\
 
&${k=0}$&${k=0}$&$k=k_{\rm max}$&$k=k_{\rm max}$\\

\hline\\

0.8535  &  4.185 & 0.263 & 3.552 & 0.306\\
3.6963  &  5.484  & 0.254 & 6.012 & 0.312\\
8.9674  &  6.877  & 0.324 & 5.482 & 0.380\\
10.4269 &  0.472  & 0.569 & 1.156 & 0.606\\
10.7409 &  4.215  & 0.535 & 2.560 & 0.597\\
7.3039 &  2.173  & 0.536 & 2.391 & 0.620\\
7.3583 &  3.004  & 0.570 & 3.276 & 0.655\\
22.9425 &  6.915  & 0.288 & 4.891 & 0.353\\
23.6918 &  4.077  & 0.273 & 3.311 & 0.348\\

\hline\\

 


 
\label{tab:recovery}
\end{tabular}
\end{table}

A closer look at the orange and green periodograms reveal that the balance of the signal separation has been altered from those in Figure \ref{fig:blind}, by modelling the orbital motion explicitly at known periods. The algorithm efficiently reduces the correlation between the modelled orbit signals and modes of velocity variation driven by line shape changes alone.

To determine the orbital velocity amplitudes that would be recovered without applying profile-shape corrections to the velocities, we reduced the number of principal components $k_{\rm max}$ to zero. We found that, when the sinusoids are fitted to the observed RVs and left uncorrected for line-profile shape variations, the semi-amplitudes of the recovered signals depart from the semi-amplitudes recovered during the simultaneous modelling (see columns 2 and 4 of Table \ref{tab:recovery}). 

We also compared the recovered semi-amplitudes of the four strongest signals when fitted with and without the 23.69 days beat signal found in the ${\ell_1}$- periodogram search, in the appendix ( See Table \ref{table:4&5sim}). The last two columns of both tables list the semi-amplitudes and associated uncertainties for the shift-driven RVs acquired in both cases. While the differences in the RV amplitudes for planets b, c and d between the 4-signal model and the 5-signal model are small when compared with their uncertainties, the 5-signal model gives a more faithful approximation to the behaviour of the 23-d activity signal between seasons. 

The 23-day peak in the shift signal is therefore best modelled as the beat pattern between the two closely spaced periods, requiring 2 frequencies to be fitted for the stellar rotation period. In the 4-signal case, a single sinusoid is fitted at this frequency, so the 23-day peak shows strong cycle-count interference. We attribute this effect to the 1990 days beat period between the 22.7 days and 22.9 days signals being greater than the entire observation span of $\sim$1195 days. When both 23-day periods found by the ${\ell_1}$- periodogram are fitted, the form of the 23-day peak in the model signal (red) is much closer in appearance to the shift signal (green) (See Figures \ref{fig:simultaneous} and \ref{fig:appsimfig}). However, there is still some power in the residuals at $P_{\rm rot}/2$ and $P_{\rm rot}/3$ (marked with pink vertical lines in Figure \ref{fig:appsimfig}), when only fitting 5 signals. This indicates that a purely sinusoidal model for the residual activity signal, even with two closely-spaced frequencies, might be insufficient. 

Therefore, it’s better to model the shift-like component of the activity-driven RV signal in such a way as to minimise the residuals and make the residuals as flat as possible. This motivated us to include pairs of sinusoids for the $P_{\rm rot}/2$ and $P_{\rm rot}/3$ as shown in Figure \ref{fig:simultaneous}. The RMS scatter is now significantly reduced to 2.37 ms$^{-1}$. The resultant orbital periods, semi-amplitudes, and uncertainties of the sinusoidal signals included in the orbit model are listed in Table \ref{tab:recovery}.


Hence, we conclude that the beat model is a better description of the long-term evolution of the 23-day signal, making it unlikely that a fourth planet signal is superposed on the activity signal at this period. The property of the 23-d signal, being a quasi-periodic shift-like change in the CCF shape produced by the activity, complicates the RV analysis.

In brief, all three candidate planet signals reported in the previous sections were recovered successfully with improved precision in their $K$ amplitudes. Particularly for the 8.97d signal, no significant counterpart was detected in the shape RVs obtained from the simultaneous modelling irrespective of the number of planet sinusoids included, strengthening the evidence for the existence of CoRoT-7d, the third planet in the system. Our inability to fit the $\sim$23-day signal satisfactorily with a single sinusoid indicates that it is likely a stellar rotation signal with variable amplitude and phase. 

\subsection{Nested Sampling using GP +{\sc scalpels}}
\label{sec:kima}
Using {\sc kima} \citep{2018JOSS....3..487F}, we sampled the posterior distribution of orbital parameters along with a complete Keplerian solver and a Gaussian Process.  The posterior distribution constrained by the RV data was then used to probe the credible number of planets given the data, the orbital parameters and the planetary masses.  
The priors on the GP model parameters hyperpriors were carefully tailored for CoRoT-7, as given in Table.\ref{table:kima}.

To account for the stellar activity, we decorrelated the {\sc scalpels} U-vectors (when $k=k_{\rm max}$) that are found to be significant contributors to the shape signal, which are $\mathbf{U_{1}}$, $\mathbf{U_{2}}$ \& $\mathbf{U_{3}}$. Knowing the parameters of one planet provides a small amount of information about the parameters of another
planet \citep{2016A&A...588A..31F}. Keeping this in mind, we treated the transiting planet CoRoT-7b as a known planet, adopting the updated transit parameters in \citet{2014A&A...569A..74B}. By putting tight priors on the orbital period P and mid-transit time $T_{0}$ we ensure that the phase of this transiting planet signal is consistent with the photometry.

For the known planet, we put a lower limit on $K$ derived from the mass it would have if it were made of compressed water, with a radius of 1.528 $R_{\oplus}$ \citep{2014A&A...569A..74B} that is around 1.56 $M_{\oplus}$ (according to the Growth model interpretation of mass distribution proposed by \citet{2016ApJ...819..127Z}), giving an RV amplitude of $K_{min}$ $\sim$ 1.08 ${\rm ms^{-1}}$. The upper limit was set to be about a physically-motivated number of $K_{max}$ $\sim$ 10 m $s^{-1}$ corresponding to the 10 $M_{\oplus}$ expected for a pure iron planet of that size.

Taking into account the existing knowledge of the rotation period from the literature \citep[e.g.,][]{2009A&A...506..303Q,2014MNRAS.443.2517H} and the present work, we restricted the GP hyperparameter $\eta_{3}$ (representing the stellar rotation period) to a sensible range between 19 and 30 days.  We adopted a prior period distribution for non-transiting planets with periods between 0.5 days and the duration of the 2008-9 data, i.e., 100 days. 

As an initial experiment, the number of planets $N_{p}$ was set to be a free parameter in the MCMC, with a uniform prior between 0 and 5. Considering $N_{p}$ as an unknown parameter, {\sc kima} uses a birth-death MCMC proposed by \citet{2014arXiv1411.3921B}, which allows estimating the improvements in evidence values when the sampler switches between solutions with different $N_{p}$ values. 

\begin{table}
\caption{Prior distribution for model parameters. $\cal{LU}$- Log Uniform ; $\cal{MLU}$- Modified Log Uniform; $\cal{K}$- Kumaraswamy ; $\cal{U}$- Uniform ; $\cal{G}$- Gaussian }
\centering
\begin{tabular}{|c| c| c|}
\hline
Notations&\textbf {Orbital parameters}&Priors \\ [0.5ex] 
\hline
$P$&Orbital Period&$\cal{LU}$(0.5,100) \\
$K$&Semi-amplitude&$\cal{MLU}$(1,25) \\
$e$&Eccentricity &$\cal{K}$(0.867,3.03) \\
$\phi$ &Orbital phase   & $\cal{U}$(0, 2$\pi$) \\
$\omega$&Longitude of line of sight   & $\cal{U}$(0, 2$\pi$)\\ [1ex]
\hline
&$\textbf {GP parameters}$& \\ [0.5ex] 
\hline
$\eta_{1}$&Amplitude of correlations&$\cal{LU}$(1.6,2.7) \\
$\eta_{2}$&Decay timescale&$\cal{LU}$(15,60) \\
$\eta_{3}$&Correlation period&$\cal{U}$(19,30) \\
$\eta_{4}$& Periodic scale &$\cal{U}$(-1,0) \\
$s$&Extra white noise&$\cal{LU}$(0.5,max $v$) \\
\hline
&\textbf {For known planet}& \\ [0.5ex] 
\hline
$P$&Orbital Period&$\cal{G}$(0.85359, 5.7e-7) \\
$K$&Semi-amplitude&$\cal{MLU}$(1.08,10) \\
$e$&Eccentricity&$\cal{U}$(0, .1) \\
$t_{0}$ &Time of mid-transit  & $\cal{G}$(54398.0776, 0.0007) \\
$\omega$&Longitude of line of sight  & $\cal{U}$(-$\pi$,$\pi$)\\ [1ex]
\hline
$\beta$&$\textbf {Activity indicators}$& $\cal{G}$(0,10)\\
\hline
\end{tabular}
\label{table:kima}
\end{table}

\begin{figure}
    \centering
    \includegraphics[width=1\columnwidth]{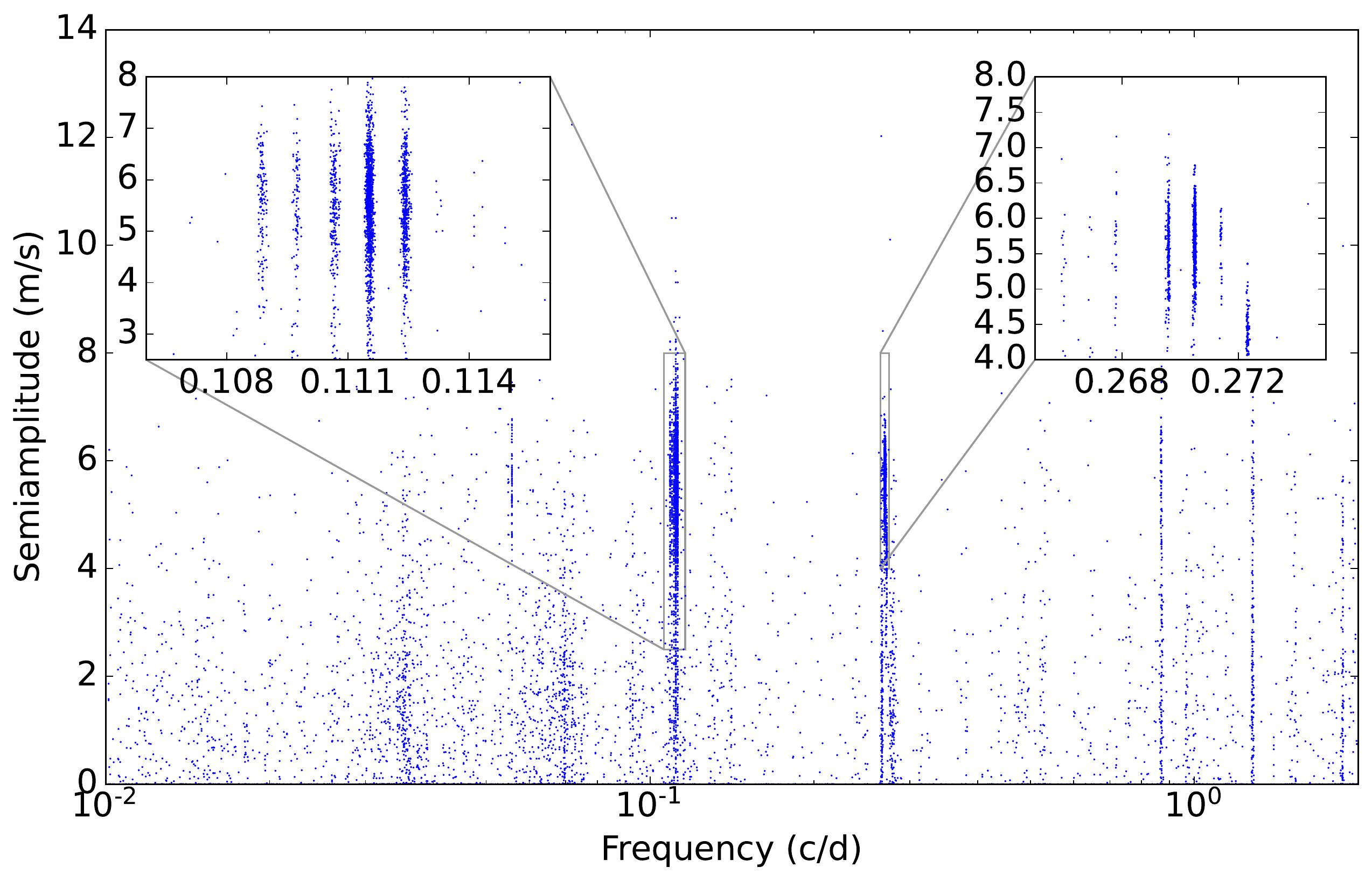}
    \includegraphics[width=1\columnwidth]{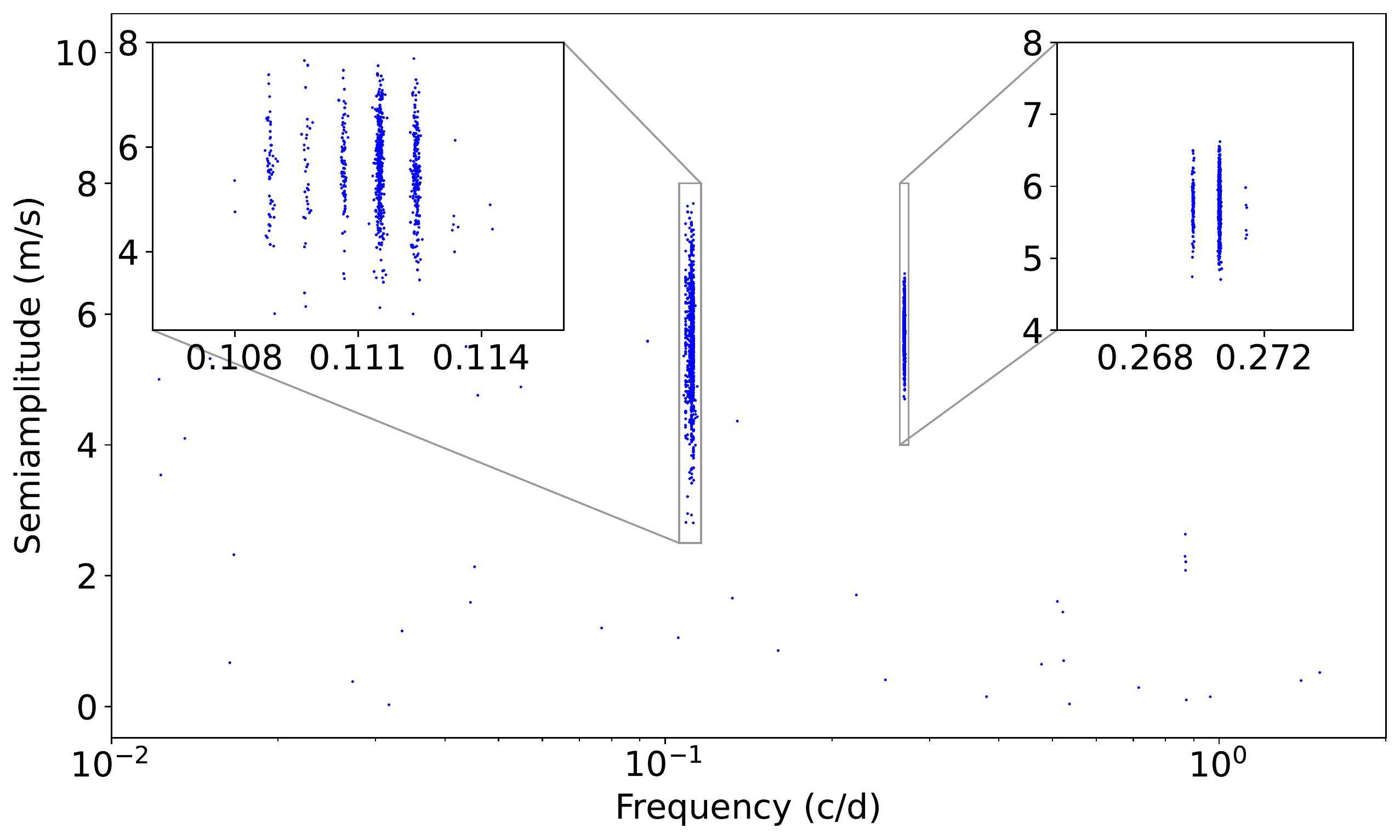}
    \caption{The posterior distribution for semi amplitudes ($\rm ms^{-1}$) in log frequency space ($\rm cd^{-1}$) is shown. The transiting 0.85 days planet is considered as 'known'. The inset plots zoom into the clusterings corresponding to the $\sim$9 days planet  and $\sim$3.7 days planet (right), showing the interference patterns occurred due to the widely separated observing seasons. $\mathbf{Top:}$ Setting number of planets, $N_{p}$ as a free parameter with uniform priors between 0 and 5. $\mathbf{Bottom:}$ $N_{p}$ is fixed at 2.}
    \label{fig:beat}
    
\end{figure}

The resultant joint posterior distributions for the semi-amplitudes of the signals are shown in the top panel of Figure \ref{fig:beat}. We found that two non-transiting planets are favoured by the data, considering the ephemeris of the transiting planet CoRoT-7b as known in the model. Amidst the posterior spread in the parameter space, these are seen as heavily populated regions at $P_{c}$ = 3.69 days and $P_{d}$ = 8.96 days. 

Reviewing this posterior distribution triggered another experiment by fixing the number of planets as $N_{p}$=2 with the rest of the priors as the same as in Table.\ref{table:kima}. The resultant posterior distribution of RV semi-amplitudes is represented in the bottom panel of Figure \ref{fig:beat}, with well-constrained peaks around $P_{c}$ = 3.69 days and $P_{d}$ = 8.96 days, within the orbital period uncertainties reported in the literature. Supporting the presence of CoRoT-7c and CoRoT-7d, the same constrained detail can be seen in the posterior distribution of eccentricities also (See Figure \ref{fig:amp_ecc}). From the 10000 effective samples in the joint posterior distribution, the evidence log$\cal{Z}$$_{2}$ for this model was calculated as log(${p(D\,|\,M)}$) = -475.46. 
\begingroup
\newcolumntype{P}[1]{>{\centering\arraybackslash}p{#1}}
\renewcommand{\arraystretch}{1.5}
\begin{table*}
\caption{The mass and mean-density (column 4 \& 5) calculated for the transiting planet CoRoT-7b  from the posterior distributions of orbital period and RV semi-amplitudes (column 2 \& 3) considering different models are listed. Important note: CoRoT-7b is considered as a known planet in all these models with transit parameters from \citet{2014A&A...569A..74B} (See Table \ref{table:kima}), who updated the planet radius as 1.528 ± 0.065 $R_{\oplus}$. $N_{p}$ is the fixed number of planets in each model. The first three rows give results from models with GP + {\sc scalpels} and different number of planets, where the last three rows are the results for different models with GP alone and different $N_{p}$. }
    \centering
    \begin{tabular}{|c|c|c|c|c|c|c|}
 \hline 
 Model    &       $P_b$[days]         &        $K_b$[$ms^{-1}$]         &     $M_b$ [$M_{\oplus}$]          & $\rho_b$[g cm$^{-1}$]&  log $\cal{Z}$ & $\cal{Z}$$_{n}$-$\cal{Z}$$_{n+1}$\\  
 \hline 
 GP +{\sc scalpels}&&&&&&\\
 \hline
$N_{p}$=1 & 0.853592 ± 6.01e-7   &  3.943 ± 0.559    & 5.595 ± 0.794 & 8.641 ± 0.269 &  -479.20 & -- \\
$N_{p}$=2 & 0.853592 ± 5.87e-7   &  4.291 ± 0.460    & 6.056 ± 0.653 & 9.355 ± 0.235 &  -475.46 & 42.31 \\
$N_{p}$=3 & 0.853592 ± 5.78e-7   &  4.278 ± 0.453    & 6.048 ± 0.643 & 9.322 ± 0.234 &  -474.58 & 2.40\\                        
\hline
GP only&&&&&&\\
 \hline
$N_{p}=1$ & 0.853592 ± 5.48e-7   &  4.007 ± 0.550   & 5.679 ± 0.781  & 8.786 ± 0.264 &  -478.58 & --\\
$N_{p}=2$ & 0.853592 ± 5.53e-7   &  4.293 ± 0.468   & 6.087 ± 0.664  & 9.394 ± 0.236 &  -474.87 & 40.89\\
$N_{p}=3$ & 0.853592 ± 5.72e-7   &  4.331 ± 0.476   & 6.146 ± 0.676  & 9.459 ± 0.236 &  -474.27 & 1.82\\
\hline 
\end{tabular}
\label{table:M&rho}
\end{table*}
\endgroup

If we fix $N_{p}$=1, we recover the orbital parameters for CoRoT-7c only, similar to those reported by \citet{2016A&A...588A..31F}. The evidence is degraded to log$\cal{Z}$$_{1}$ = -479.20 for this model. The difference in log evidence between the model with 2 non-transiting planets and the model with 1 non-transiting planet is 3.8. This is in the substantial support range of the Jeffreys scale for Bayesian evidence comparison (Jeffreys 1961). These results therefore indicate a 2-planet model to be 42.31 times ($\cal{Z}$$_{1}$-$\cal{Z}$$_{2}$) more probable than a 1-planet model. We emphasize that, when saying a 2-planet model, we are referring to CoRoT-7c and CoRoT-7d, as the model considers the orbital parameters of  transiting planet CoRoT-7b as known.

\begin{figure}
    \centering
    \includegraphics[width=1\columnwidth]{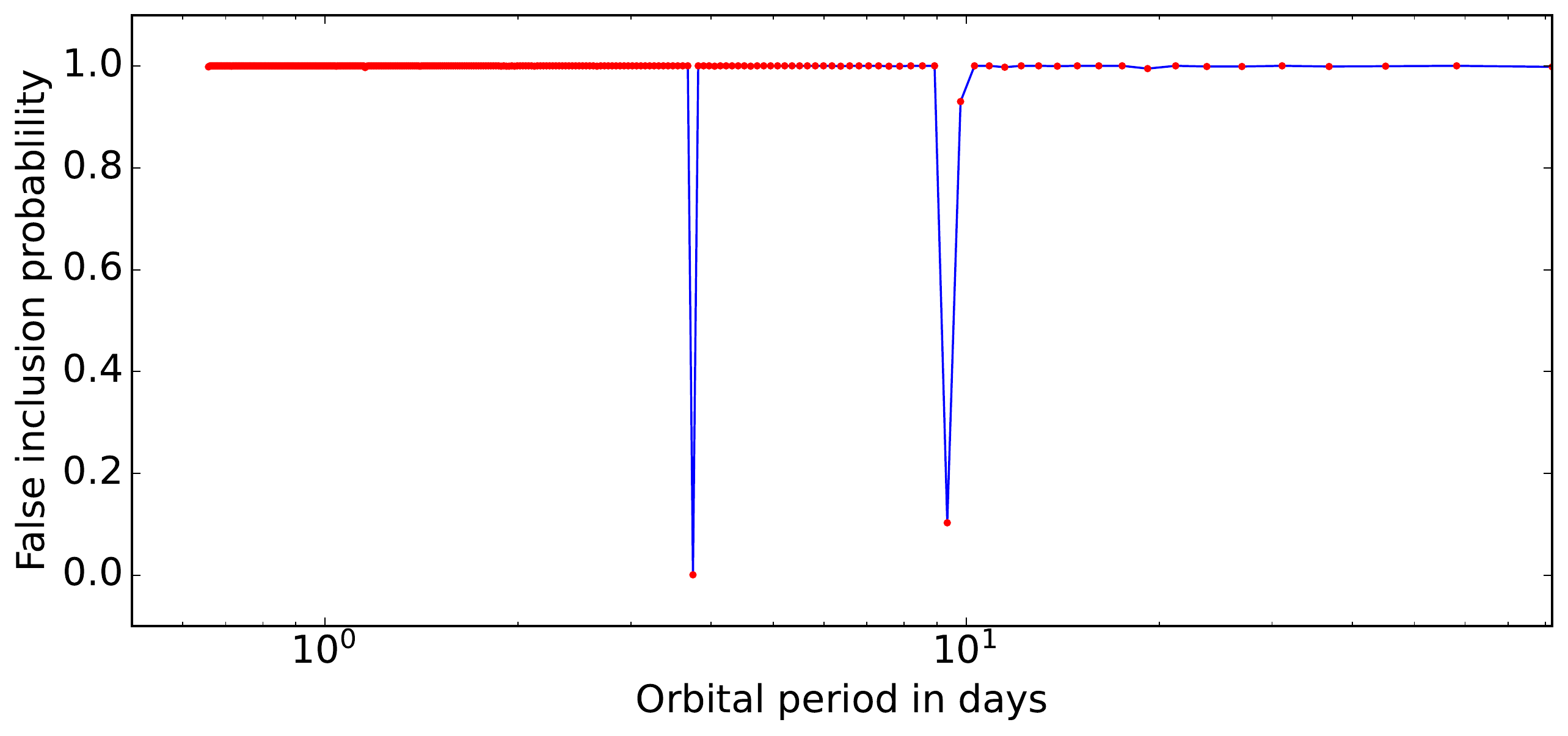}
    \includegraphics[width=1\columnwidth]{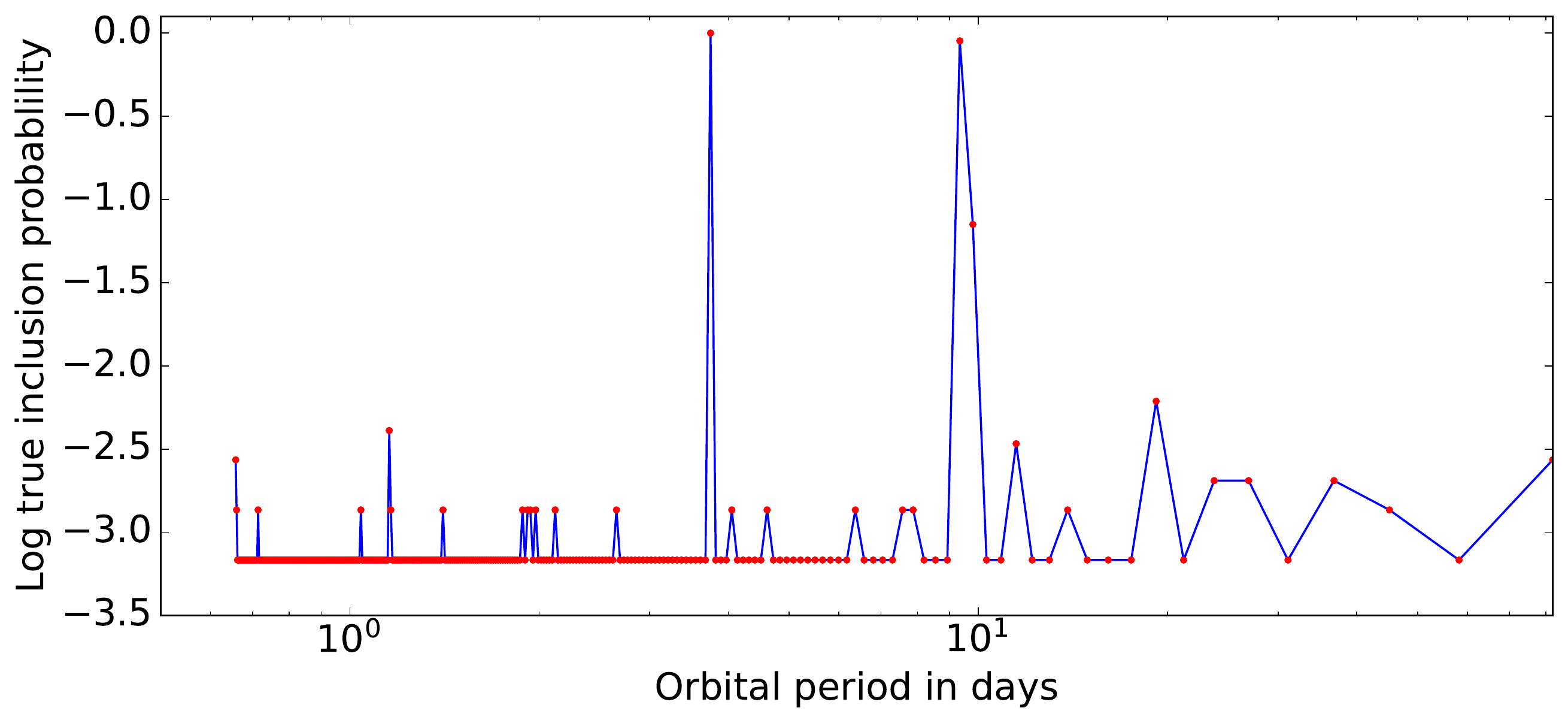}
    \caption{{\bf Top}: The FIP periodogram of CoRoT-7 HARPS data, computed using the joint posterior distributions obtained from KIMA, modelled with{\sc scalpels}, GP and $P_{b}$ as known. {\bf Bottom}: TIP periodogram. Both periodograms strongly favour 2 planets at orbital periods 3.69 days and 8.96 days, additional to the known transiting 0.85 day signal.}
    \label{fig:fip_tip}
    
\end{figure}

\begin{figure*}
  \centering
  \begin{minipage}[b]{1.5\columnwidth}
    \includegraphics[width=\textwidth]{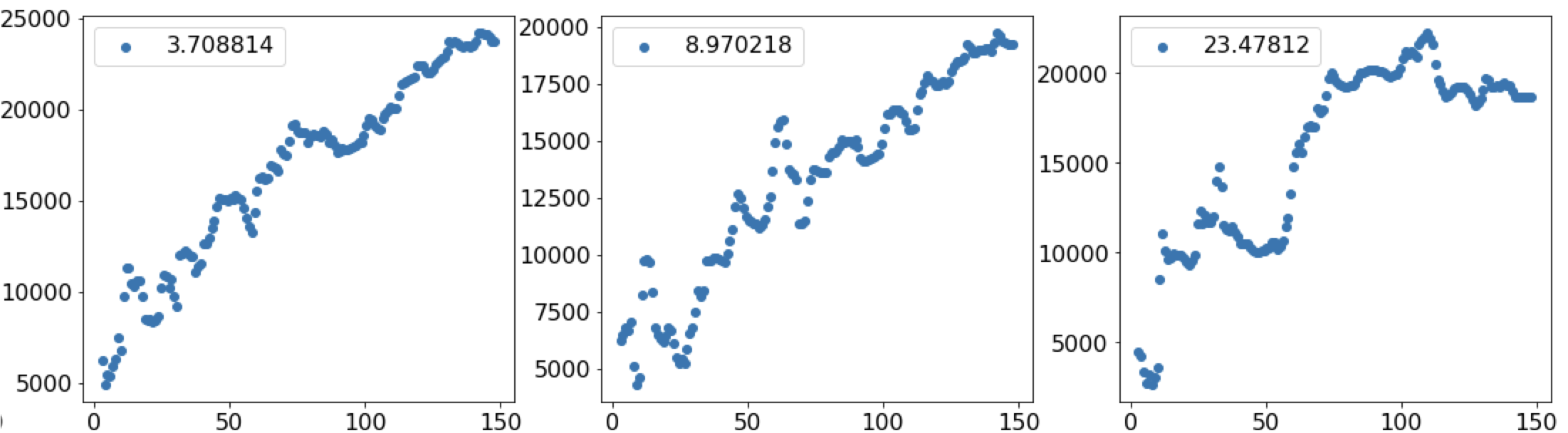}
  \end{minipage}
  \caption{$\mathbf{ x-axis}$: Number of observations, $\mathbf{ y-axis}$: signal-to-noise ratio. The SNR plotted against the number of observations for the optimal periods of individual candidate signals at periods 3.70 days, 8.97 days and 23.47 days.  }
  \label{fig:snr}
\end{figure*}
\begin{figure}
    \centering
    \includegraphics[width=1.05\columnwidth]{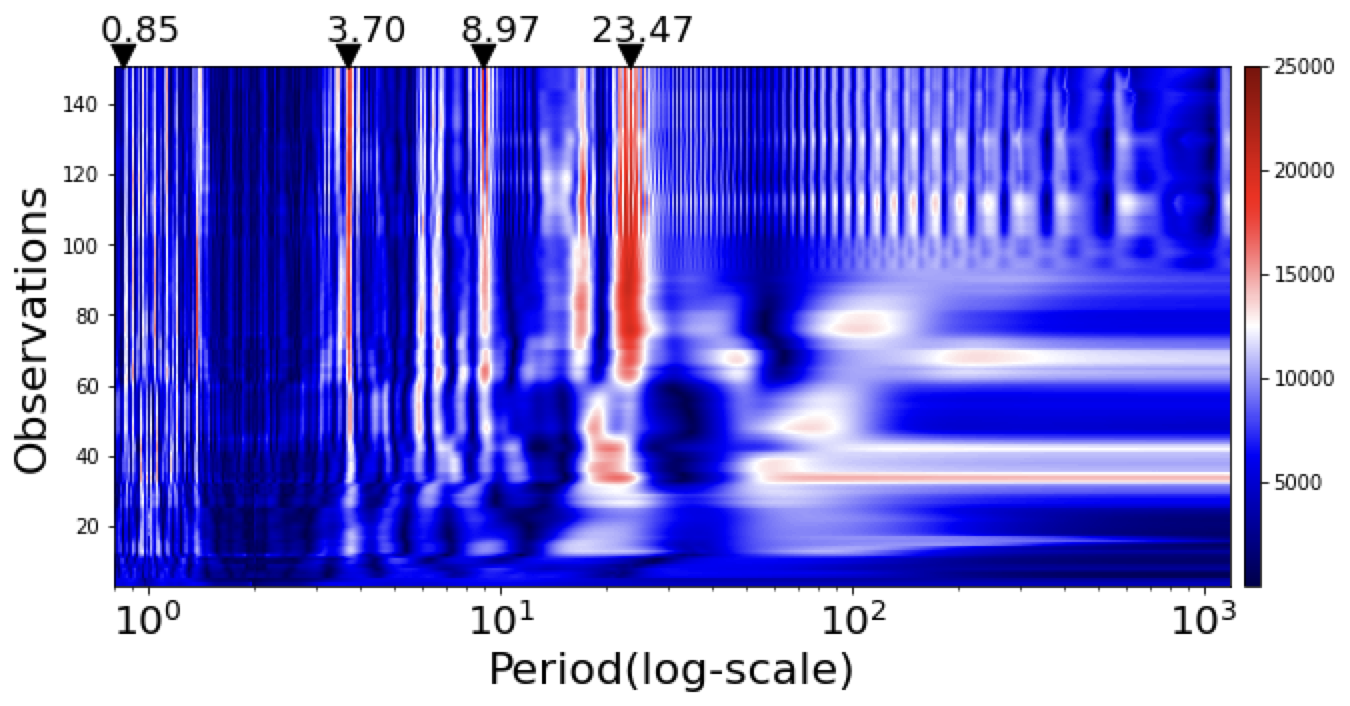}
    \caption{The Stacked Bayesian General Lomb-Scargle periodogram for CoRoT-7 with number of observations in y-axis plotted against period in days displayed in logarithmic scale. The colour gradient designates the logarithm of probability, where redder is more promising.}
    \label{fig:BGLS}
    
\end{figure}

After `confidently detecting' CoRoT-7b and CoRoT-7c, \citet{2016A&A...588A..31F} pointed out that it is more likely that there are four planets according to the posterior distribution. Their posterior showed smaller peaks around $\sim$9 days and $\sim$2 days. However, adding the {\sc scalpels} basis vectors to the model for activity decorrelation enabled us to confidently recover the orbital signal of CoRoT-7d at 8.96 days. We were unable to recover any signal around 2 days, as it is removed as being some artefact arising from activity, aliasing or stellar rotation harmonics. In an attempt to validate the improvement in the model with the inclusion of the {\sc scalpels} shape-signal decorrelation vectors along with a GP, we compared the run with the {\sc scalpels} against the one without. The results are given in Table \ref{table:M&rho}. We found that de-trending the RVs for line shape variations using the {\sc scalpels} basis vectors yields a model that is 4.14 times more probable than a model that does not include these activity variations.

\subsubsection{FIP and TIP calculation}
\label{sec:fip_tip}

The zoomed-in view of the over density regions in Figure \ref{fig:beat} revealed interference patterns associated with CoRoT-7c and CoRoT-7d at 3.69 days and 8.96 days. It is interesting to consider the two seasons of observations in 2008-9 and 2012 as a 'double-slit experiment', in which the path lengths between the light waves from two slits give rise to an interference pattern due to the phase shift. We see an analogous scenario here, with sharp peaks separated by a frequency $\sim$ $\frac{1}{1200}$ $cd^{-1}$, accounting for the time separation (1200 days) between two observing campaigns. This clearly indicates that both signals are detected unambiguously in both seasons.

We performed the FIP analysis in the frequency space, setting the bin size equal to the Nyquist frequency resolution of the 2008-9 season rather than the full duration of the data. Since the probability of each planet is spread over several interference fringes, we should marginalize over the uncertainty in the cycle count to establish the existence or otherwise of the planet. A frequency bin size of 0.005 cycles per day captures all probabilities while fully resolving the structure of the 2008-9 periodogram.

\citet{2021arXiv210506995H} presented FIP and TIP testing as good practice for the analysis of RV data, given that assumptions on the priors and the noise model (log$\cal{L}$) might be incorrect. Following this idea, we searched for various planets simultaneously using a frequency window sliding along the entire frequency space  spanned by the posterior distribution of the 2-planet model (with CoRoT-7b being known). Consequently, two strong peaks were found, at 3.697 days and 8.965 days, both with significantly minimal FIP and maximal TIP, reinforcing their planetary nature (Figure \ref{fig:fip_tip}). To avoid being misled by the shared power among adjacent interference peaks linked with relevant signals, a basic histogram was used to find these most probable periods, See Figure \ref{fig:fip_beat}.

Being a trans-dimensional sampling algorithm, {\sc kima} enabled us to directly sample the joint posterior of the number of planets ($N_{p}$) and other orbital elements.  Several ambiguous detections are prompted at spurious periods due to the cross-talk between different aliases. FIP and/or TIP attenuated the invalid detections associated with these periods. Interesting is the enfeebled TIP and very large FIP ($\sim$1) at the stellar rotation period of $\sim$23 days and its harmonics, providing more evidence against the planetary nature of a body at this period. From Figure \ref{fig:simultaneous} alone, one could argue that since the stellar rotation period ($\sim$23 days) is seen in the shift signal, any other period that also survives when the shape signals are subtracted could also be activity-driven. However, a better validation for the dynamical origin of the candidate signals is achieved here through incorporating a GP to account for any remaining quasi-periodic signals. It is promising to see that both the non-transiting planet signals with orbital periods 3.69 and 8.96 days survives the GP and produces a 2-slit interference pattern as shown in Figure \ref{fig:beat}.

\subsection{Stacked Bayesian General Lomb-Scargle periodogram}




To investigate the stability of the periodic signals found in the previous analyses, we used a stacked normalized BGLS periodogram \citep{2017A&A...601A.110M} of the shift-driven RV data. 
Following the method of \citet{2017A&A...601A.110M}, we derived the stacked periodograms of CoRoT-7 from the cleaned shift-driven RVs obtained after the {\sc scalpels} shape-signal separation.  The BGLS periodogram was initially obtained from the first {\em 4} data points and then refined the progressively by adding more one by one. This analysis was done for different subsets of data, obtained from the two observing runs (2008-9 \& 2012) separately (See Section \ref{sec:sbgls_appendix}) and also for the entire data set as shown in Figure \ref{fig:BGLS}. Figure \ref{fig:BGLS} is colour coded based on the log probability of the periods in the BGLS periodogram. One should expect to see the SNR of a planetary signal increasing asymptotically in proportion to the square root of the number of data points added.

\begin{figure*}
    \centering
    \includegraphics[width=1.5\columnwidth]{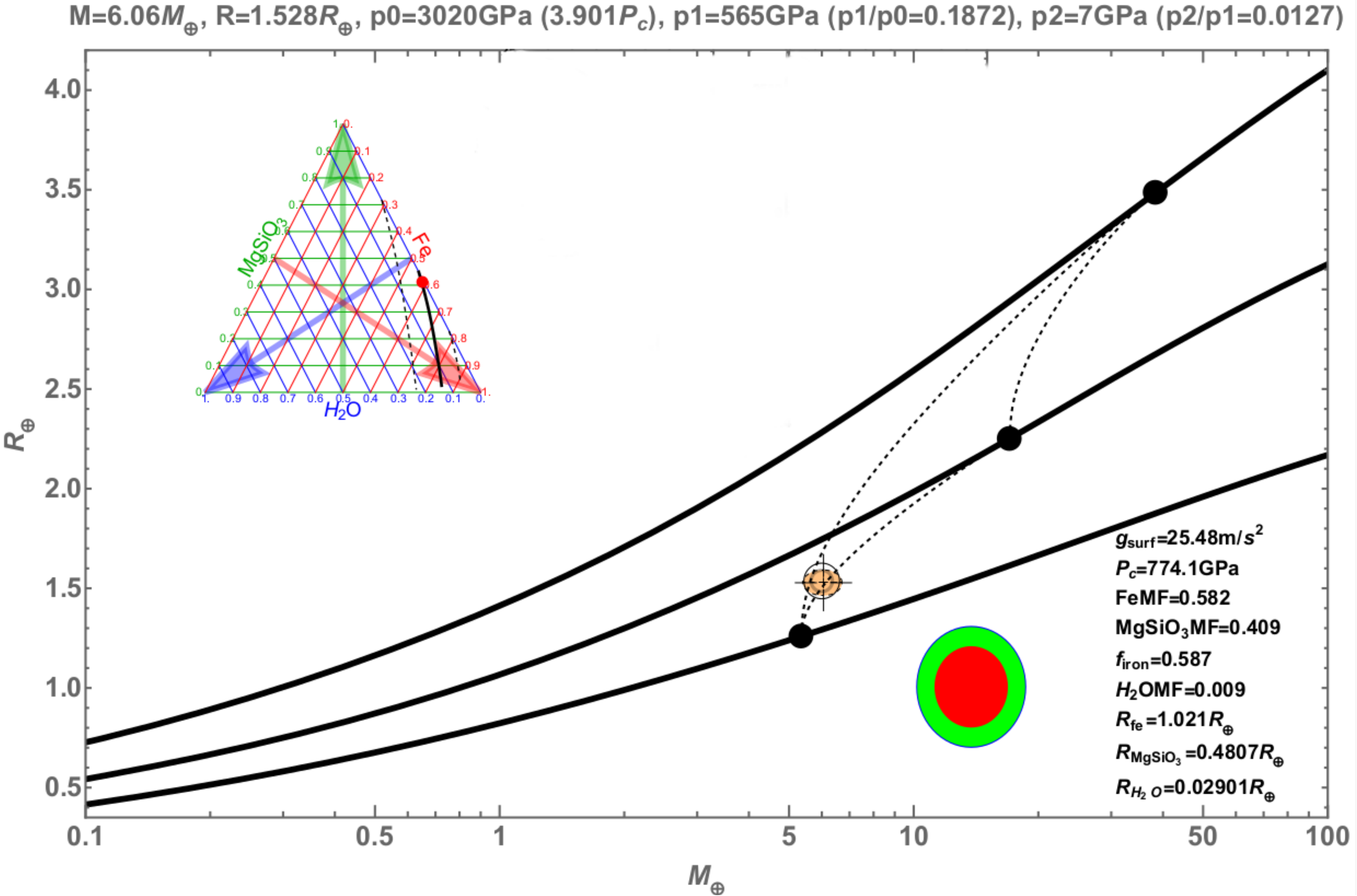}
    \caption{The most plausible composition of the transiting planet CoRoT-7b backed by the derived mass and radius for minimum core pressure is shown as obtained from the $\emph{manipulateplanet}$ tool developed by \citet{2016ApJ...819..127Z}. The thick black curve in the ternary diagrams shows the degeneracy of tri-layer model due to the trade-off between the three components, but not due to the mass or radius measurement uncertainties.}
    \label{fig:manipulateplanet}
    
\end{figure*}
Fringes began to appear in the stacked BGLS periodogram when analysed the both data sets together, as seen from Figure \ref{fig:BGLS}. These are caused by the uncertainty in the cycle count elapsed across the three-year gap separating the 2008-9 and 2012 RV campaigns \citep{2017A&A...601A.110M}. However, the combined data set works well for tracking the candidate signals.

Except the 23 days signal, all other candidate signals stand out with clearly growing probability features at periods 0.85, 3.69 and 8.97 days, supporting the coherent nature of these signals. It is worth noting that the 23.47 days signal grows unsteadily, with the strength maximizing around 70 to 100 observations and decreasing later on, see also Figure \ref{fig:snr}. We looked particularly into the 8.97 days signal to check its behaviour after adding more data points from the 2012 campaign and found that the SNR constantly increases with more observations, as expected for a real and coherent signal.


\subsubsection{Tracking the significance of detection}
\label{sec:snplot}

  We tracked the significance of each strong signal in the Figure \ref{fig:BGLS} by plotting the SNR against the number of observations (Figure \ref{fig:snr}). The optimal frequency for each candidate signal was obtained from the dominant peaks in the top line of the stacked BGLS periodogram. 

Each point in Figure \ref{fig:snr} represents the maximum SNR for the set of all RVs up to that point in time. 
We particularly looked into the 3.69 and 8.97 days signals and found that they exhibit a monotonic rise towards asymptotic square-root behaviour, confirming that they are strictly periodic and were present throughout the observations, hence substantiating the existence of CoRoT-7c and CoRoT-7d. As a result of the steady growth, their detections reach $\sim$25$\sigma$ and $\sim$20$\sigma$ respectively. 

On the other hand, the detection significance of the 23.47 days signal first grows rapidly but then levels out and even drops at some point when more observations are added. This is a clear sign of an incoherent signal. We have an advantage of having longer-baseline which make this analysis relevant, which would not be the case for shorter baselines and longer orbital periods. This evidence again strengthens the activity origin of the ~23-day signal, due to its unstable nature throughout the period of observation. 
\section{Discussion}
\label{sec:discussion}

\subsection{ Mass \& Mean density of CoRoT-7b}

As discussed in Section \ref{sec:whycorot}, the mass of the transiting planet CoRoT-7b has been a subject of debate for more than a decade. Table \ref{tab:parameters} shows the wide range of masses as reported in the literature. This can be attributed to the relatively high level of activity in the host star. Different posterior distributions for semi-amplitudes, considering different models, are listed in Table.\ref{table:kima}. We consider the model with $\emph{GP + SCALPELS + $N_{p=2}$ + CoRoT-7b$_{known}$}$ as the optimal one. Several results motivated us to reach this conclusion. The major motivation was the significant improvement of evidence ($\cal{Z}$) when fixing the number of non-transiting planets at 2.  This model yielded a RV semi-amplitude of 4.291 ± 0.46 ms$^{-1}$ for CoRoT-7b. 

The planetary mass was calculated using this semi-amplitude, assuming a circular orbit. The orbital inclination $i$ and stellar mass $M_{*}$ were obtained from \citet{2014A&A...569A..74B} as 80.98 ± 0.51 degree and 0.915 ± 0.017$M_{\odot}$ respectively. We determined the mass of CoRoT-7b to be 6.056 ± 0.653$M_{\oplus}$, making it a super-Earth. This mass measure is within the uncertainty limit of several previous studies \citep[e.g.,][]{2016A&A...588A..31F,2010A&A...520A..93H}. Incorporating with the planet radius of 1.528 ± 0.065 $R_{\oplus}$ from \citep{2014A&A...569A..74B}, the planet bulk density was calculated as 9.355 ± 0.235 g cm$^{-1}$.

We used the {\sc manipulateplanet} tool developed by \citet{2016ApJ...819..127Z} to get an estimate on the planetary composition. The derived value of mass and a minimum core pressure (3020 GPa) supports a planet with an iron core (Fe : 58.7\% by mass), Silicate mantle (MgSiO$_3$ : 40.9\%) and a thin layer of water on the surface    (H$_2$O : 0.9\%), favouring the rocky composition discussed in the literature. 

\begin{figure*}
    \centering
    \includegraphics[width=1.5\columnwidth]{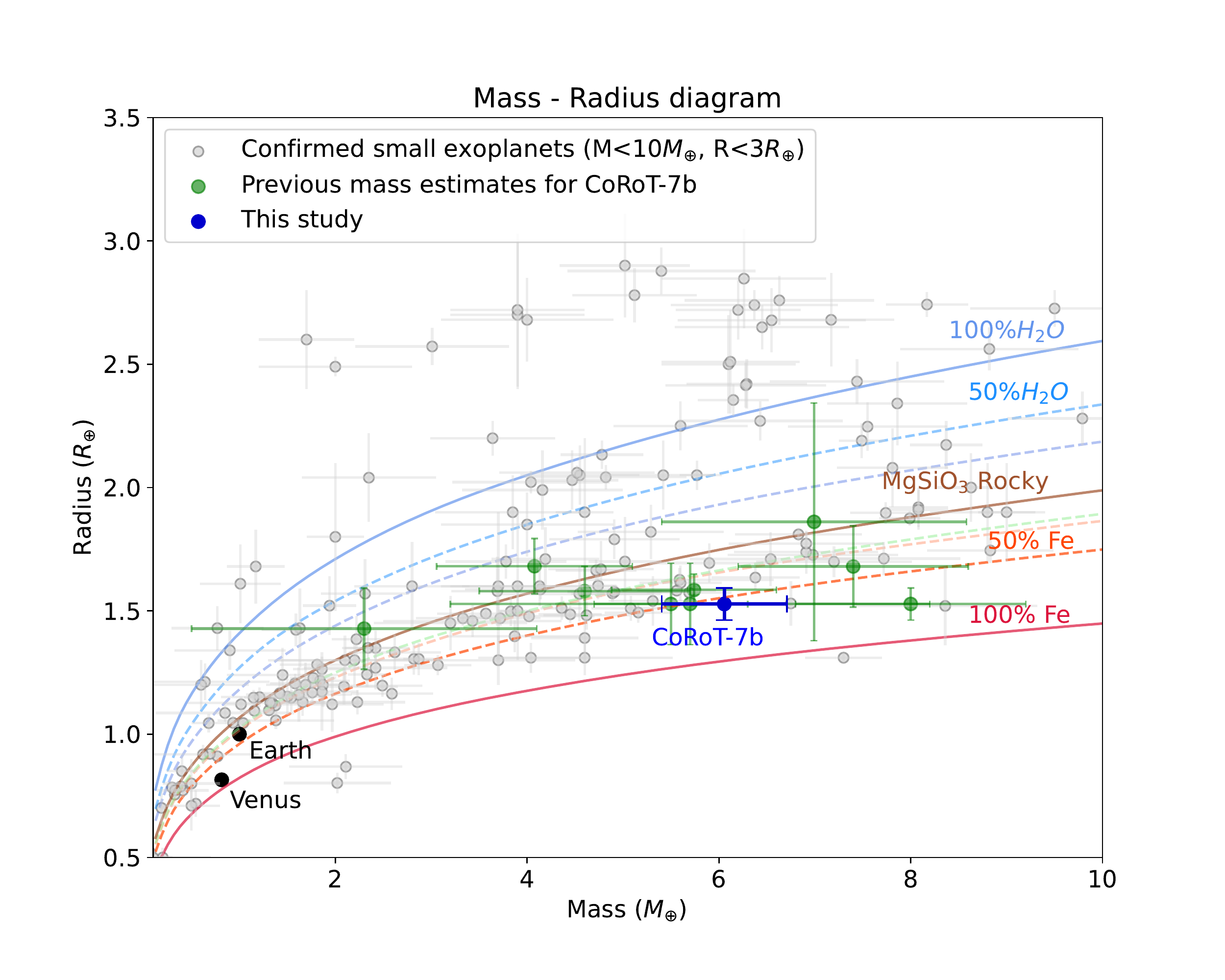}    \caption{Mass–radius relation for planets with masses <10 $M_{\odot}$ represented by grey circles. The plot also includes Earth and Venus, for reference. The lines show models of different compositions, with solid lines indicating single composition planets (either H2O, MgSiO3, i.e., rock, or Fe). The dashed and dotted lines indicate Mg-silicate planets with different amounts of H$_2$O and Fe. The green circles show the widely different mass estimates for the transiting planet CoRoT-7b from the literature, where the refined mass estimate from this study is marked in blue, with significant improvement in uncertainty. We want to point out the reason for a few points falling in a horizontal line with our estimate being the similar value of planetary radius considered in the corresponding analyses.}
    \label{fig:m-r}
    
\end{figure*}

For the given value of the mass and radius of the planet, we varied the core pressure in the model to investigate the degeneracy in the composition of the planet up to a maximum supported core pressure of 3346 GPa. This maximum core pressure proposes a dense iron core (Fe : 96.3 \%) immersed in deep ocean mantle (H$_2$O : 12.6 \%) with a compromised silicate layer(MgSiO$_3$ : 3.2 \%).  Also, the lower value of planet mass (M-1$\sigma$) suggests a composition that can be up to 78.6 \% Fe, 6.2\% MgSiO$_3$ and 15.1\% H$_2$O. However, none of the above compositions seem to be feasible, considering the high equilibrium temperature (1756 K) of CoRoT-7b estimated assuming a zero albedo \citep{2014A&A...569A..74B}. Therefore, the most plausible composition occurs to be the one discussed earlier in this section, shown in Figure \ref{fig:manipulateplanet} with little or no surface layer of water. It should be borne in mind that the {\sc manipulateplanet} tool assumes a condensed volatile layer typical of a cold planet. More recent studies, \citep[e.g.][]{2021ApJ...923..247Z} show that a substantial water layer would establish a deep steam atmosphere in a planet as strongly-irradiated as CoRoT-7b, increasing the radius well beyond the observed value.

We also explored the masses of known planet CoRoT-7b for different models, as tabulated in Table \ref{table:M&rho}. As seen from the table, the mass estimation improves and settles for the transiting + 2-planet model incorporated with GP regression and the {\sc scalpels} decorrelation. This approach takes into account the contributions of all additional planets whose existence may not be established conclusively, but whose RV variations influence the mass determination of the known planet(s).

\subsection {Masses of non-transiting planets}

As listed in Table \ref{table:ourmasses}, we also computed the masses of non-transiting planets from the RV semi-amplitudes obtained from the analysis described in Section \ref{sec:kima}. The inner non-transiting planet CoRoT-7c, with an orbital period of 3.69 days, was found to have a mass of 13.289 ± 0.689 $M_\oplus$. This planet could be structurally similar to Uranus ($\sim$ 14.5 $M_\oplus$) in our Solar System. 
The RV semi-amplitude corresponding to the second non-transiting planet CoRoT-7d at 8.97 days suggests a planetary mass of 17.14 ± 2.55 $M_\oplus$. The mass of this long-period planet is comparable to that of Neptune ($\sim$ 17.1 $M_\oplus$). However, the bulk density and composition of these non-transiting planets are unknown, due to the lack of radius information from the photometry.

\begin{table}
\caption{Planetary parameters for CoRoT-7b, CoRoT-7c and CoRoT-7d obtained from the best model.}
\centering

\begin{tabular}{|c c c c|}
\hline
Planet&P[days]& K[ms$^{-1}$]& M [$M_{\oplus}$] \\ [0.5ex] 
\hline
CoRoT-7b & 0.8535±5.87e-7 & 4.291±0.46 & 6.056±0.653 \\
CoRoT-7c & 3.697 ± 0.005 & 5.757 ± 0.298 & 13.289 ± 0.689 \\
CoRoT-7d & 8.966 ± 1.546 & 5.525 ± 0.792 & 17.142 ± 2.552 \\
\hline

\end{tabular}
\label{table:ourmasses}
\end{table}

\subsection{ Areas for further improvement}

The quasi-periodic signal with P$\sim$23 days showed up in the shift-driven periodogram is very close to the stellar rotation period that stands out as a clear peak in CoRoT photometry, activity-indicator periodograms and the {\sc scalpels} basis-vector periodograms (Figure \ref{fig:bgls_act} and \ref{fig:bgls_u},  marked with the rightmost dotted lines). This points out that the $\sim$23 days shift signal is more likely to be a stellar rotation artefact than a planetary reflex-motion signal. However, the presence of this signal in both shift periodograms (Figure \ref{fig:blind} and Figure \ref{fig:simultaneous}) and BGLS periodograms of activity indicators strengthens the conclusion reached in Section \ref{sec:blindpgm}: the property of this signal that produces a shape change in the profile that also closely mimics a shift. A note of caution is due here, since this confusing signal survived the {\sc scalpels} signal separation by imitating the behaviour of a real planet signal.
We anticipate this to be the effect of time-lags between the temporal coefficients of $\mathbf{U}$ vectors and the RVs, which makes a linear decorrelation difficult and incomplete.


\section{Conclusion}
\label{sec:conclusion}

In this paper, we present the results of applying the {\sc scalpels} to  extract precise RV estimates from the high-resolution spectroscopic planet surveys without using time-domain information. {\sc scalpels} construct an orthogonal basis containing the coefficients of the first few principal components of shape-induced CCF variations, then project the raw RVs onto this basis to obtain a time series of shape-induced RV variations. These are then subtracted from the original RVs to leave shift-only RV variations, where the planet signals are sought. 

By separating the activity-induced line-shape changes from the HARPS archival RV information of an active-star CoRoT-7, {\sc scalpels} isolated the planet-induced shift-driven signals. The consequent shift-driven RVs suggest that CoRoT-7 is best modelled as a 3-planet system, with quasi-periodic harmonics of the stellar rotation period also present. An ultra-short period (0.85 days) planet CoRoT-7b; which is transiting, a second planet CoRoT-7c with orbital period of 3.69 days together with a third companion CoRoT-7d at a long-period orbit of 8.97 days, make CoRoT-7 an ultra-compact planetary system. Both the long-period planets (CoRoT-7c \& CoRoT-7d) are non-transiting.

We used the $\ell_1$ periodogram, a compressed-sensing method developed by \citet{2017MNRAS.464.1220H}, to identify the sparsest set of orbital periods that could simultaneously fit the {\sc scalpels} shift signal. The resulting sparse solution eliminated unwanted alias signals and revealed three dominant candidate signals. This analysis favoured the results from the {\sc scalpels} projection, that CoRoT-7 is better modelled as a 3-planet system than a 2-planet system. Because the $\sim$23-d signal varies slowly in phase and amplitude, we noted the usefulness of a split-period representation for reproducing the detailed frequency structure of the periodogram and for determining the orbital velocity amplitudes of the three dominant planet signals.

A simultaneous modelling of stellar variability and planetary motion was performed to study the impact of the {\sc scalpels} signal separation on the CoRoT-7 system, considering planet signals of known periods. A good match was observed for a 5-signal model comprising three planet signals + two closely-spaced periods near 23\,d representing the quasi-periodic stellar rotation signal, as suggested by the $\ell_1$ periodogram. 
This showed the effectiveness of the {\sc scalpels} in reducing the contribution of stellar variability to the RV signal and enabling the detection of exoplanets' signals in data from active stars. 

The posterior probability distributions for the orbital parameters were determined along with a complete Keplerian solver and a Gaussian Process using {\sc kima} \citep{2018JOSS....3..487F}. This GP model, incorporating  the {\sc scalpels} basis vectors for line-shape decorrelation, is found to be 4 times more probable than a model without {\sc scalpels}. The improvement in the evidence again suggested a system with two non-transiting planets and the known transiting planet.

In addition to improving the reliability of planet detection, our method also allows more rigorous determination of the masses of known, transiting planets whose bulk densities reveal their interior composition. This approach takes into account the contributions of all additional (non-transiting) planets whose existence may not be established conclusively, but whose RV variations influence the mass determination of the known planet(s). Combined with modelling of any surviving shift-like signal from stellar activity, this approach offers a significant improvement in the precise characterization of  exoplanet systems, especially for the low-mass planets whose RV amplitudes  are close to the detection threshold.

\section*{Acknowledgements}
AAJ acknowledges the support from World leading St Andrews Doctoral Scholarship. ACC and TGW acknowledge support from STFC consolidated grant numbers ST/R000824/1 and ST/V000861/1, and UKSA grant number ST/R003203/1. AAJ, ACC and TGW thank Nathan C. Hara for the productive discussions on $\emph{l}$1-periodogram. We are also grateful to Joao P. Faria whose algorithm ({\sc kima}) played a significant role in validating our results.

\section*{Data availability}

This paper uses of data from the Data \& Analysis Center for Exoplanets (DACE) database dedicated to the visualization, exchange and analysis of extrasolar planets' data. The HARPS data products used for the analyses are publicly available in the DACE platform (https://dace.unige.ch). The {\sc PYTHON} codes and notebooks used to generate the results and diagrams in this paper will be made available through the University of St Andrews Research Portal.





\begin{thebibliography}{}


\bibitem[\protect\citeauthoryear{} .] A Adler, Roy \& Konheim, Alan(1962). PROC AMER MATH SOC. 13. 425-425. 

\bibitem[\protect\citeauthoryear{Aigrain, Pont, \& Zucker}{2012}]{2012MNRAS.419.3147A} Aigrain S., Pont F., Zucker S., 2012, MNRAS, 419, 3147. doi:10.1111/j.1365-2966.2011.19960.x



\bibitem[\protect\citeauthoryear{Barros et al.}{2014}]{2014A&A...569A..74B} Barros S.~C.~C., Almenara J.~M., Deleuil M., Diaz R.~F., Csizmadia S., Cabrera J., Chaintreuil S., et al., 2014, A\&A, 569, A74.



\bibitem[\protect\citeauthoryear{Bedell et al.}{2019}]{2019AJ....158..164B} Bedell M., Hogg D.~W., Foreman-Mackey D., Montet B.~T., Luger R., 2019, AJ, 158, 164.


\bibitem[\protect\citeauthoryear{Brewer}{2014}]{2014arXiv1411.3921B} Brewer B.~J., 2014, arXiv, arXiv:1411.3921

\bibitem[\protect\citeauthoryear{Brewer, P{\'a}rtay, \& Cs{\'a}nyi}{2010}]{2010ascl.soft10029B} Brewer B.~J., P{\'a}rtay L.~B., Cs{\'a}nyi G., 2010, ascl.soft. ascl:1010.029






Chib \& Ivan Jeliazkov (2001) Marginal Likelihood From the Metropolis–Hastings Output, Journal of the American Statistical Association, 96:453, 270-281


\bibitem[\protect\citeauthoryear{Collier Cameron et al.}{2021}]{2021MNRAS.505.1699C} Collier Cameron A., Ford E.~B., Shahaf S., Aigrain S., Dumusque X., Haywood R.~D., Mortier A., et al., 2021, MNRAS, 505, 1699. 

\bibitem[\protect\citeauthoryear{Cretignier et al.}{2021}]{2021A&A...653A..43C} Cretignier M., Dumusque X., Hara N.~C., Pepe F., 2021, A\&A, 653, A43. 

\bibitem[\protect\citeauthoryear{Dai et al.}{2017}]{2017AJ....154..226D} Dai F., Winn J.~N., Gandolfi D., Wang S.~X., Teske J.~K., Burt J., Albrecht S., et al., 2017, AJ, 154, 226. 

\bibitem[\protect\citeauthoryear{Davis et al.}{2017}]{2017ApJ...846...59D} Davis A.~B., Cisewski J., Dumusque X., Fischer D.~A., Ford E.~B., 2017, ApJ, 846, 59. 

\bibitem[\protect\citeauthoryear{de Beurs et al.}{2021}]{2021AAS...23733204D} de Beurs Z.~L., Vanderburg A., Shallue C.~J., Harps-N Collaboration, 2021, AAS

\bibitem[\protect\citeauthoryear{Dobbs-Dixon, Lin, \& Mardling}{2004}]{2004ApJ...610..464D} Dobbs-Dixon I., Lin D.~N.~C., Mardling R.~A., 2004, ApJ, 610, 464. doi:10.1086/421510

\bibitem[\protect\citeauthoryear{Dorn et al.}{2015}]{2015A&A...577A..83D} Dorn C., Khan A., Heng K., Connolly J.~A.~D., Alibert Y., Benz W., Tackley P., 2015, A\&A, 577, A83.

\bibitem[\protect\citeauthoryear{Dumusque et al.}{2011}]{2011IAUS..276..527D} Dumusque X., Santos N.~C., Udry S., Lovis C., Bonfils X., 2011, IAUS, 276, 527.

\bibitem[\protect\citeauthoryear{Dumusque, Boisse, \& Santos}{2014}]{2014ApJ...796..132D} Dumusque X., Boisse I., Santos N.~C., 2014, ApJ, 796, 132. 

\bibitem[\protect\citeauthoryear{Dumusque et al.}{2015}]{2015ApJ...814L..21D} Dumusque X., Glenday A., Phillips D.~F., Buchschacher N., Collier Cameron A., Cecconi M., Charbonneau D., et al., 2015, ApJL, 814, L21.

\bibitem[\protect\citeauthoryear{Dumusque}{2018}]{2018A&A...620A..47D} Dumusque X., 2018, A\&A, 620, A47.

\bibitem[\protect\citeauthoryear{Faria et al.}{2016}]{2016A&A...588A..31F} Faria J.~P., Haywood R.~D., Brewer B.~J., Figueira P., Oshagh M., Santerne A., Santos N.~C., 2016, A\&A, 588, A31.

\bibitem[\protect\citeauthoryear{Faria et al.}{2018}]{2018JOSS....3..487F} Faria J.~P., Santos N.~C., Figueira P., Brewer B.~J., 2018, JOSS, 3, 487. 

\bibitem[\protect\citeauthoryear{Ferraz-Mello et al.}{2011}]{2011A&A...531A.161F} Ferraz-Mello S., Tadeu Dos Santos M., Beaug{\'e} C., Michtchenko T.~A., Rodr{\'\i}guez A., 2011, A\&A, 531, A161.

\bibitem[\protect\citeauthoryear{Figueira et al.}{2010}]{2010A&A...513L...8F} Figueira P., Marmier M., Bonfils X., di Folco E., Udry S., Santos N.~C., Lovis C., et al., 2010, A\&A, 513, L8.


\bibitem[\protect\citeauthoryear{Fischer et al.}{2016}]{2016PASP..128f6001F} Fischer D.~A., Anglada-Escude G., Arriagada P., Baluev R.~V., Bean J.~L., Bouchy F., Buchhave L.~A., et al., 2016, PASP, 128, 066001.


\bibitem[\protect\citeauthoryear{Gregory}{2016}]{2016MNRAS.458.2604G} Gregory P.~C., 2016, MNRAS, 458, 2604.

\bibitem[\protect\citeauthoryear{Guillot \& Gautier}{2014}]{2014arXiv1405.3752G} Guillot T., Gautier D., 2014, arXiv, arXiv:1405.3752.

\bibitem[\protect\citeauthoryear{Hara et al.}{2017}]{2017MNRAS.464.1220H} Hara N.~C., Bou{\'e} G., Laskar J., Correia A.~C.~M., 2017, MNRAS, 464, 1220. 

\bibitem[\protect\citeauthoryear{Hara et al.}{2021}]{2021arXiv210506995H} Hara N.~C., Unger N., Delisle J.-B., D{\'\i}az R., S{\'e}gransan D., 2021, arXiv, arXiv:2105.06995

\bibitem[\protect\citeauthoryear{Hatzes et al.}{2010}]{2010A&A...520A..93H} Hatzes A.~P., Dvorak R., Wuchterl G., Guterman P., Hartmann M., Fridlund M., Gandolfi D., et al., 2010, A\&A, 520, A93.

\bibitem[\protect\citeauthoryear{Hatzes et al.}{2011}]{2011ApJ...743...75H} Hatzes A.~P., Fridlund M., Nachmani G., Mazeh T., Valencia D., H{\'e}brard G., Carone L., et al., 2011, ApJ, 743, 75. 

\bibitem[\protect\citeauthoryear{Hatzes}{2013}]{2013AN....334..616H} Hatzes A.~P., 2013, AN, 334, 616.

\bibitem[\protect\citeauthoryear{Haywood et al.}{2014}]{2014MNRAS.443.2517H} Haywood R.~D., Collier Cameron A., Queloz D., Barros S.~C.~C., Deleuil M., Fares R., Gillon M., et al., 2014, MNRAS, 443, 2517. 

\bibitem[\protect\citeauthoryear{Haywood et al.}{2018}]{2018AJ....155..203H} Haywood R.~D., Vanderburg A., Mortier A., Giles H.~A.~C., L{\'o}pez-Morales M., Lopez E.~D., Malavolta L., et al., 2018, AJ, 155, 203.

\bibitem[\protect\citeauthoryear{Howard et al.}{2011}]{2011ApJ...726...73H} Howard A.~W., Johnson J.~A., Marcy G.~W., Fischer D.~A., Wright J.~T., Henry G.~W., Isaacson H., et al., 2011, ApJ, 726, 73.

\bibitem[\protect\citeauthoryear{Jeffreys}{1961}]Jeffreys S. H., 1961, The Theory of Probability. Oxford Univ. Press, Oxford


\bibitem[\protect\citeauthoryear{Jurgenson et al.}{2016}]{Jurgenson2016} Jurgenson C., et al., 2016, SPIE,  99086T, SPIE.9908 

\bibitem[\protect\citeauthoryear{Klein et al.}{2022}]{2022MNRAS.512.5067K} Klein B., Zicher N., Kavanagh R.~D., Nielsen L.~D., Aigrain S., Vidotto A.~A., Barrag{\'a}n O., et al., 2022, MNRAS, 512, 5067. doi:10.1093/mnras/stac761

\bibitem[\protect\citeauthoryear{Ksanfomaliti}{1999}]{1999SoSyR..33..482K} Ksanfomaliti L.~V., 1999, SoSyR, 33, 482



\bibitem[\protect\citeauthoryear{L{\'e}ger et al.}{2009}]{2009A&A...506..287L} L{\'e}ger A., Rouan D., Schneider J., Barge P., Fridlund M., Samuel B., Ollivier M., et al., 2009, A\&A, 506, 287.

\bibitem[\protect\citeauthoryear{Lomb}{1976}]{1976Ap&SS..39..447L} Lomb N.~R., 1976, Ap\&SS, 39, 447

\bibitem[\protect\citeauthoryear{L{\'o}pez-Morales et al.}{2016}]{2016AJ....152..204L} L{\'o}pez-Morales M., Haywood R.~D., Coughlin J.~L., Zeng L., Buchhave L.~A., Giles H.~A.~C., Affer L., et al., 2016, AJ, 152, 204.

\bibitem[\protect\citeauthoryear{Martins et al.}{2018}]{2018MNRAS.478.5240M} Martins J.~H.~C., Figueira P., Santos N.~C., Melo C., Mu{\~n}oz A.~G., Faria J., Pepe F., et al., 2018, MNRAS, 478, 5240

\bibitem[\protect\citeauthoryear{Meunier}{2021}]{2021arXiv210406072M} Meunier N., 2021, arXiv, arXiv:2104.06072

\bibitem[\protect\citeauthoryear{Mayor et al.}{1995}]{1995IAUC.6251....1M} Mayor M., Queloz D., Marcy G., Butler P., Noyes R., Korzennik S., Krockenberger M., et al., 1995, IAUC, 6251

\bibitem[\protect\citeauthoryear{M{\'e}gevand et al.}{2014}]{Megevand2014} M{\'e}gevand D., et al., 2014, SPIE,  91471H, SPIE.9147 


\bibitem[\protect\citeauthoryear{Mortier \& Collier Cameron}{2017}]{2017A&A...601A.110M} Mortier A., Collier Cameron A., 2017, A\&A, 601, A110. 

\bibitem[\protect\citeauthoryear{Motalebi et al.}{2015}]{2015A&A...584A..72M} Motalebi F., Udry S., Gillon M., Lovis C., S{\'e}gransan D., Buchhave L.~A., Demory B.~O., et al., 2015, A\&A, 584, A72. 

\bibitem[\protect\citeauthoryear{Nava et al.}{2020}]{2020AJ....159...23N} Nava C., L{\'o}pez-Morales M., Haywood R.~D., Giles H.~A.~C., 2020, AJ, 159, 23. 



\bibitem[\protect\citeauthoryear{Pepe et al.}{2000}]{2000SPIE.4008..582P} Pepe F., Mayor M., Delabre B., Kohler D., Lacroix D., Queloz D., Udry S., et al., 2000, SPIE, 4008, 582. 

\bibitem[\protect\citeauthoryear{Pepe et al.}{2010}]{2010SPIE.7735E..0FP} Pepe F.~A., Cristiani S., Rebolo Lopez R., Santos N.~C., Amorim A., Avila G., Benz W., et al., 2010, SPIE, 7735, 77350F

\bibitem[\protect\citeauthoryear{Pont, Aigrain, \& Zucker}{2011}]{2011MNRAS.411.1953P} Pont F., Aigrain S., Zucker S., 2011, MNRAS, 411, 1953.


\bibitem[\protect\citeauthoryear{Queloz et al.}{2001}]{2001A&A...379..279Q} Queloz D., Henry G.~W., Sivan J.~P., Baliunas S.~L., Beuzit J.~L., Donahue R.~A., Mayor M., et al., 2001, A\&A, 379, 279.

\bibitem[\protect\citeauthoryear{Queloz et al.}{2009}]{2009A&A...506..303Q} Queloz D., Bouchy F., Moutou C., Hatzes A., H{\'e}brard G., Alonso R., Auvergne M., et al., 2009, A\&A, 506, 303.


\bibitem[\protect\citeauthoryear{Robertson, Roy, \& Mahadevan}{2015}]{2015ApJ...805L..22R} Robertson P., Roy A., Mahadevan S., 2015, ApJL, 805, L22. 

\bibitem[\protect\citeauthoryear{Saar \& Donahue}{1997}]{1997ApJ...485..319S} Saar S.~H., Donahue R.~A., 1997, ApJ, 485, 319

\bibitem[\protect\citeauthoryear{Salabert et al.}{2016}]{2016A&A...596A..31S} Salabert D., Garc{\'\i}a R.~A., Beck P.~G., Egeland R., Pall{\'e} P.~L., Mathur S., Metcalfe T.~S., et al., 2016, A\&A, 596, A31.

\bibitem[\protect\citeauthoryear{Santos et al.}{2014}]{2014A&A...566A..35S} Santos N.~C., Mortier A., Faria J.~P., Dumusque X., Adibekyan V.~Z., Delgado-Mena E., Figueira P., et al., 2014, A\&A, 566, A35. 

\bibitem[\protect\citeauthoryear{Santos \& Buchhave}{2018}]{2018haex.bookE.181S} Santos N.~C., Buchhave L.~A., 2018, haex.book, 181. 

\bibitem[\protect\citeauthoryear{Scargle}{1982}]{1982ApJ...263..835S} Scargle J.~D., 1982, ApJ, 263, 835

\bibitem[\protect\citeauthoryear{Seager et al.}{2007}]{2007ApJ...669.1279S} Seager S., Kuchner M., Hier-Majumder C.~A., Militzer B., 2007, ApJ, 669, 1279.

\bibitem[\protect\citeauthoryear{Schwab et al.}{2016}]{Schwab2016} Schwab C., et al., 2016, SPIE,  99087H, SPIE.9908  

\bibitem[\protect\citeauthoryear{Su{\'a}rez Mascare{\~n}o et al.}{2020}]{2020A&A...639A..77S} Su{\'a}rez Mascare{\~n}o A., Faria J.~P., Figueira P., Lovis C., Damasso M., Gonz{\'a}lez Hern{\'a}ndez J.~I., Rebolo R., et al., 2020, A\&A, 639, A77

\bibitem[\protect\citeauthoryear{Thompson et al.}{2016}]{2016SPIE.9908E..6FT} Thompson S.~J., Queloz D., Baraffe I., Brake M., Dolgopolov A., Fisher M., Fleury M., et al., 2016, SPIE, 9908, 99086F. 


\bibitem[\protect\citeauthoryear{Tuomi et al.}{2014}]{2014arXiv1405.2016T} Tuomi M., Anglada-Escude G., Jenkins J.~S., Jones H.~R.~A., 2014, arXiv, arXiv:1405.2016

\bibitem[\protect\citeauthoryear{Udry \& Mayor}{2001}]{2001ESASP.496...65U} Udry S., Mayor M., 2001, ESASP, 496, 65


\bibitem[\protect\citeauthoryear{Vogt et al.}{2015}]{2015ApJ...814...12V} Vogt S.~S., Burt J., Meschiari S., Butler R.~P., Henry G.~W., Wang S., Holden B., et al., 2015, ApJ, 814, 12

\bibitem[\protect\citeauthoryear{Wilson}{1978}]{1978ApJ...226..379W} Wilson O.~C., 1978, ApJ, 226, 379. doi:10.1086/156618

\bibitem[\protect\citeauthoryear{Wilson et al.}{2022}]{2022MNRAS.511.1043W} Wilson T.~G., Goffo E., Alibert Y., Gandolfi D., Bonfanti A., Persson C.~M., Collier Cameron A., et al., 2022, MNRAS, 511, 1043. 

\bibitem[\protect\citeauthoryear{Wright \& Robertson}{2017}]{2017RNAAS...1...51W} Wright J.~T., Robertson P., 2017, RNAAS, 1, 51

\bibitem[\protect\citeauthoryear{Wu et al.}{2021}]{2021MNRAS.503.3032W} Wu Y., Chen S., Wang P., Zhou S., Feng Y., Zhang W., Wei R., 2021, MNRAS, 503, 3032 

\bibitem[\protect\citeauthoryear{Zechmeister \& K{\"u}rster}{2009}]{2009A&A...496..577Z} Zechmeister M., K{\"u}rster M., 2009, A\&A, 496, 577

\bibitem[\protect\citeauthoryear{Zeng, Sasselov, \& Jacobsen}{2016}]{2016ApJ...819..127Z} Zeng L., Sasselov D.~D., Jacobsen S.~B., 2016, ApJ, 819, 127. 

\bibitem[\protect\citeauthoryear{Zeng et al.}{2021}]{2021ApJ...923..247Z} Zeng L., Jacobsen S.~B., Hyung E., Levi A., Nava C., Kirk J., Piaulet C., et al., 2021, ApJ, 923, 247. doi:10.3847/1538-4357/ac3137

\bibitem[\protect\citeauthoryear{Zhao \& Ford}{2022}]{2022arXiv220103780Z} Zhao J., Ford E.~B., 2022, arXiv, arXiv:2201.03780

\bibitem[\protect\citeauthoryear{Zhao \& Tinney}{2020}]{2020MNRAS.491.4131Z} Zhao J., Tinney C.~G., 2020, MNRAS, 491, 4131. doi:10.1093/mnras/stz3254


\end{thebibliography}


\appendix
\section{}
\subsection{SCALPELS }
 \label{sec:appscalpels}
\begin{figure*}
    \centering
    \includegraphics[width=2\columnwidth]{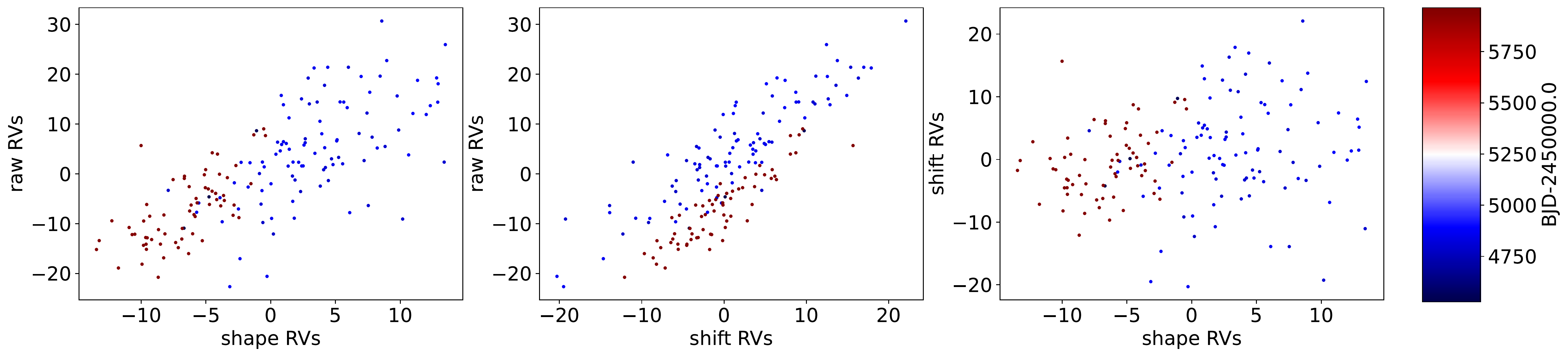}
    
    \caption{The correlation plots for the observed velocities ($v_{obs}$), shape-driven component ($v_\parallel$) and the shift component ($v_\perp$) }
    \label{fig:appcorrelation}
\end{figure*}

The {\sc scalpels} identified shape-driven perturbations are strongly correlated with the observed RVs, as seen from the first panel of Figure \ref{fig:appcorrelation}, which could reliably reproduce the stellar activity dominated long-term and short-term fluctuations. 

\section{}
\subsection{Blind search periodograms }
\begin{figure*}
    \centering
    \includegraphics[width=0.6\columnwidth]{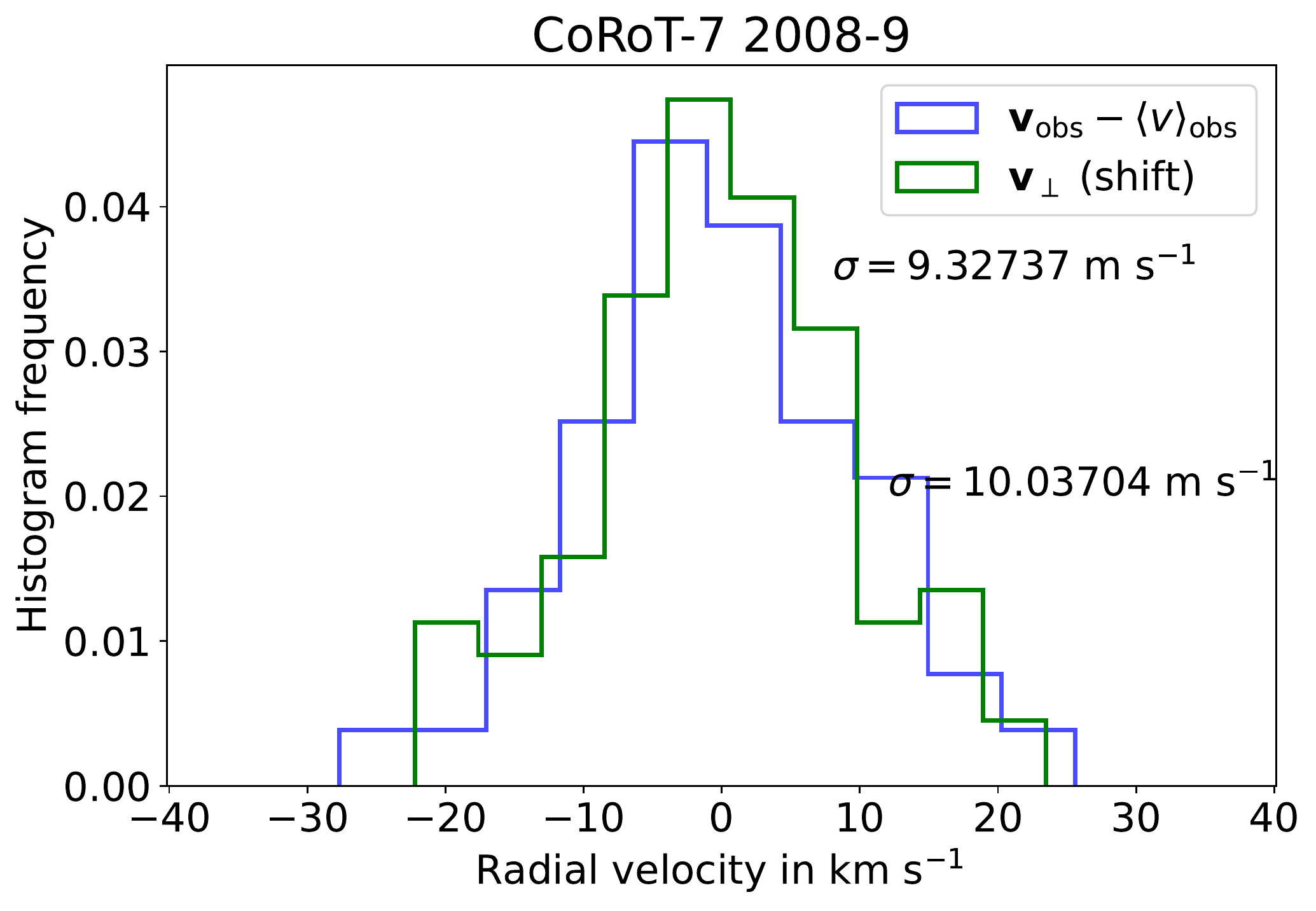}
    \includegraphics[width=0.6\columnwidth]{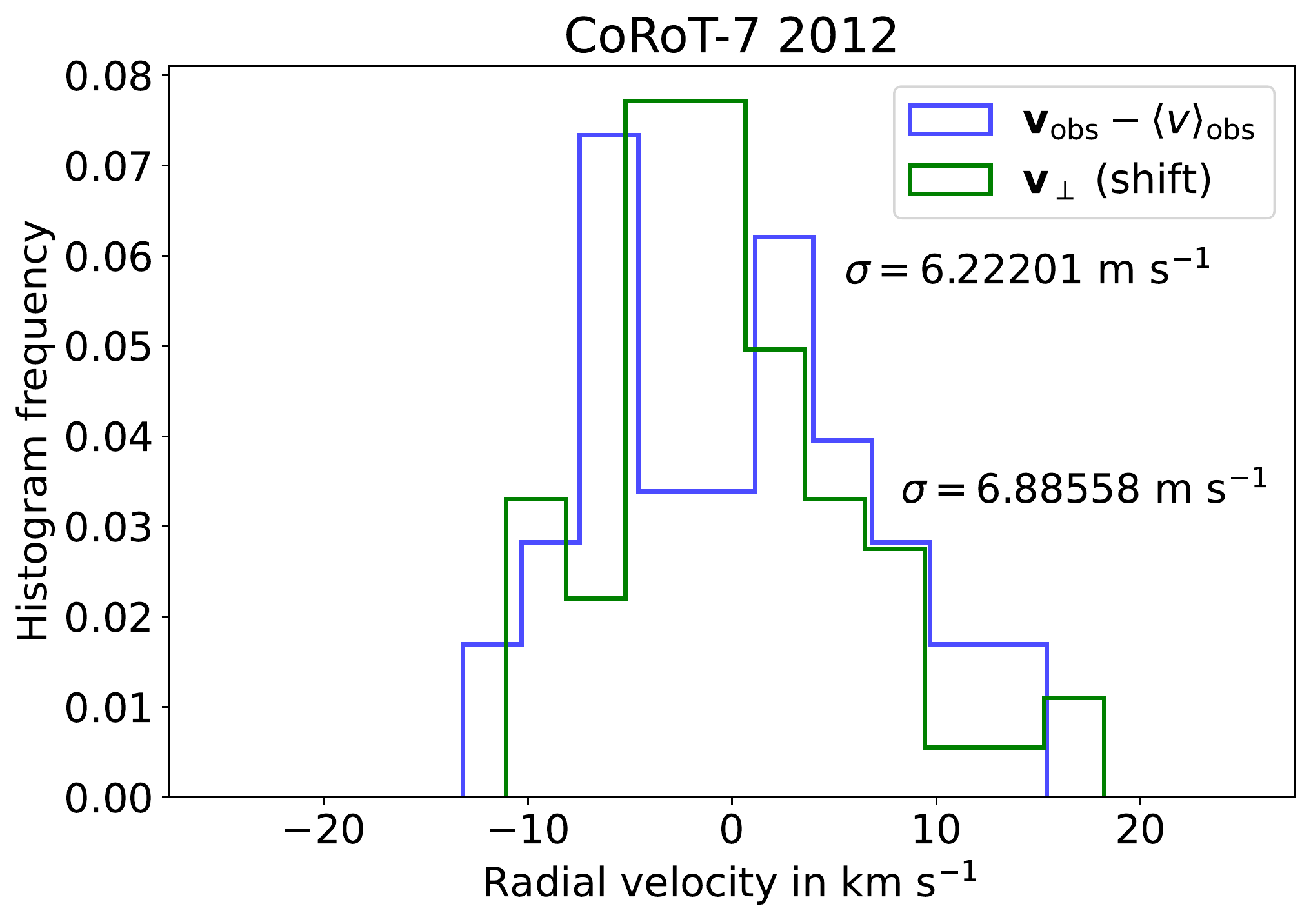}
    \includegraphics[width=0.6\columnwidth]{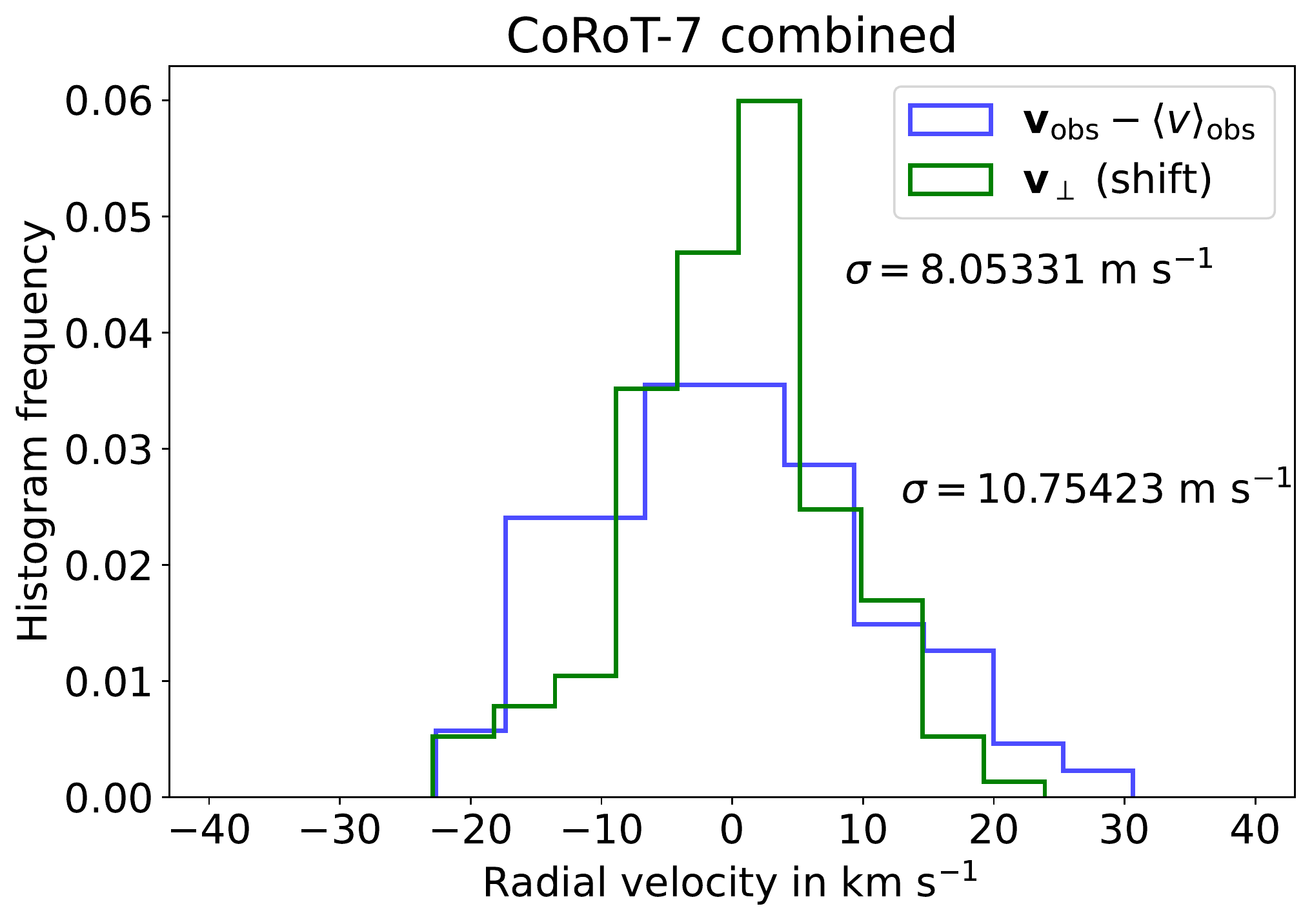}
    \caption{Histograms showing the decrease in RMS scatter in the shift-driven velocities (shown in green) during the blind-period search for 2008-9, 2012 and combined data respectively.}
    
    \label{fig:hist}
\end{figure*}

We applied {\sc scalpels} to the barycentric RVs of CoRoT-7, and examined the results using a Generalized Lomb-Scargle periodogram \citep{2009A&A...496..577Z} for different subsets of the data (2008-9, 2012 and combined). Histograms showing the reduction of RMS scatter after separating the shape-driven signals are presented for all three analyses in Figure \ref{fig:hist}.
\label{sec:appblind}

\subsection{Simultaneous sinusoidal fit : Steps }
\label{sec:appsimul}

After computing the Singular Value Decomposition(SVD) of the ACF rows;
\begin{enumerate}
    \item Compute $\mathbf{F} = \{\cos\omega_1 t_j, \sin\omega_1 t_j, \cdots, \cos\omega_n t_j, \sin\omega_n t_j\}$.
    \item Concatenate $\mathbf{A}_{(m\times(2n+l))}=\left[\mathbf{F}_{m\times 2n}\ \ \  \mathbf{U}_{A,(m\times l)}\right]$ using reduced rank $l$.
    \item Compute reduced-rank covariance matrix using:
    \begin{equation}
    \frac{1}{m}\mathbf{R}^T_{k_{\rm max}}\cdot \mathbf{R}_{k_{\rm max}} =
    \frac{1}{m}(\mathbf{P}_{C,k_{\rm max}} \cdot {\rm diag}(\mathbf{S}_{C,k_{\rm max}}^2)\cdot \mathbf{P}_{C,k_{\rm max}}^T).
    \label{eq:covreduced}
    \end{equation}
    \item Compute row variances $\sigma^2_j$ of $\mathbf{R}-\mathbf{R}_{k_{\rm max}}$
    \item Compute model of full covariance matrix as:
    \begin{equation}
    \Sigma_{\xi,\eta}(t_j) \simeq 
 \sigma^2_j\times {\rm max}\left(1-\frac{|v_\xi-v_\eta|}{\delta v},0\right)
 + \frac{1}{m}\mathbf{R}^T_{k_{\rm max}}\cdot \mathbf{R}_{k_{\rm max}}.
    \label{eq:covmatrix}
    \end{equation}
    \item Compute $\mathbf{C'}$ and  and ${\rm Var}(\mathbf{v}(t_j))$ using:
    \begin{equation}
    {\rm Var}(\mathbf{v}(t_j)) = \frac{1}
    {\mathbf{C'}(v,t_j)^T \cdot \mathbf{\Sigma}^{-1}(t_j) \cdot \mathbf{C'}(v,t_j)}.
    \label{eq:getvar}
    \end{equation}
    \item Construct $\Sigma = {\rm Diag}({\rm Var}(\mathbf{v}(t_j)))$.
    \item Solve \begin{equation}
    (\mathbf{A}^T \cdot\mathbf{\Sigma}^{-1}\cdot\mathbf{A})\cdot\mathbf{\theta} =
    \mathbf{A}^T \cdot\mathbf{\Sigma}^{-1}\cdot\mathbf{\mathbf{v}_{\rm obs}} 
    \label{eq:solvetheta}
    \end{equation} to obtain $\mathbf{\theta}$ and ${\rm Var}(\mathbf{\theta})=1/{\rm Diag}(\mathbf{A}^T \cdot\mathbf{\Sigma}^{-1}\cdot\mathbf{A})$.
    \item Partition $\mathbf{\theta}$ into coefficients of columns of $\mathbf{F}$ and $\hat{\mathbf{\alpha}}$.
    
    \item Compute $\mathbf{v}_\parallel= \hat{\mathbf{\alpha}}\cdot{\mathbf{U}_A}$ in reduced-rank basis.
    \item Compute $\mathbf{v}_\perp=\mathbf{v}_{\rm obs}-\mathbf{v}_\parallel$.
\end{enumerate}
{\bf Return:} RV amplitudes and variances, $\mathbf{v}_\parallel$, $\mathbf{v}_\perp$.

\begin{figure*}
    \centering
    \includegraphics[width=1.2\columnwidth]{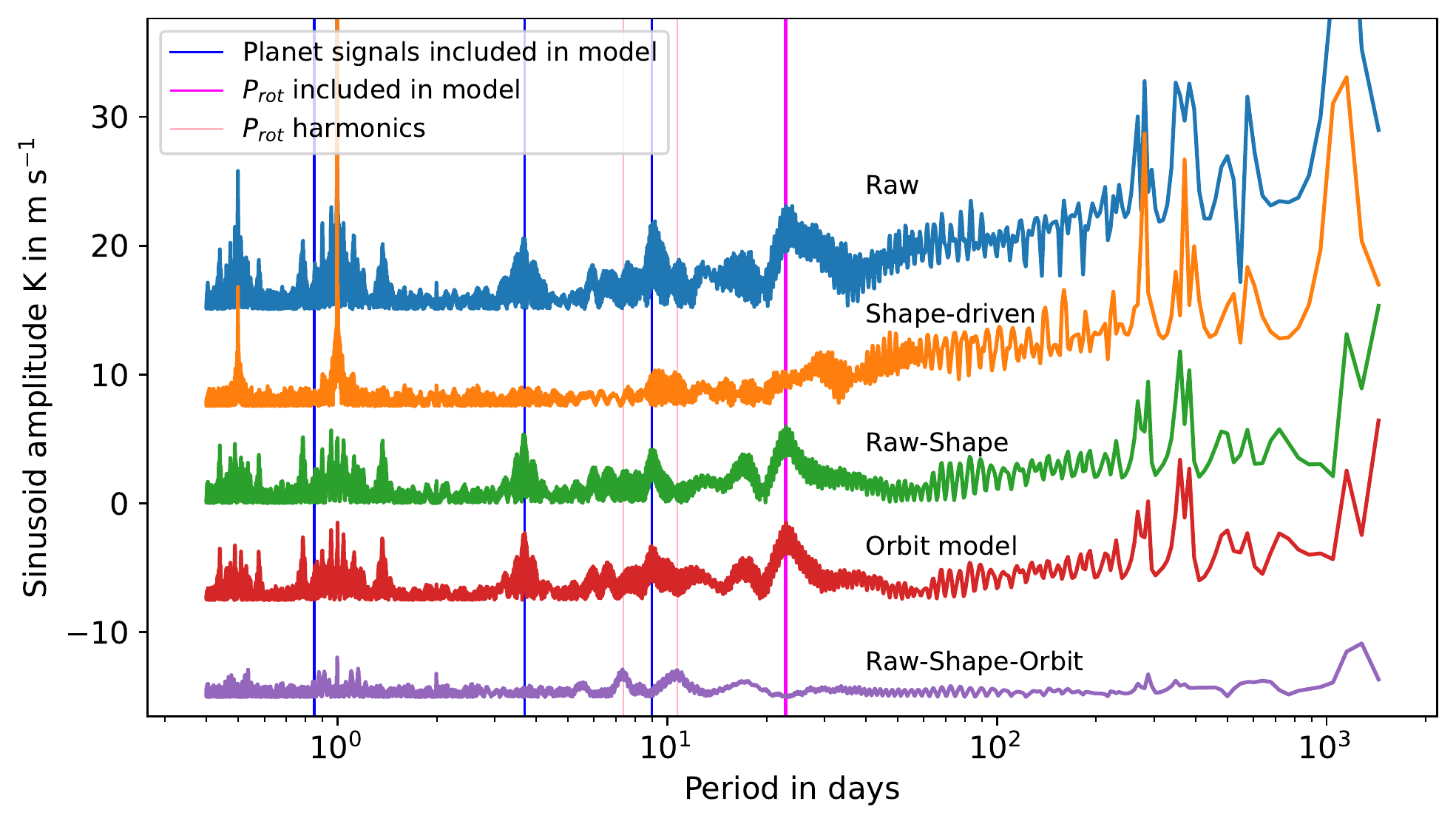}
    \includegraphics[width=0.8\columnwidth]{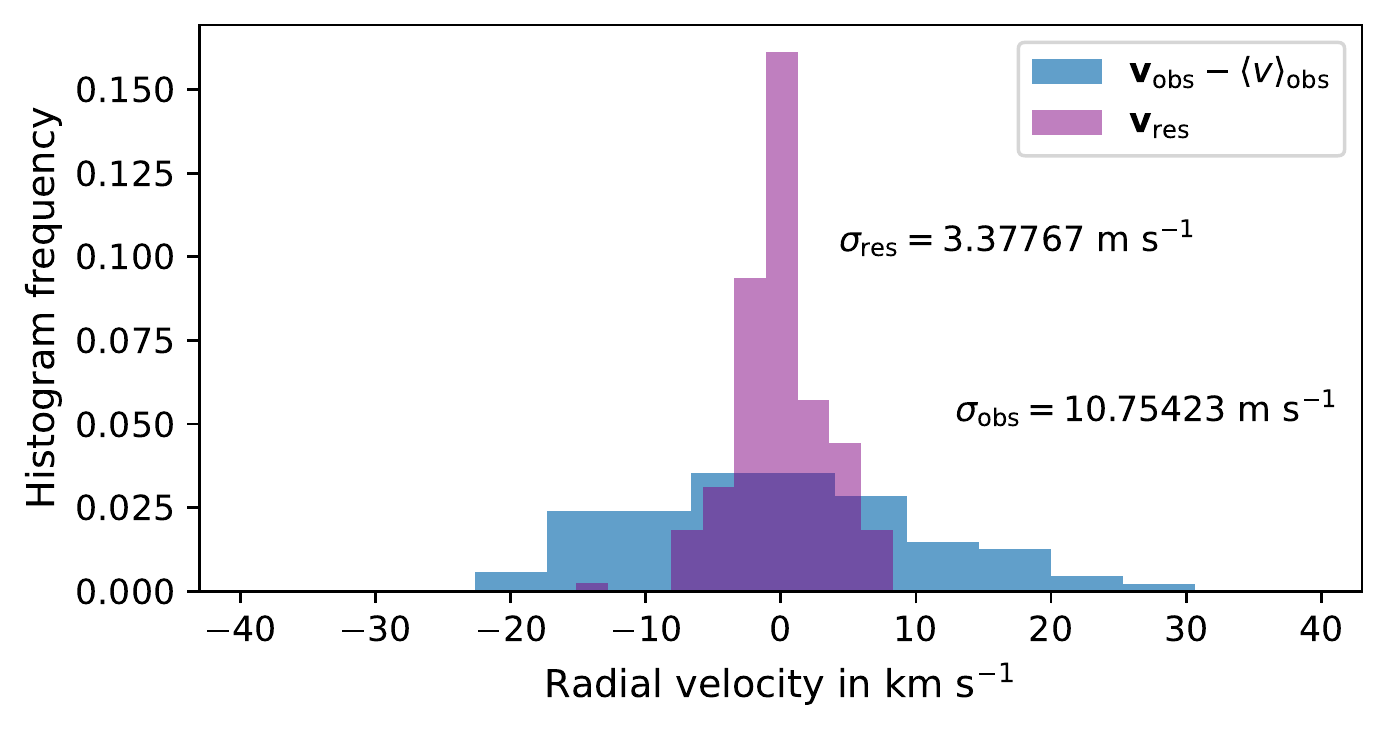}
    \includegraphics[width=1\columnwidth]{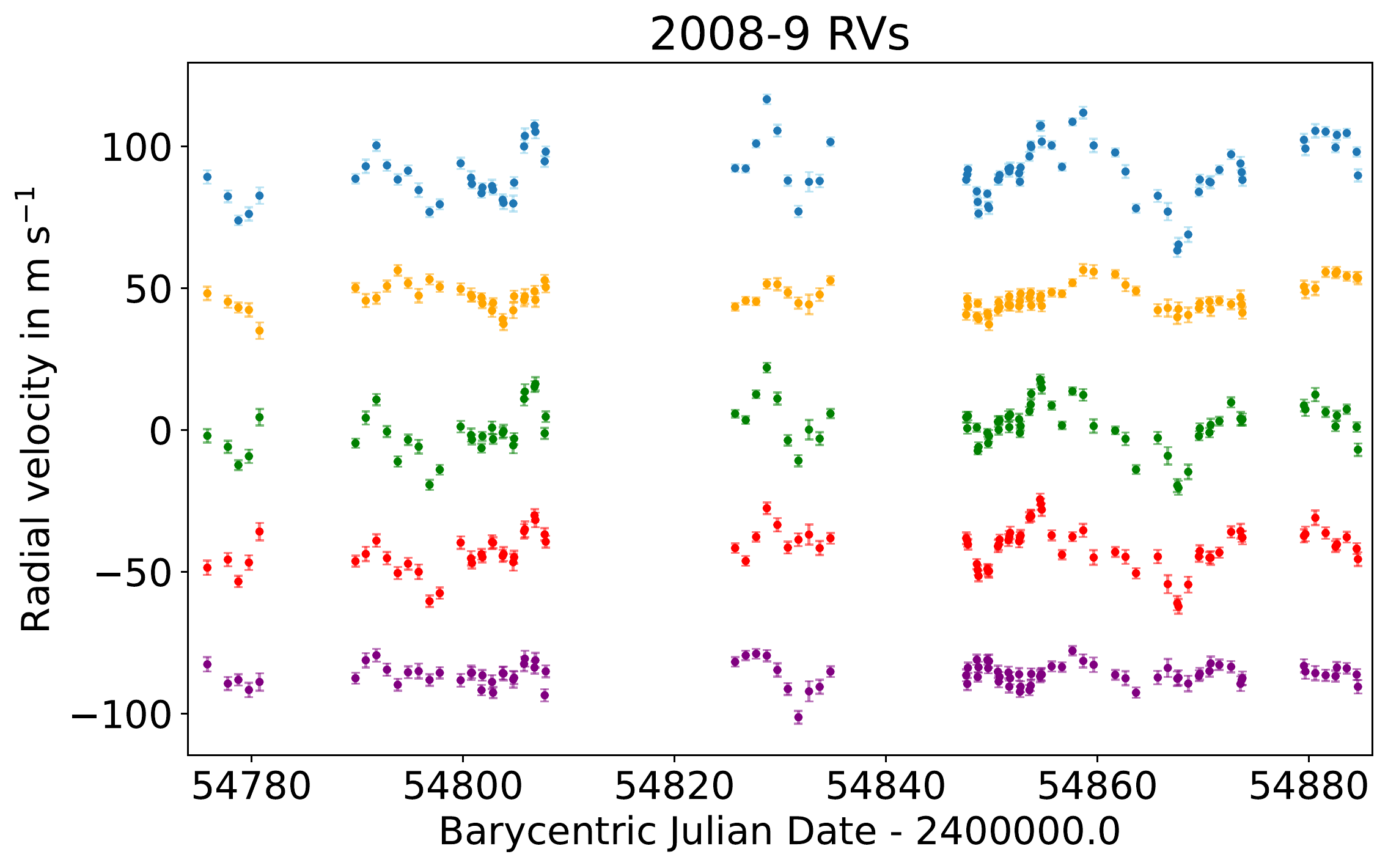}
    \includegraphics[width=1\columnwidth]{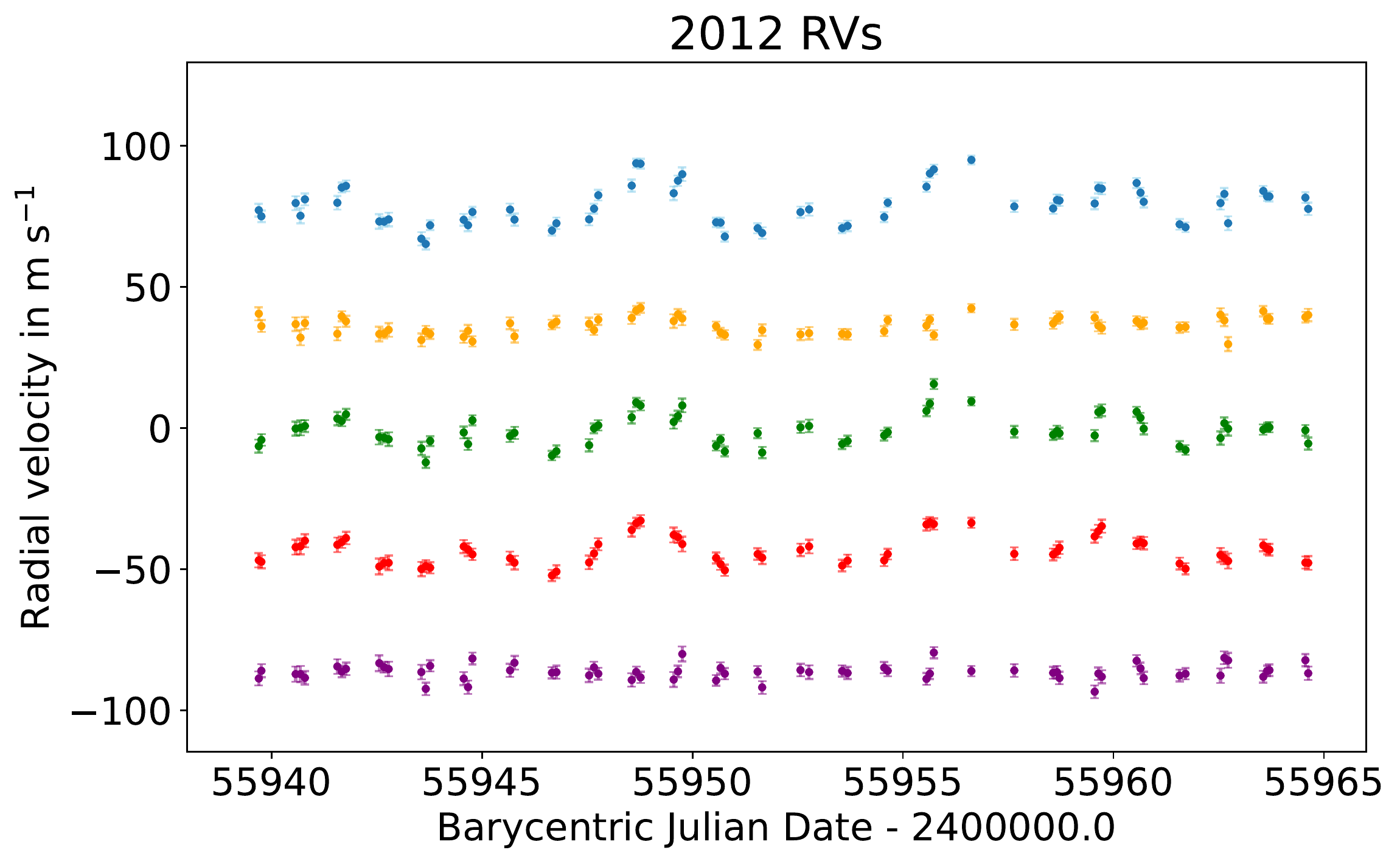}
    \caption{$\mathbf{Left}$: All the signals found by the ${\ell_1}$- periodogram are fitted including the 3 planet candidate signals (blue vertical lines) and 1 pairs of closely spaced beat period associated with the rotation period (magenta vertical lines). The orbital model signal (red) matches with the shift signal (green), except for the harmonics (light pink vertical bars) of the stellar rotation period. $\mathbf{Right}$:The resultant histogram of residuals with significantly reduced RMS scatter is shown.}
    \label{fig:appsimfig}
    
\end{figure*}

\begin{table}
\caption{$\mathbf{Top}$: The periods and semi-amplitudes of the strongest signals in the periodograms from simultaneous modelling of CCF shape changes and planetary motion, made with prior knowledge of the periods (from l1) of 3 planet candidates an a single period representing the stellar rotation are listed. $\mathbf{Bottom}$ A pair of beat period corresponding to the stellar rotation is included instead of the single period, considering the varying phase and amplitude.}
\centering
\begin{tabular}{|c c c|}
\hline
P(days) &K(ms$^{-1}$) &$\sigma_K$(ms$^{-1}$)\\
\hline
 


0.8535   &  2.915 & 0.280\\
3.6963  &  5.112 & 0.302\\
8.9674   &  3.223 & 0.292\\
22.9425  &  5.437 & 0.321 \\
\hline
0.8535 & 3.099 & 0.284 \\
3.6963 & 5.343 & 0.303\\
8.9674 & 3.320 & 0.293 \\
22.9425  &  5.142 & 0.326 \\
23.6968 &  2.923 & 0.337 \\
\hline 
\end{tabular}
\label{table:4&5sim}
\end{table}
\section{}
\subsection{\texorpdfstring{$l_{1}$}{l1}-periodograms for shape RVs \& activity indicators}
\begin{figure}
    \centering
    \includegraphics[width=0.83\columnwidth]{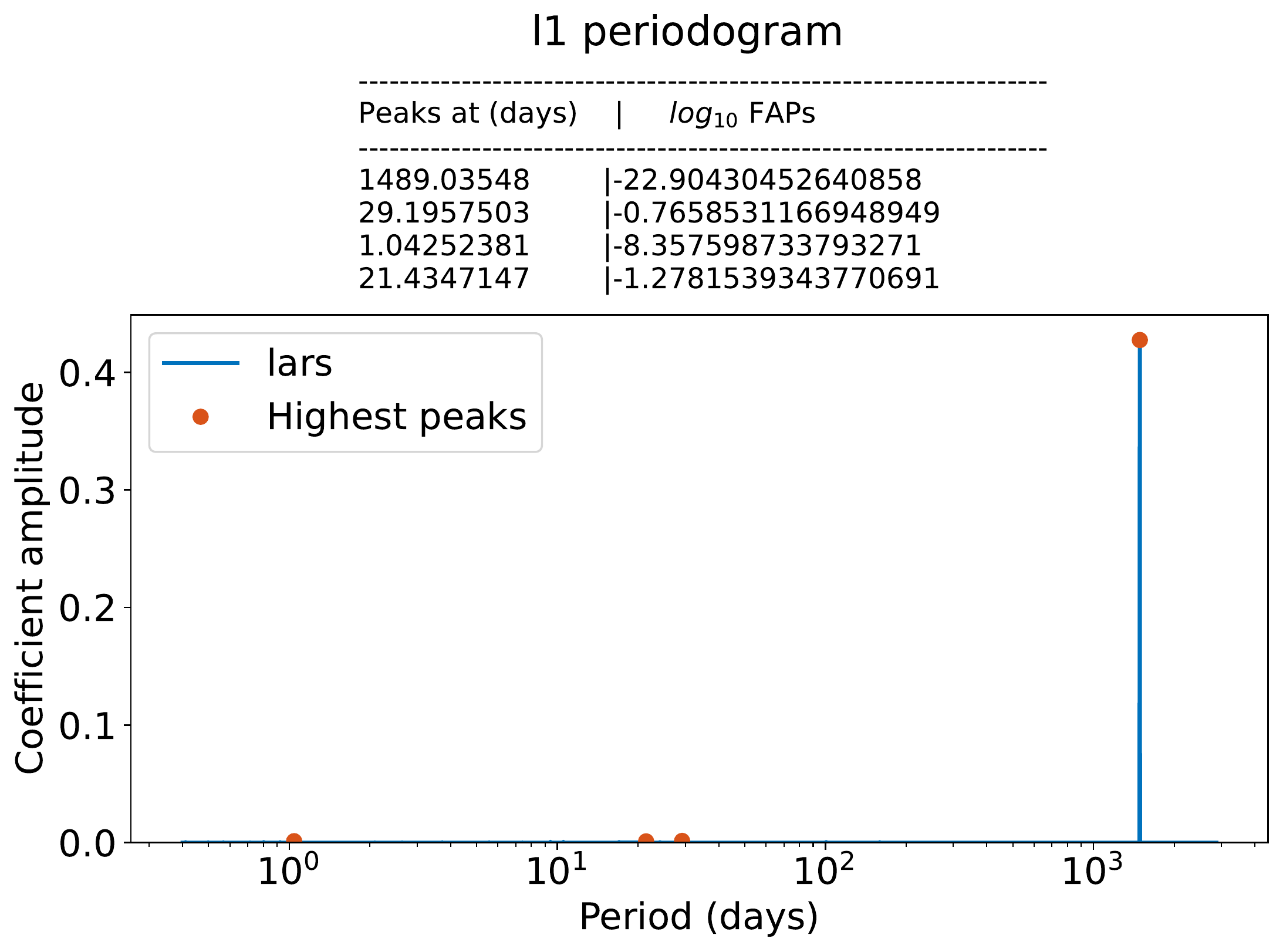}
    \includegraphics[width=0.83\columnwidth]{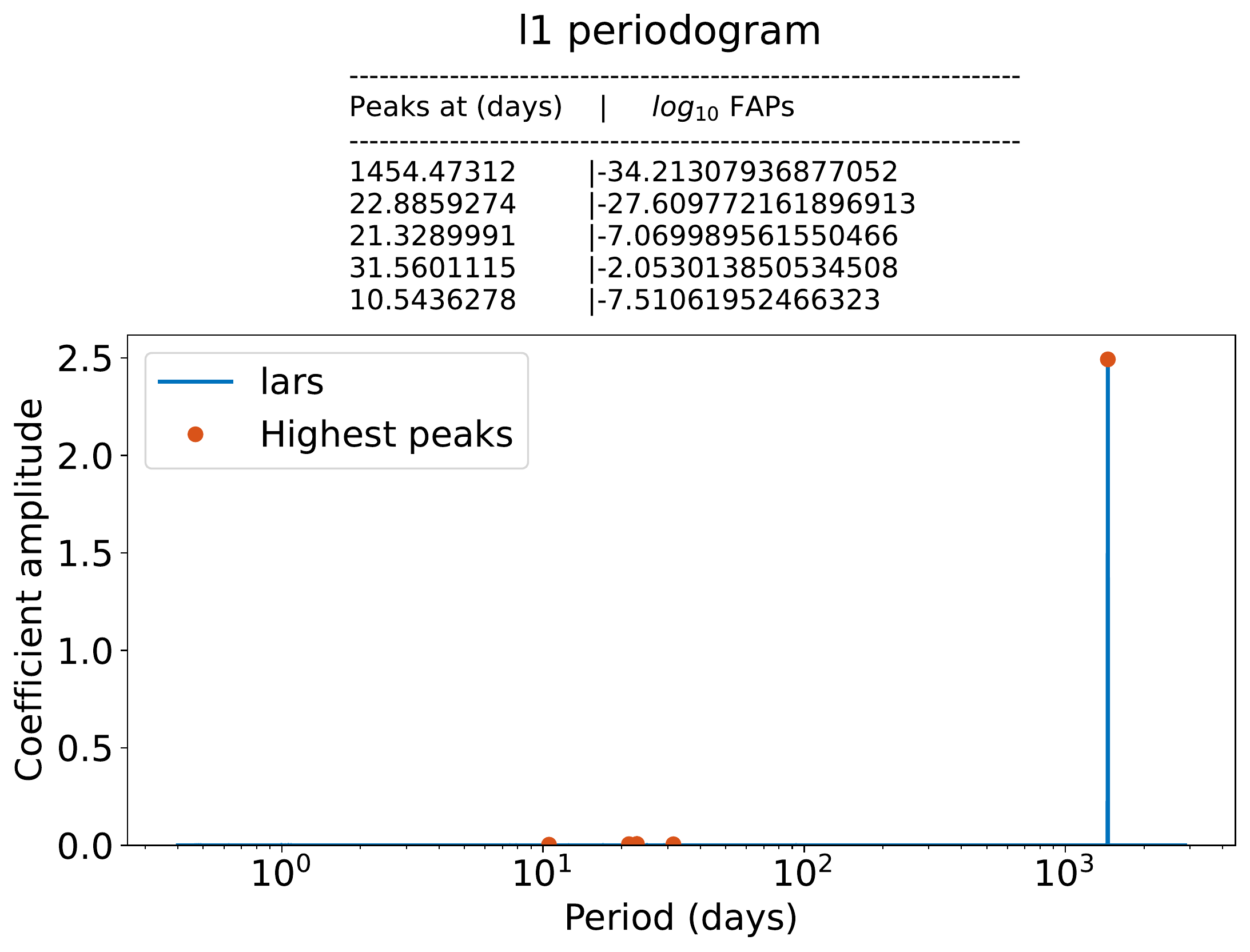}
    \includegraphics[width=0.83\columnwidth]{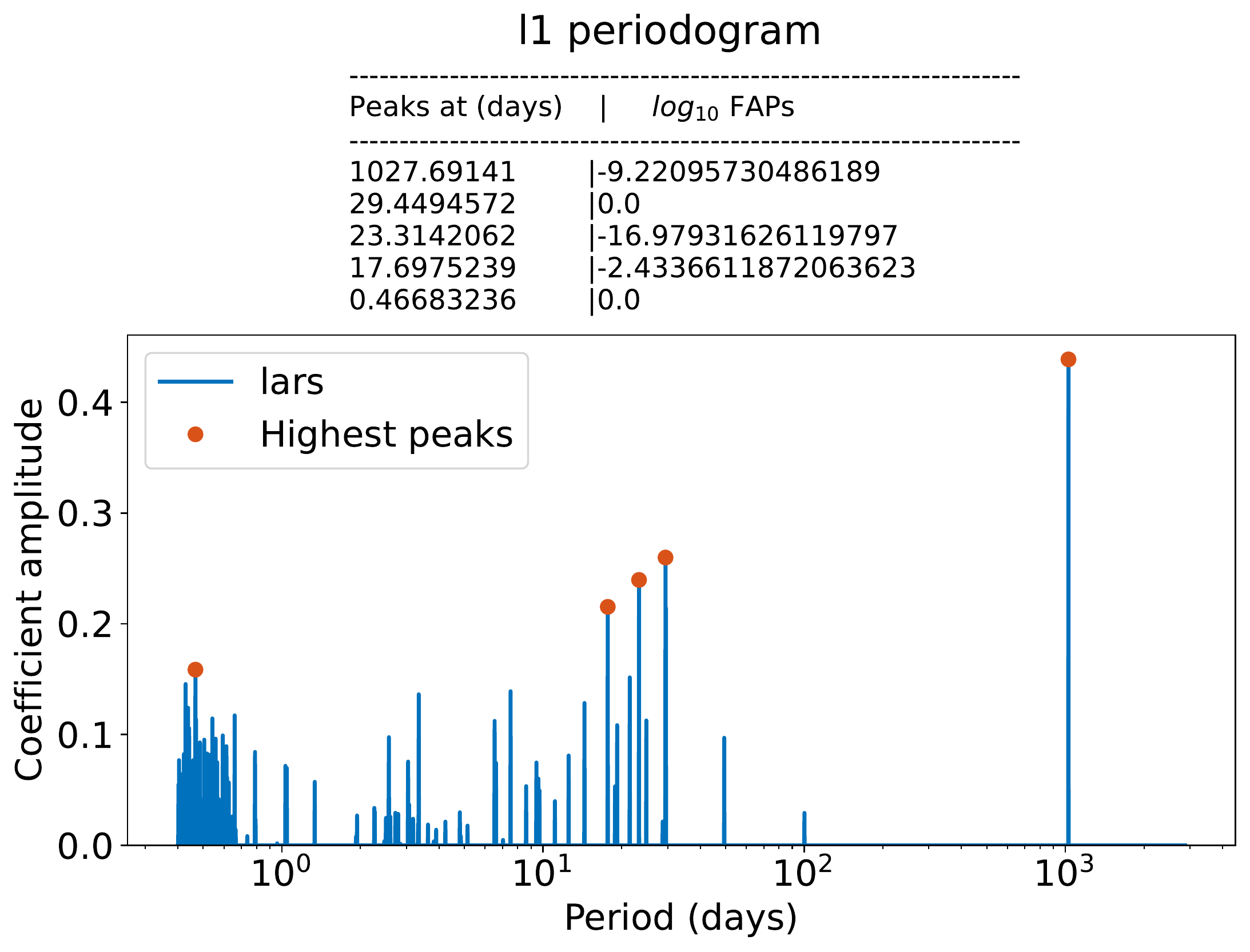}
    \includegraphics[width=0.83\columnwidth]{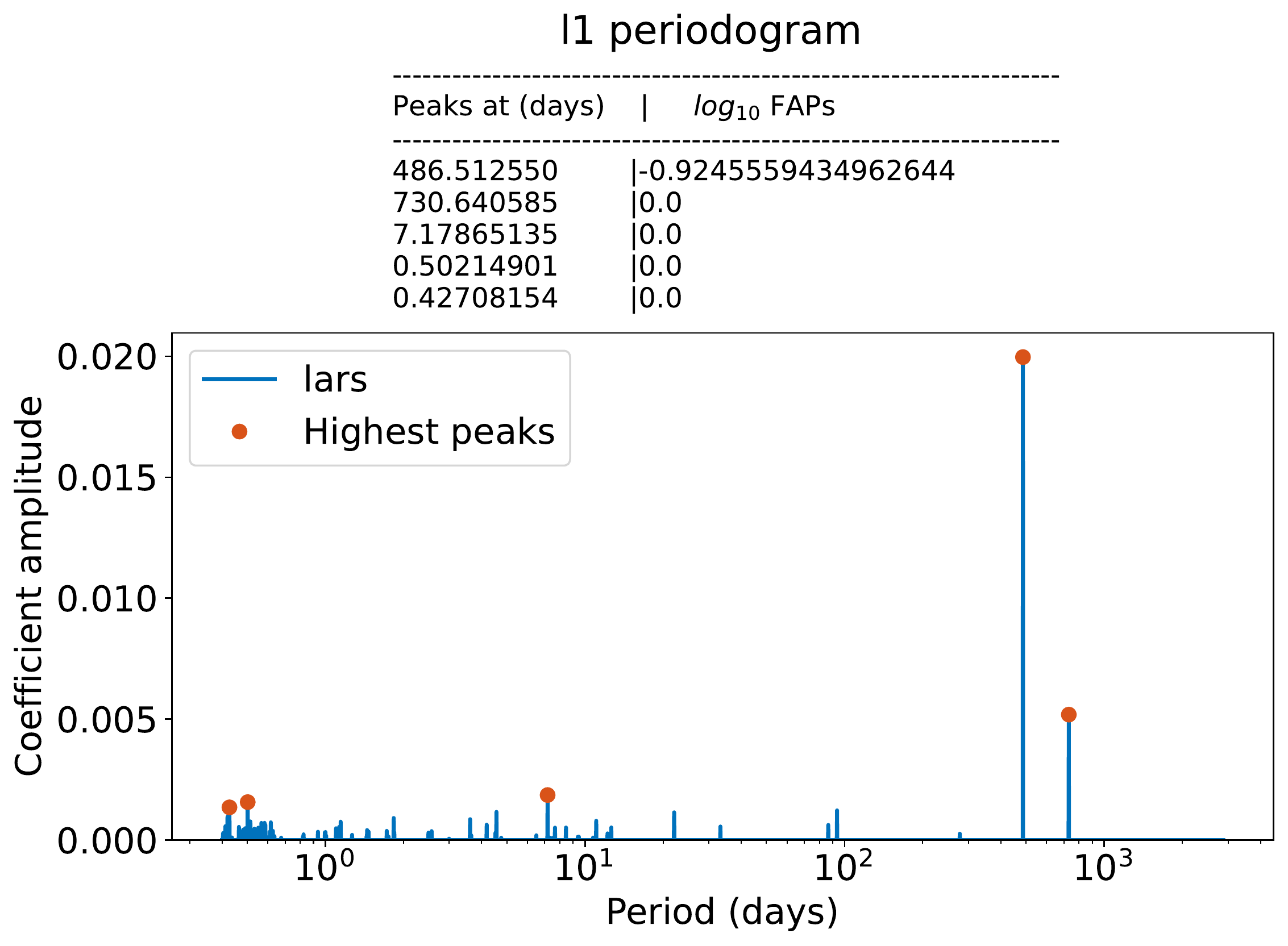}
    \caption{ $\mathbf{Top}$: The ${\ell_1}$- periodogram for the shape-driven RVs and traditional activity indicators like FWHM, Area and the BIS are shown from top to the bottom panels. None of these show significant signals at 0.85 days, 3.69 days and 8.96 days, adding to the case for a dynamical origin. The corresponding False Alarm Probabilities (FAP) are also listed above each panel.}
    \label{fig:l1_act}
    
\end{figure}

The physical origin of the planet candidate signals, were validated by computing additional $\ell_1$ periodograms for the shape-driven RVs and the activity-indicators, as shown in Figure \ref{fig:l1_act}.
\label{sec:l1_act_appendix}

\section{}
\subsection{BGLS periodogram for activity indicators}
\label{sec:bgls_act_appendix}

We studied the shape changes of CCFs comprehensively to investigate intrinsic stellar variability. The spectral parameters such as FWHM, BIS, and Area are used for this diagnostic, as these offer different measures of shape changes in the CCF profile. The Ca\,{\sc ii}\,$H$\&\ $K$ S-index is also used as a proxy for active regions on the stellar surface. While the FWHM carries information about  change in the width of the line profile, the BIS records the profile asymmetry. The Area is obtained by taking the product of FWHM and the central line depth. The time series information of these all these indicators for different observing runs are detailed in Section \ref{sec:act} along with Figure \ref{fig:activity}.

\begin{figure}
  \centering
  \includegraphics[width=1\columnwidth]{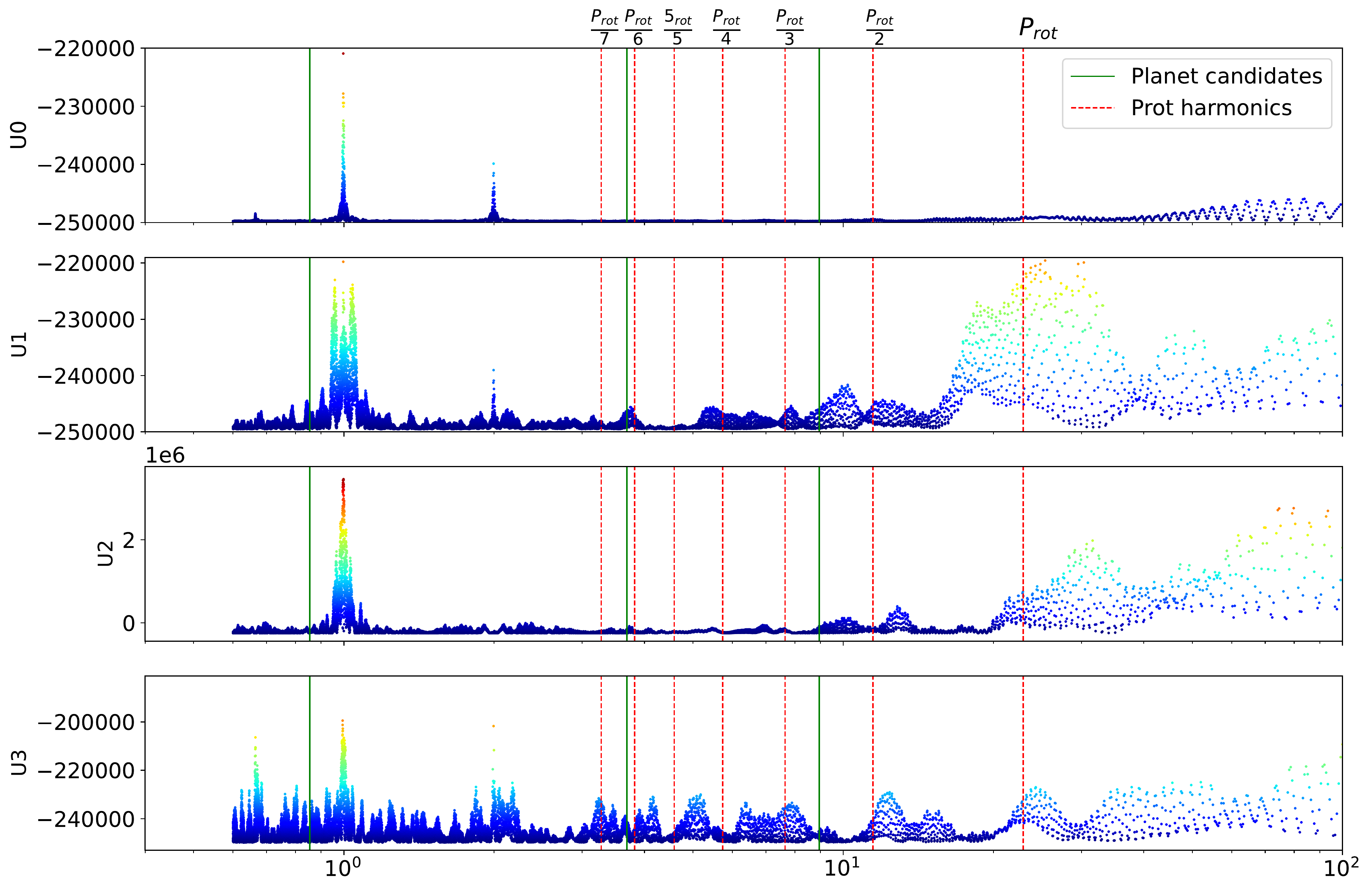}
  
  \caption{Bayesian Generalized Lomb-Scargle periodograms of the 4 leading principal components of the residual CCF of RV time series. Except the first, all other principal components exhibit power around the rotation period and the first and second harmonics, denoted by red-dotted vertical bars at $\sim$23 d, $\sim$11.5 days and $\sim$7.6 d. Similar vertical bars in green marks the periods of candidate planet signals.}
\label{fig:bgls_u}
\end{figure}

\begin{figure}
  \centering
  \includegraphics[width=1\columnwidth]{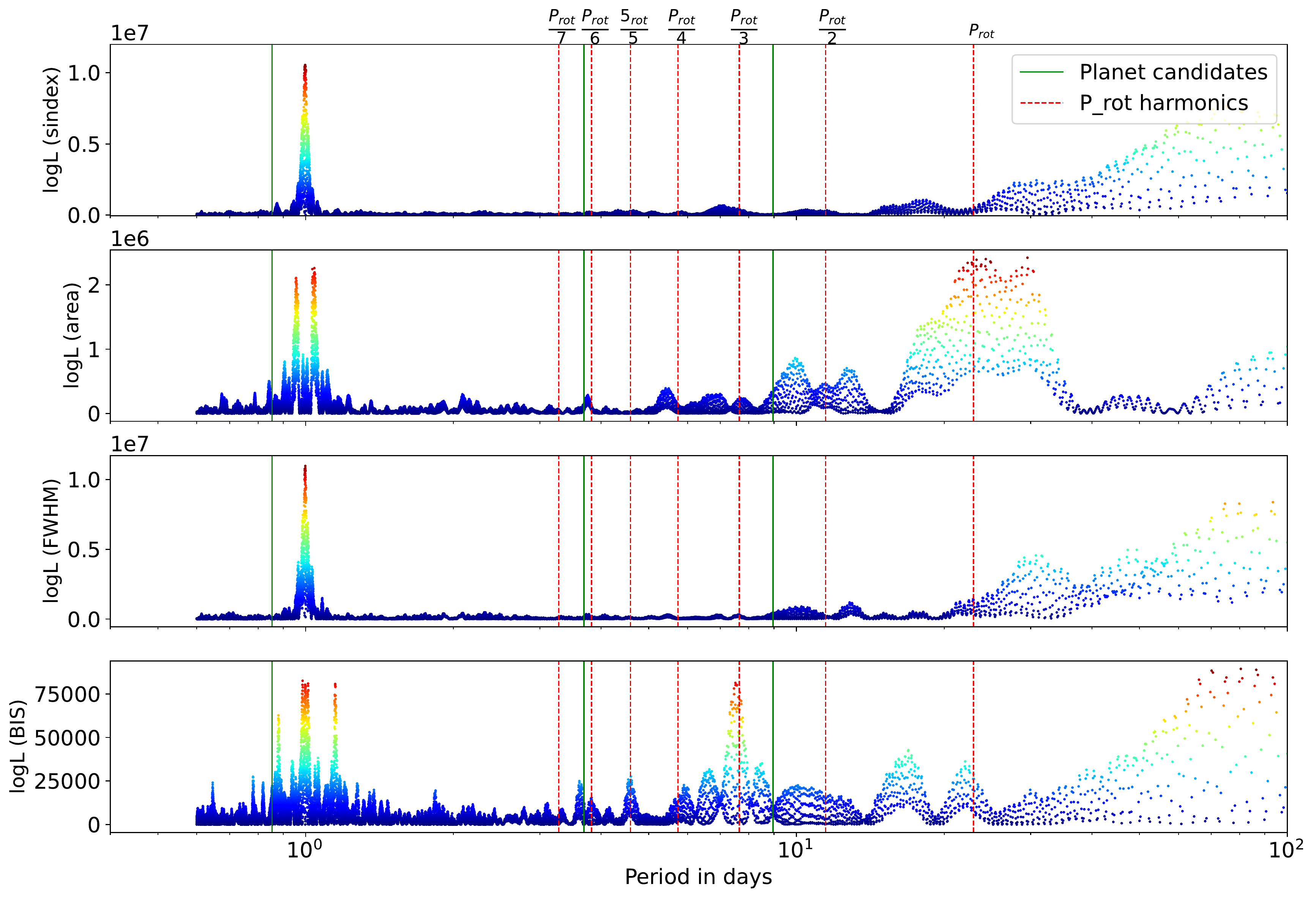}
  \caption{ Bayesian Generalized Lomb-Scargle periodograms of activity proxies such as S-index, Area, Full Width Half Maximum (FWHM) and Bisector span (BIS). The vertical lines have the same meaning as in the panel above.}
  \label{fig:bgls_act}
\end{figure}

Here, we analyse the Bayesian Generalized Lomb-Scargle periodograms of the above-mentioned activity indicators (Figure \ref{fig:bgls_act}). The periods corresponding to the stellar rotation period and its harmonics are represented by red dotted vertical lines, while the candidate signal periods are marked 
in green. The primary intention is to examine whether any of these periodograms has peaks at or near the candidate signals of interest. We also examined the BGLS periodograms of the first 4 leading principal components that contributed significantly in modelling the stellar activity (of Figure \ref{fig:bgls_u}), to understand how efficiently the Principal Component Analysis of the ACF correlate with the known aforesaid activity proxies.

We find that the second principal component ($\mathbf{U}_1$) exhibits a strong resemblance to the variability of CCF area. Another correlation is found between the third principal component($\mathbf{U}_2$) and the FWHM, indicating that the changes in the width of the profile are also reliably considered while modelling the shape-driven component. 

Except the first principal component and the S-index, all other principal components and activity indicators trace down the stellar rotation period efficiently and hence show significant power at $\sim$23d. Unsurprisingly, no traces of power are found in any of the periodograms at 0.85 days, 3.70 days and 8.97 days , thus ruling out activity as a plausible origin for these signals. We also noticed that, while $\mathbf{U}_3$ traces some stellar rotation harmonics very efficiently($P_{\rm rot}$,$\frac{P_{\rm rot}}{2}$,$\frac{P_{\rm rot}}{3}$,$\frac{P_{\rm rot}}{6}$ and $\frac{P_{\rm rot}}{7}$), the Bisector inverse Slope(BIS) has peaks at almost every harmonics ($P_{\rm rot}$,$\frac{P_{\rm rot}}{2}$,$\frac{P_{\rm rot}}{3}$,$\frac{P_{\rm rot}}{4}$,$\frac{P_{\rm rot}}{5}$, $\frac{P_{\rm rot}}{6}$ and $\frac{P_{\rm rot}}{7}$). In the BIS periodogram, the second harmonic ($\frac{P_{\rm rot}}{3}$ = 7.8d) has a significantly greater power than the original stellar rotation period ($P_{\rm rot}$ $\sim$ 23d).

\section{}
\subsection{Stacked BGLS periodogram- significance check}
\label{sec:sbgls_appendix}
The top panel shows the Stacked BGLS periodogram for the 2008-9 observing season \citep{2009A&A...506..303Q} with 106 observations, where the periods of four candidate signals are marked using arrows. As anticipated, the region around period $\sim$1 day appears to be crowded due to the presence of aliases of several signals, which makes it difficult to spot the comparatively weaker signal at 0.8549 days corresponding to the transiting planet. However, the zoomed-in plot clearly shows the stable nature of this signal. The 1-day alias of 0.854 days signal is clearly seen at 5.92 d, appearing to share power with the real signal. 

\begin{figure}
  \centering
   \includegraphics[width=1\columnwidth]{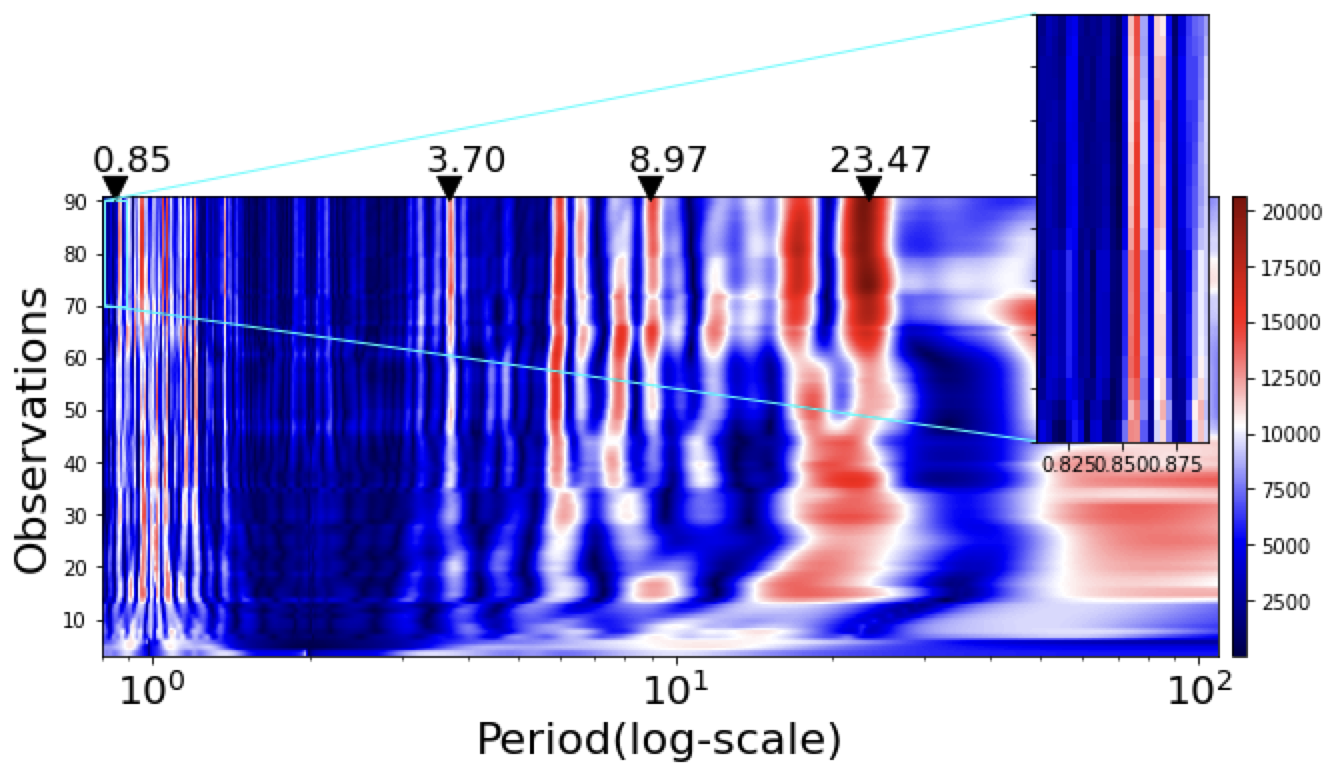}
    \includegraphics[width=1\columnwidth]{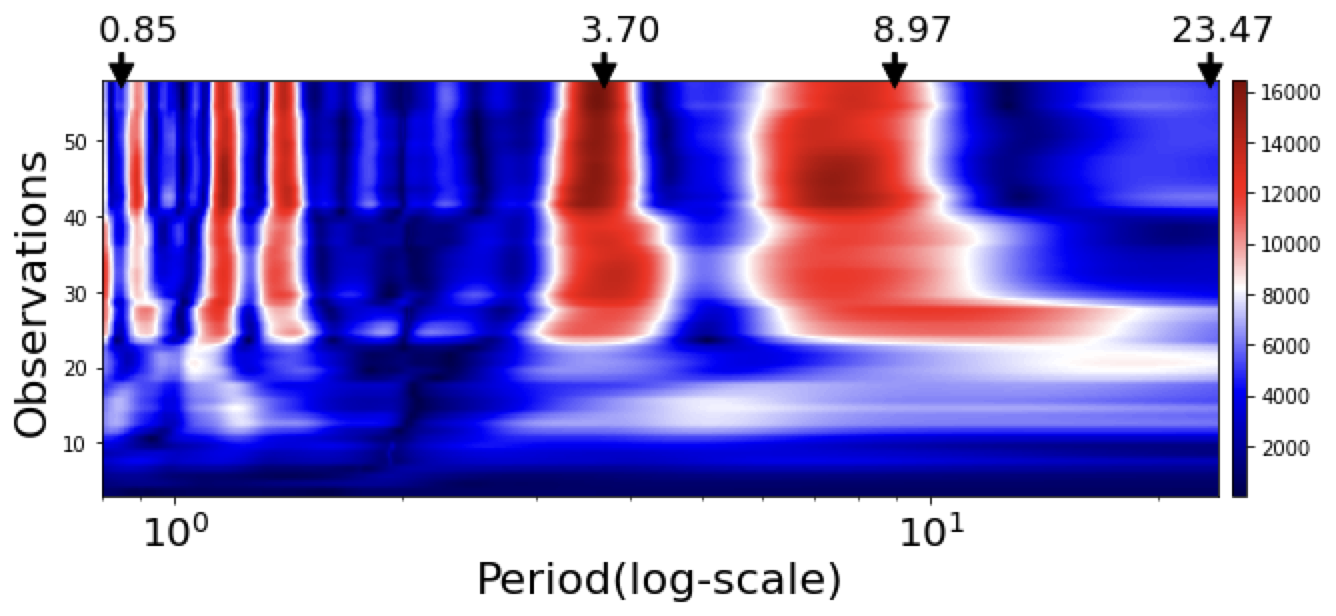}
  \caption{SBGLS periodograms for the 2008-9 and 2012 data subsets are given in the left and right panels respectively.}
  \label{fig:appsbgls}
\end{figure}

Another strong signal is present at 3.689 days which becomes apparent after about 30 observations, growing steadily with more observations (Figure \ref{fig:appsbgls}\&\ref{fig:snr}). The 8.97 days signal appears slightly different, with the probability passing through a short-lived maximum at $\sim$ 60$^{th}$ observation and then decaying briefly before resuming a monotonic increase thereafter, as expected for a real planetary signal. We expect to get a clear picture of this signal by analysing the entire data set. There is a broad strong signal around 23.47 days with the SNR showing strong fluctuations over the span of observations, which is not expected for a coherent signal. Its initial structure is also confusing with a single sparse signal bifurcating into two strong signals with periods 17.35 days and 23.47 d, where the 17.35 day signal can be explained as the beat period occurring from the 8.97 days signal and the 5.92 days alias of the transiting signal. 

The observations from 2012 \citep{2014MNRAS.443.2517H} returned a stacked BGLS periodogram, as shown in the bottom panel of Figure \ref{fig:appsbgls}. We noticed that the transit signal does not appear to have the same period and stability as seen from the previous figure. Instead, two other strong signals are visible at 0.78d and 0.88 days probably sharing the power from the real planet signal. We can explain the former and the latter ones as the ${|\frac{1}{P}\pm 1|}^{-1}$ aliasing of the 0.85d signal and the 8.97 days signal, respectively. Unsurprisingly, the 3.68 days signal again presents itself, with the strength steadily increasing towards the maximum number of observations. The broadly spread signal spanning from 6 to 13 days complicates the identification of the 8.97 days signal. The shorter observation baseline during the 2012 observing run probably is insufficient to capture the 23.47 days signal. 
\citet{2017A&A...601A.110M} also analysed the stacked BGLS periodograms for different subsets and for the entire CoRoT-7 data from both RV campaigns, with  particular interest in investigating the planetary significance of the $\sim$9 days signal. They reported the origin of the $\sim$9 days signal to be more likely from stellar activity than a third planet based on the unstable nature recorded in the original dataset \citep{2009A&A...506..303Q}, the follow-up campaign \citep{2014MNRAS.443.2517H} and in the combined data set. 
In the present analysis, we have the advantage of handling the {\sc scalpels}-treated data, providing us with less complicated, activity corrected,  RV information spanning over a longer baseline. Having effectively eliminated the short-term and long-term activity and trends, we are left with well-defined signals whose nature is less ambiguous. 

It is also worth noting that most of the aliasing and beat period signals that appeared in the individual data set analyses (Section \ref{sec:sbgls_appendix}) are no more obvious in the stacked BGLS periodogram for the entire data set.

\section{}
\subsection{KIMA posteriors}
\label{sec:kima_appendix}


\begin{figure}
    \centering
    \includegraphics[width=1\columnwidth]{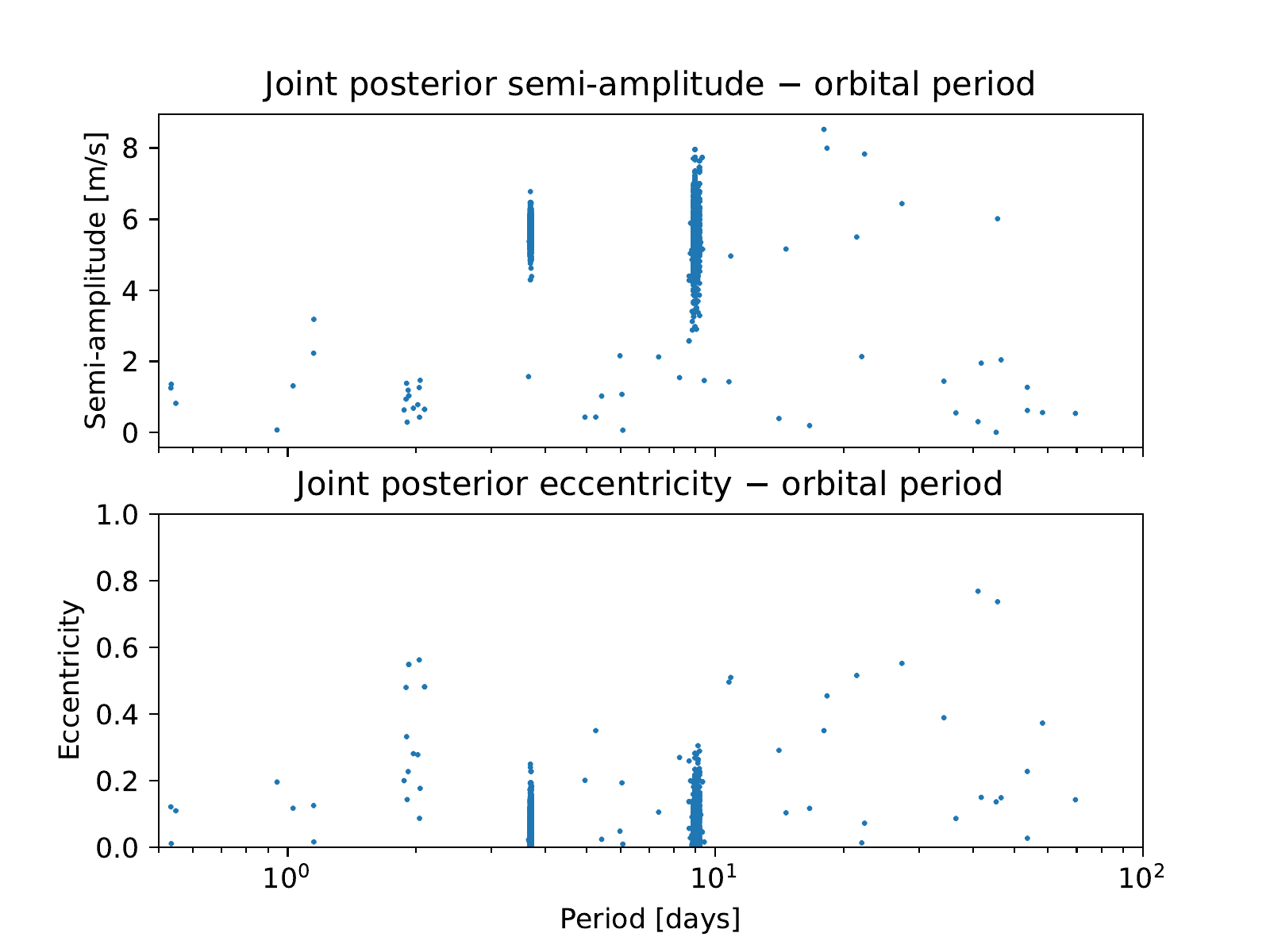}
    \caption{Joint posterior distribution for the semi-amplitudes and eccentricities along with the orbital periods in the x-axes}
    \label{fig:amp_ecc}
\end{figure}

\begin{figure}
    \centering
    \includegraphics[width=0.8\columnwidth]{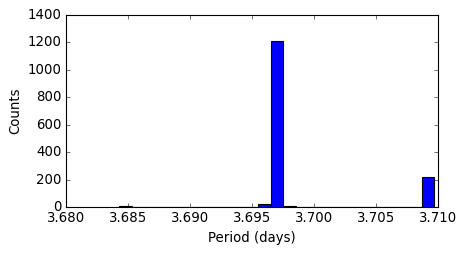}
    \includegraphics[width=0.78\columnwidth]{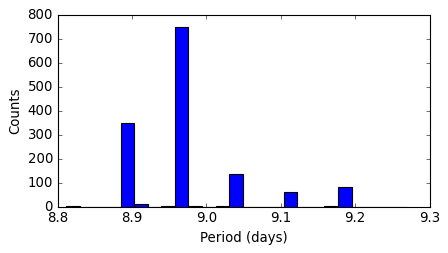}
    \caption{The FIP histograms showing the most probable periods for CoRoT-7c and CoRoT-7d from the corresponding interference patterns.}
    \label{fig:fip_beat}
\end{figure}


\begin{figure}
    \centering
    \includegraphics[width=1\columnwidth]{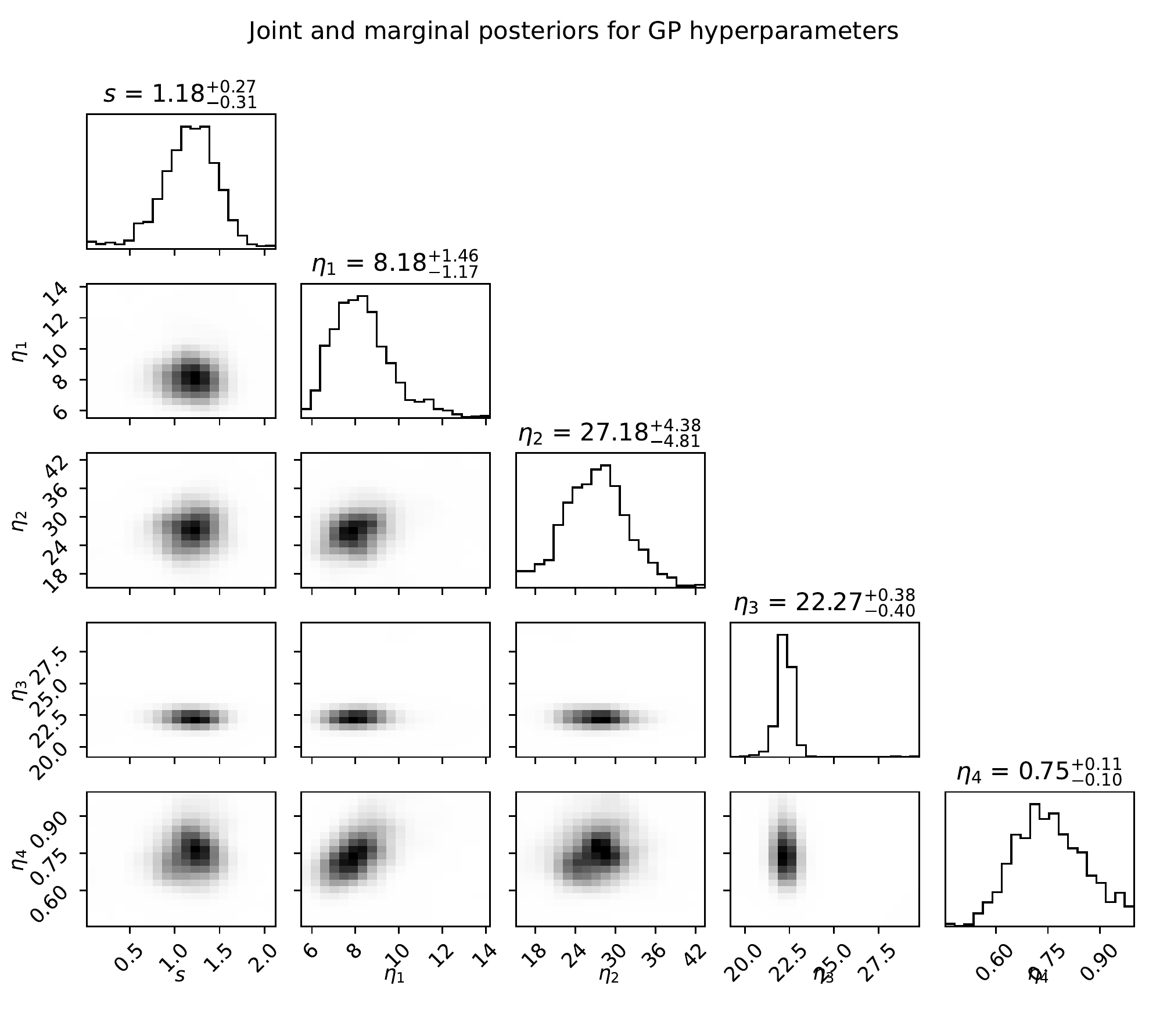}
    \caption{Posterior distributions for the GP parameters and the extra white noise. The samples for all values of Np were combined. The median of the posterior and the uncertainties calculated from the 16\% and 84\% quantiles are given in respective titles. }
    \label{fig:GP}
\end{figure}

\bsp	
\label{lastpage}
\end{document}